\documentclass[12pt,preprint]{aastex}
\usepackage{graphicx,epsfig,times}

\newcommand {\kms}{km s$^{-1}$}
\newcommand {\lya}{Ly$\alpha$}
\newcommand {\lyb}{Ly$\beta$}
\newcommand {\lyc}{Ly$\gamma$}
\newcommand {\lyd}{Ly$\delta$}
\newcommand {\lye}{Ly$\epsilon$}
\newcommand {\lyf}{Ly$\zeta$}
\newcommand {\lyg}{Ly$\eta$}

\shorttitle{Lyman-$\alpha$ Absorption in a Quasar Triplet}
\shortauthors{Petry et al.}

\begin{document}

\title{LYMAN-$\alpha$ ABSORPTION IN THE QUASAR TRIPLET Q0107$-$025A, B, 
Q0107$-$0232:\\ DATA CALIBRATION AND LINE SELECTION}

\author{C. E. Petry, C. D. Impey, and J. L. Fenton}
\affil{Steward Observatory, University of Arizona, 933 North Cherry Avenue, Tucson, AZ 85721}
\email{cpetry@as.arizona.edu, cimpey@as.arizona.edu, josie\_fenton@yahoo.co.uk}
\and
\author{C. B. Foltz\altaffilmark{1}}
\affil{MMT Observatory, University of Arizona, Tucson,
AZ 85721}
\email{cfoltz@as.arizona.edu}

\altaffiltext{1}{Currently at the National Science Foundation, cfoltz@nsf.gov}

\begin{abstract}

We present spectroscopic observations for quasars Q0107$-$025A, Q0107$-$025B, and Q0107$-$0232. For the first time, all data obtained with the {\it Hubble Space Telescope's} GHRS and FOS for these three quasars are combined.   Ly$\alpha$ absorbers selected from the spectra can be used to trace the coherence of the intergalactic medium (IGM) on scales of 1$-$2 $h_{70}^{-1}$ Mpc over the redshift range $0.4 < z_a < 0.9$. A new analysis of double and triple absorber coincidences derived from the improved spectra is performed. Automated line-selection algorithms were used to detect absorption lines in the QSO spectra; coincident absorbers were identified across two and three sight lines based on their proximity to each other in velocity space. A control simulation designed to generate artificial absorption-line spectra was used to gauge the significance of the coincidences. Four intervening metal line absorption systems are detected, with three of the four coincident to the A and B sight lines and one coincident across all three sight lines. This is evidence for substantial clustering among the objects responsible for the metal-enriched gas. By contrast, large scale coherence of the IGM gas is detectable but weak. Fewer than half of the absorbers are coincident on scales of 1$-$2 $h_{70}^{-1}$ Mpc, a result that is significant at the 99\% confidence level, and coherence arises preferentially in the higher column density gas. Triple absorber coincidences occur at a level that is significant at the 99\% confidence level, which indicates that the higher column density gas has a planar geometry.

\end{abstract}

\keywords{quasars: absorption lines - intergalactic medium - cosmology:  observations - quasars: individual (Q0107$-$025A,B, Q0107$-$0232)}

\section{INTRODUCTION}

The absorption observed in quasar spectra affords a great opportunity to study the dim and diffuse baryonic components of the universe. In particular, the use of low column density hydrogen absorbers of the Ly$\alpha$ forest has been a versatile cosmological tool \citep{rau98}. Given a bright enough background source and a combination of observations from the ground and space, the properties of the intergalactic medium (IGM) can be studied over 90\%$-$95\% of the Hubble time. Cosmological evolutionary theory and gas dynamical simulations indicate that Ly$\alpha$ absorbers are important because they represent most of the baryons in the universe at $z > 1$, and they trace the dark matter potential more reliably than galaxies \citep{cen94, zha95, her96, mir96, dav99}. The simplest physical picture of Ly$\alpha$ absorption is referred to as the fluctuating Gunn-Peterson approximation---the spectrum represents a continuous opacity map, where the IGM is highly ionized, and the \ion{H}{1} opacity scales as a power law with gas density. At low redshift, this approximation breaks down because a larger proportion of the absorbing gas is shocked as it migrates into collapsed structures.

The statistical properties of quasar absorbers are usually deduced from the combination of observations along single, widely separated sight lines. Working below a redshift of $z\sim1.6$ requires the spectrographs aboard the {\it Hubble Space Telescope}\footnotemark[2]\
({\it HST}; see papers from the {\it HST} Quasar Absorption Line Key Project, especially \cite{bah96, jan98, wey98}). Quasar pairs or groups can be used to measure the transverse size of the absorbers -- if coincident Ly$\alpha$ absorption is seen at a similar velocity in two or more lines of sight and the probability of such a line coincidence is small, then the transverse separation is a firm lower limit on the characteristic size of the cloud. Yet, guided by the insights of hydrodynamic simulations, it is clear that the idea of the Ly$\alpha$ absorbers as clouds with any simple geometry is simplistic. The absorbing gas traces the complex topology of the ``cosmic web,'' so absorber coincidences are best interpreted in terms of a generic coherence length scale rather than a characteristic size or radius.

\footnotetext[2]{Based on observations made with the NASA/ESA {\it Hubble Space Telescope}, obtained at the Space Telescope Science Institute, which is operated by the Association of Universities for Research in Astronomy, Inc., under NASA contract NAS 5-26555. These observations are associated with programs --- 5172, 5320, 6260, 6592, 6100, and 7752. Support for all programs was provided by NASA through a grant from the Space Telescope Science Institute.}

The first measurements of Ly$\alpha$ coherence used gravitational lenses as probes. With almost all the absorption features in common, only lower limits to the coherence length could be deduced \citep{sha82, sha83, wey83, fol84, sme92, sme95}. The first evidence of large absorber sizes came from studies of the quasar pair Q1343$+$264A,B \citep{din94, bec94}. Since then, a number of pairs have been observed to extend these experiments to larger transverse separations \citep{elo95, fan96, pet98, dod98, wil00} and lower redshifts \citep{din95, din98, mon99, ara02}. At low redshift, these experiments have all been limited by the relatively small number of absorption lines available for any analysis of coincidence. The strongest lines detected in these experiments correspond to high peaks in the \ion{H}{1} opacity; given sufficiently high signal-to-noise ratio (S/N) data, a more powerful analysis could use the full pixel information in the spectra. The addition of a third sight line to any quasar pair adds important information about the two-dimensional structure of absorbers in the plane of the sky \citep{cro98, you01}.

Spectroscopy of quasar groups is a first step towards three dimensional tomography of the intervening IGM. Maps of absorbers and their relation to large-scale structure have been attempted at high \citep{wil00, lis00} and low redshift \citep{imp99}, limited primarily by the low surface density of sufficiently bright quasars, which affects both the quality of data that can be obtained and the minimum transverse separation that can be probed. For Ly$\alpha$ absorbers at high redshift, the observational situation will improve as multi-object spectrographs are deployed on large telescopes, allowing better resolutions and S/Ns on faint targets. The prospects for low-redshift Ly$\alpha$ mapping will improve dramatically with the launch of the Cosmic Origins Spectrograph (COS) for the {\it HST} \citep{gre01}. Ly$\alpha$ spectroscopy can be compared directly with {\it N}-body simulations that incorporate gas dynamics \citep{kat96}. With a dense grid of sight lines, inversion methods have been developed to recover the three-dimensional topology of the underlying dark matter density field \citep{nus99, pic01, vie02}. The most efficient use of the data involves continuous flux statistics such as auto-correlation and cross-correlation power, but many {\it HST} spectra are in the regime of low resolution and modest S/N, where line-counting techniques are equally effective \citep{hui03}.

There are very few asterisms in the sky that can be used to probe transverse scales of $\sim1\ h^{-1}_{70}$ Mpc at $z < 2$. The pair Q0107$-$025A,B was first studied by \citet{din95, din97}, who used a maximum likelihood technique to infer characteristic absorber sizes of 700$-$1000 $h^{-1}_{70}$ Mpc at $z \sim0.7$ and a distribution of coincident and anti-coincident lines that indicated a nonspherical absorber geometry. \citet{you01} added observations of a third nearby quasar, Q0107$-$0232, creating baselines with up to 2.5 times the transverse separation of the original pair. The detection of a strong absorber coincidence in all three sight lines was good evidence of at least one sheet-like structure. 

In this paper we present the combination of spectroscopic data from {\it HST}'s Faint Object Spectrograph (FOS) and Goddard High Resolution Spectrograph (GHRS) for Q0107$-$025A,B and Q0107$-$0232, substantially boosting the sensitivity of the experiment, to measure absorber coherence. The chronology and details of the observations are presented in \S~2. In \S\S~3 and 4 the data calibration and reduction is discussed, with particular emphasis on the effort to derive the best possible wavelength calibration and registration. The process of line selection and identification is discussed in \S~5. In \S~6 we present samples and methods used to measure the absorber coherence, followed by a brief summary of the results. A direct comparison with extractions from simulations will be presented in a later paper.

\section{OBSERVATIONS}

The data presented in this paper were gathered by our group over three {\it HST} cycles, 6 yr, and six different GO programs. The observations included an acquisition failure, an instrument failure, and the longest ever time to completion for an {\it HST} cycle. The names of the members of the quasar triplet are presented in Table~\ref{tbl-1}, with their J2000.0 coordinates, magnitudes, and emission redshifts and the angular separations between the lines of sight of each quasar pair. Q0107$-$025A and Q0107$-$0232 were originally discovered in the Large Bright Quasar Survey (LBQS), they are also known as LBQS~0107$-$0235 and LBQS~0107$-$0232, respectively \citep{hew95}. It is noted that Q0107$-$025A is a FIRST radio source with a 4.85 GHz flux density of 122 mJy, while Q0107$-$025B is undetected by FIRST \citep{bec91}; thus, the two quasars with similar redshifts are definitely not gravitationally lensed, because lensing is achromatic and the radio flux ratio would have to match the optical flux ratio.  This distinction is important because, compared to a physical pair, a lens probes smaller transverse separations as the beams converge on the source. 

This is the fourth paper in a series by our group on this set of quasars, following \citet{din95}, \citet{din97}, and \citet{you01}. Table~\ref{tbl-2} gives a summary of all the {\it HST} observations that have been obtained to date. All data were obtained post-COSTAR. The first paper presented only the 1994 FOS G190H observations of the pair Q0107$-$025A, B \citep{din95}. The second paper added 1994 GHRS G140L observations and ground-based observations of the pair from the MMT \citep{din97}. The most recent paper extended the wavelength coverage of the pair Q0107$-$025A, B redward of Ly$\alpha$ with 1996 FOS G270H observations, and it presented 1997 FOS G190H observations of the third sight line, Q0107$-$0232 \citep{you01}, called component C in this paper. This recent paper also presented newer 1997 FOS G190H observations of Q0107$-$025A (but they were not combined with the earlier 1994 data) and adopted the line lists of \citet{din97} for Q0107$-$025B. \citet{you01} also developed a methodology for identifying line coincidences that does not make {\it a priori} assumptions about the velocity difference for a ``matched'' pair of lines.

In this paper we consider for the first time 1996 GHRS G140L observations of the pair Q0107$-$025A, B. The coaddition of this far UV data with earlier data more than triples the integration time on component A and more than doubles the integration time on component B. We also coadd the 1994 and 1997 observations from FOS G190H for Q0107$-$025A, doubling the integration time.  Two attempts were made to observe Q0107$-$025B with the Space Telescope Imaging Spectroscopic (STIS) MAMA G230L. The first attempt in 1999 produced no useful data, and in 2000, despite accumulating over 5 hr of integration time, the spectra were of insufficient quality to substantially improve on the existing FOS spectra, especially when issues associated with matching the different spectral resolutions are considered. Lastly, we present 1996 observations with FOS G190H of Q0107$-$0232. When combined with the published 1997 data \citep{you01}, the result is a doubling of the exposure time on this third sight line. No other quasar group has UV spectroscopy of comparable quality.

\section{REDUCTION AND CALIBRATION OF THE FOS DATA}

\subsection{Processing with CALFOS}

The FOS data were obtained from the Space Telescope-European Coordinating Facility (ST-ECF) archive and were re-calibrated using the Space Telescope Science Institute (STScI) standard pipeline processing package, CALFOS (ver. 3.4, 2002 January). The paper products relevant to each science observation were visually examined to ensure that the target had been properly acquired and was well positioned in the aperture throughout the observation. The paper products were also used to confirm that all of the CALFOS pipeline calibration steps had been successfully completed. For each science observation, CALFOS returns arrays of the calibrated wavelengths, fluxes, statistical errors, and data quality flags. The error arrays contain the statistical errors associated with the original data, which have been calibrated in an identical manner to the science data files. The raw data are in the form of photon counts, so the Poisson errors were estimated by calculating the square root of the raw counts per pixel. Zero error was assigned to pixels with zero raw counts. Systematic errors caused by sky and background subtraction or inappropriate flat fields and inverse sensitivity files are not included in these error estimates.

CALFOS uses data quality flag arrays to mark pixels that are known to be problematic and additional flags are added for problems detected in the calibration process. The data were flat-fielded with the default CALFOS calibration files. The default pipeline wavelength solution is a set of dispersion relations for each grating based on a single long exposure of the Pt-Cr-Ne comparison lamp. For the remainder of the paper, we refer to each of these calibration exposures as a ``super-wavecal.'' The rms error in the dispersion fit is approximately 0.16 pixels. The nonlinearity of the diode array is about 0.08 pixels. The line centroiding error for FOS arc lines is 0.08 pixels. These three numbers, combined in quadrature to give 0.20 pixels, represent the physical limit for the absolute accuracy of wavelength calibration for the FOS (Voit 1997). The accuracy of the relative wavelength scale is 0.1 diode (Keyes et al. 1995). Note that for the FOS Digicon cameras, one diode is equivalent to 4 pixels. The dispersions of the various gratings used in this project are listed in Table~\ref{tbl-2}.

All of the data used in this project were obtained with the red side of the FOS. The blue side of the FOS has been re-calibrated by the ST-ECF staff as part of an effort to create a post-operational archive for selected {\it HST} instruments \citep{kri91, key97, ale01}. By using physical modeling rather than the traditional empirical corrections, the new CALFOS pipeline for Blue side data corrects geomagnetically-induced image motion, electronic offsets in the detector, and temperature variations in the FOS optical bench. It also includes improved models for the grating dispersion relations, the flat field illumination, and the normalization of the dark current. The new CALFOS pipeline reduces the uncertainty in the wavelength zero point from 5 to $\sim$0.75 pixels, assuming oversampled data. Unfortunately, no red-side recalibration is available or planned, so we had to rely on empirical methods to achieve the best calibration. The discussion that follows is fairly extensive because several of the scientific goals of this experiment are constrained in practice by wavelength calibration.

\subsection{Wavelength Calibration}

One of the aims of this project is to measure the coherence length and geometry of Ly$\alpha$\ forest absorbers by searching for coincident (or common) absorption features among the three lines of sight in the triple quasar system. In order to assess whether an individual absorption feature is coherent across more than one line of sight, we need to know the degree of uncertainty in the zero point of the wavelength scale of each individual quasar spectrum. The size of this uncertainty limits the accuracy to which the velocity splitting across lines of sight can be measured and this is essential to gaining a deeper knowledge of the geometry and kinematics of the absorbers. The absorbing gas seems to be strikingly quiescent on megaparsec scales \citep{din95}, and we hope to tighten this constraint and then use it as a cosmological tool. This is possible because rms velocity differences on megaparsec scales represent the cumulative effect of gravitational collapse and so are sensitive to the cosmic matter density.  There are several sources of uncertainty in the relative zero points of the FOS wavelength scales, which we discuss, in turn, below. All spectra in this paper are calibrated to in vacuo wavelengths.

\subsubsection{Filter-Grating Wheel Non-Repeatability}

In principle, there can be a difference of as much as one full diode (4 pixels) between the zero point defined by a super-wavecal and the zero point appropriate to any individual science observation. This difference corresponds to $\sim$225~ \kms\ at the mid-point of the wavelength coverage of each of the gratings, where the bulk of this potential uncertainty is caused by non-repeatability in the positioning of the filter-grating wheel (Keyes 1998). To quantify this  uncertainty, the acquisition of some of the {\it HST} data was designed to include a contemporaneous calibration spectrum (or ``wavecal'') with each epoch of observations. This special calibration was obtained for each of the five visits of data taken with the G190H grating but not for the two data sets that were taken with the G270H grating. A zero-point offset was measured for each epoch of G190H data. Each offset was calculated by cross-correlating the appropriate wavecal with the super-wavecal, which forms an excellent template, since it is much better exposed than wavecals attached to individual science observations. The measured offsets and 1 $\sigma$ error are presented in Table~\ref{tbl-3}. These offsets were used to place all the G190H spectra on a common wavelength scale, with the heliocentric correction included.

\subsubsection{Cross-Calibration of G190H and G270H}

With no wavecals available for the G270H observations, a different procedure had to be used to register the G270H spectra to the G190H wavelength scale. There is an $\sim$100 \AA\ overlap region in the coverage of the two gratings. This overlap region was cross-correlated to find offsets between the spectra. The resulting offsets were applied to the G270H data and are presented in Table~\ref{tbl-4}. The cross-correlation offsets were checked in two ways. First, we compared the positions of the strongest lines in the overlap region. The difference in the measured line centers, as well as the 1 $\sigma$ error returned by the software, is given in Table~\ref{tbl-4}. In addition, we measured the positions of six Galactic lines in the G270H wavelength range. The first two techniques use only the blue end of the G270H spectra, so this third technique is a cross-check against systematic errors in the wavelength solution, which is a cubic. The six positions were compared to the wavelengths from \citet{mor88}. The intensity-weighted mean of the differences is shown in Table~\ref{tbl-4}, along with the rms scatter. We rejected the idea of using Ly$\alpha$-Ly$\beta$ pairs that spanned the two gratings as a tether, since the Ly$\beta$ lines are in a noisy and crowded region of the spectrum. We adopted the cross-correlation offsets as the most reliable, since they use the largest portion of the data, but all three techniques yielded the same results within the errors.

\subsubsection{Geomagnetically Induced Image Motion}

The FOS was affected by the changing geomagnetic field during an {\it HST} orbit which resulted in a time-dependent shift of the zeropoint in the spectrum.  This effect is known as geomagnetically induced image motion (GIM). At the time of launch and instrument verification, the 1 $\sigma$\ uncertainty in the zero point was quoted to be no more than 2 pixels (or 0.5 diodes). However, the more recent study showed that occasionally this error could be much larger \citep{ros98}. Since the GIM effect has a period of the {\it HST} orbit, we looked for it by phasing subexposures and averaging. No periodic structure was detected (this would have corresponded to a quarter sinusoidal wave, if observed). The measured shifts were scattered around a mean of zero (Table~\ref{tbl-5}). We used the same technique to search for a GIM effect over the timescale of a visit (a visit lasted anywhere from three to eight orbits). These results are presented in Table~\ref{tbl-6}. To summarize, no evidence for GIM on a timescale of hours was found in any data set. Thus, no shifts were applied to the individual sub-exposures within an observation. With this null result we put an rms upper bound on it of less than 20\% of the $\sim$215 \kms (1.39 \AA\ for G190H and 1.97 \AA\ for G270H) resolution of each of the gratings.

\subsubsection{Target Location in the Aperture}

All of the FOS observations presented in this paper were made with the 1 arcsec science aperture, which actually has a circular diameter of 0.86 arcsec. The location of the target in the FOS aperture affects the centering of the collimated beam on the disperser parallel to dispersion and hence the wavelength calibration accuracy. Wavecals are entirely internal to the FOS, so this procedure cannot correct for any wavelength shift due to target miscentering errors. Calibration data confirmed that jitter in the target position was always extremely small, in the range 0.003-0.005 arcsec, or 2$-$4 \kms. Objects Q0107$-$025A and B were acquired in BINARY mode using the same guide star, so the 1 $\sigma$ uncertainty of their relative positions in the 0.86 arcsec aperture is 0.12 arcsec, which converts to an uncertainty of 1.6 pixels (0.4 diodes) or $\sim$90~\kms\ (0.6 \AA) in the direction of dispersion (M. Rosa 2002, private communication). The third quasar, Q0107$-$0232, was acquired using a different guide star, and the 1 $\sigma$ uncertainty in its relative position in the aperture with respect to Q0107$-$025A or B might be significantly larger. 

Target centering contributes the largest single component of uncertainty to the error budget for aligning the wavelength zero points between the G190H spectra for the three quasars. As a result, it is highly desirable to have an empirical way to check and possibly adjust the zero-points. There are three Galactic lines that fall within the wavelength coverage of the G190H grating: \ion{Al}{2} $\lambda$1671, \ion{Al}{3} $\lambda$1854, and \ion{Al}{3} $\lambda$1863; this is also the order of decreasing oscillator strengths. Careful inspection of the spectra shows the following situation: In Q0107$-$025A, \ion{Al}{2} $\lambda$1671 is a very strong (10 $\sigma$) detection, there is no line detected within a resolution element of the expected position of \ion{Al}{3} $\lambda$1854, and a weak feature at the expected position of \ion{Al}{3} $\lambda$1863 is not a significant detection. In Q0107$-$025B, \ion{Al}{2} $\lambda$1671 is also a strong (8 $\sigma$) detection, there is no line detected within a resolution element of the expected position of \ion{Al}{3} $\lambda$1854, and there is a marginal 4 $\sigma$ detection of a line at the expected position of \ion{Al}{3} $\lambda$1863. In Q0107$-$0232, \ion{Al}{2} $\lambda$1671 is a moderately strong (5 $\sigma$) detection, there is no line detected within a resolution element of the expected position of \ion{Al}{3} $\lambda$1854, and there is a marginal 4 $\sigma$ detection of a line at the expected position of \ion{Al}{3} $\lambda$1863.

Thus, it would seem that the \ion{Al}{2} $\lambda$1671 line is of the most assistance in tying down the zero point. Unfortunately, for all three quasars, it is potentially blended to within 0.32 \AA\ (or a quarter of a resolution element) with the Ly$\gamma$ line corresponding to one of the strongest Ly$\alpha$\ lines in each of the spectra. Despite this misfortune, we can investigate whether there is any useful constraint. 

In Q0107$-$025A, \ion{Al}{2} $\lambda$1671 is clearly wider than the instrumental resolution and is better fitted in a $\chi^{2}$ sense by two components (the line-fitting software is described in \S 4). While we are confident that the feature is not a single unresolved line, the significance of lines in a multi-component fit places one of them below the significance threshold. The redward component centered at 1672.33 $\pm$ 0.33 \AA\ is 2.3 $\sigma$ from the predicted wavelength of Ly$\gamma$ and is within 2.7 $\sigma$ of the predicted strength. The agreement in strength is plausible given that we must rely on a statistical prediction based on composite quasar spectra, which is $W$(Ly$\gamma$)/$W$(Ly$\alpha$) = 0.35, close to the value anticipated from atomic physics \citep{pre93}. Within the blended feature, the blue component is centered at 1671.01 $\pm$ 0.14 \AA, which is only 0.22 \AA, or 40 \kms, from the predicted \ion{Al}{2} position at 1670.79 \AA. There is secondary support for the conclusion of undetectable zero-point shift from the weak \ion{Al}{3} $\lambda$1863 line. The measured feature is centered at 1862.68 $\pm$ 0.54 \AA, which is only 0.11 \AA, or 20 \kms, from the predicted \ion{Al}{3} position at 1862.79 \AA.

In Q0107$-$025B, the line at the expected position of \ion{Al}{2} $\lambda$1671 is not resolved, so if Ly$\gamma$ is present it must be completely blended with the Galactic line. The simple scaling given above predicts that about two-thirds of the strength of the feature should be Ly$\gamma$. If the central wavelength of the feature also reflects the centroid of the Galactic line, then at 1671.18 $\pm$ 0.11 \AA\ it is 0.39 \AA, or 70 \kms, from the predicted \ion{Al}{2} position. As with component A, the weak \ion{Al}{3} 1862 line gives confirmation of an undetectable zero-point shift, since the measured feature is centered at 1862.61 \AA, which is only 0.18 \AA, or 30 \kms, from the predicted \ion{Al}{3} position. For the two quasars that were acquired with the same guide star, we see no evidence for a significant zero-point shift.

The third quasar in the triplet, Q0107$-$0232, was acquired with a different guide star and has a potentially larger zero-point shift. The strong line at the expected position of \ion{Al}{2} $\lambda$1671 is not resolved. As with component B, any Ly$\gamma$ that is present would be blended with the Galactic line. The strength of the line is 1 $\sigma$ from the predicted strength of Ly$\gamma$, so, if present, the Galactic line may be weak. The line center is 1670.42 $\pm$ 0.16 \AA, 22 \kms\ redward of the predicted position of Ly$\gamma$ and 66 \kms\ blueward of \ion{Al}{2} $\lambda$1671. The weak \ion{Al}{3} $\lambda$ 1862 line may be present, but it is in a severely blended region. If the line at 1670.42 \AA\ is mostly \ion{Al}{2} and Ly$\gamma$ is unexpectedly weak, we can say that the zero-point shift is approximately 66 \kms. However, if the line at 1670.42 \AA\ is mostly Ly$\gamma$, then there is no constraint on the zero-point shift for this sight line.

\subsection{Combination of the Data}

Within each of the seven visits of FOS G190H and G270H data (see Table \ref{tbl-2}), all of the exposures were combined into a single spectrum. As no significant GIM was measured over the timescale of an individual visit, no shifts were applied to the exposures before combination. For each pixel, the flux values from the different exposures were combined into an error-weighted mean, and a propagated error on this mean was calculated. The error arrays used for this calculation are those generated by CALFOS. As mentioned earlier, a pixel that originally has zero raw counts is assigned an error value of zero. This value is carried through to the final CALFOS-generated error array. Thus, before the error-weighted mean could be calculated, these null error values were replaced by an interpolated value, corresponding to the mean of the two nearest pixels that had nonzero error values (or, for a pixel that was at either end of the spectrum, by the single nearest pixel that had nonzero error). 

During the combination process, the data quality flag arrays were also used to reject any pixel with a flag specifying that some uncertainty had not been indicated in the error array, with the exception of any pixel that was flagged as being in an ``intermittent noisy channel.'' 
Most pixels flagged as noisy were in continuum regions of the quasar spectra. The rare cases in which a real absorption line appeared to be contaminated by a data problem of this kind are noted with a comment in the absorption-line tables that follow; those features were excluded from any subsequent analysis. In order to reject any spurious pixels that had not been rejected by the data quality flag arrays, the following rejection algorithm was applied: pixels that deviated from the unweighted mean by more than 3.5 $\sigma$ were rejected. The unweighted mean of all the spectra being co-added was then recalculated, and any pixels that deviated from the new mean by more than 4 $\sigma$ were rejected from the calculation of the final combined spectrum. Thus, each of the seven visits produced one final spectrum plus a corresponding error array. 

\begin{figure}[hp]
\vskip 2truecm
\centering
\mbox{
\includegraphics[scale=.7, angle=270]{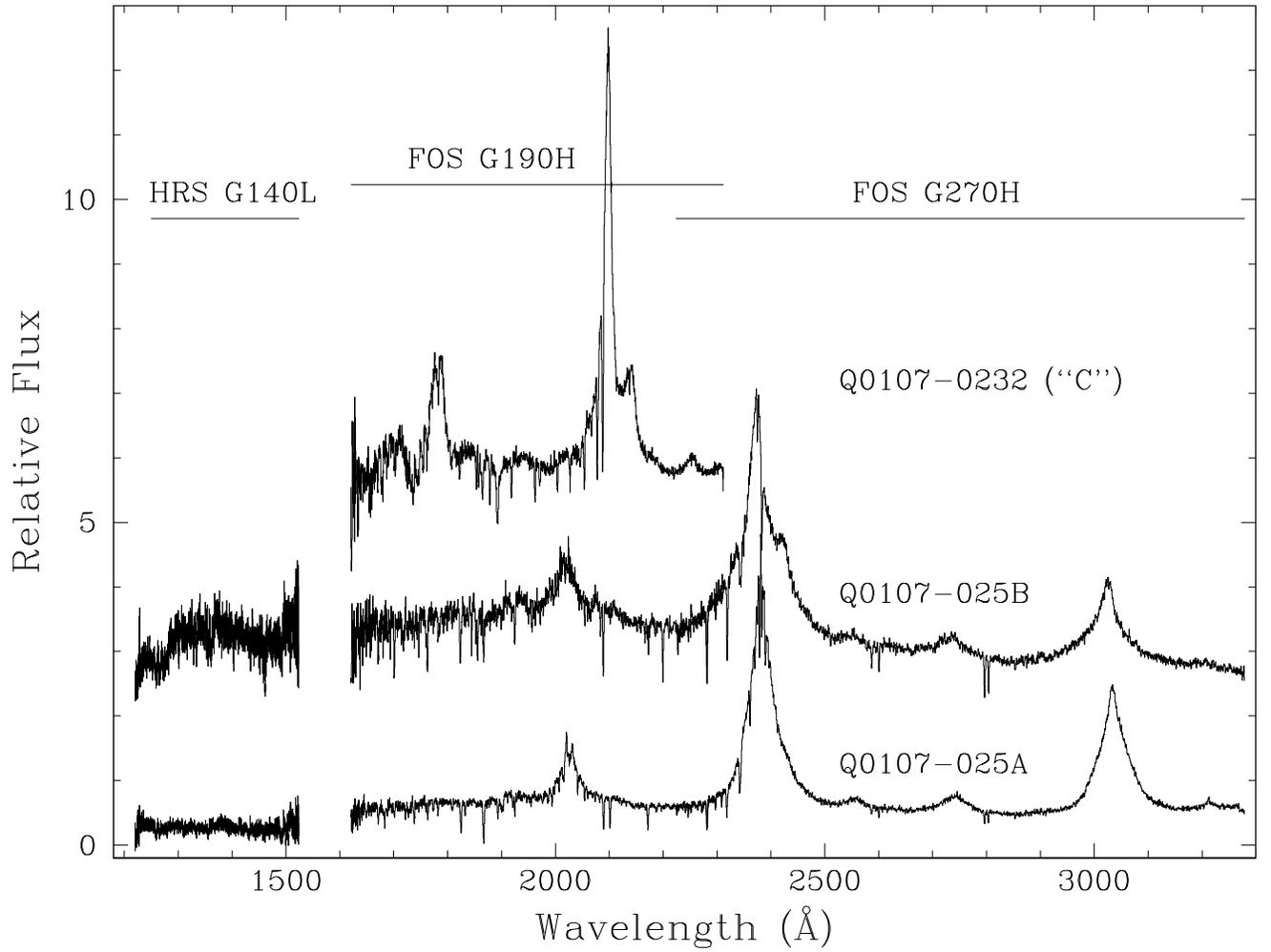}
}
\caption{
Final combined spectra for all instrumental setups for each quasar. The flux has been scaled arbitrarily. The wavelength coverage of each grating is noted by the labelled solid line, and where the G190H and G270H gratings overlap the data is simply overplotted, not combined.
}
\label{allspec}
\end{figure}

The most accurate wavelength-scale calibration for the entire data set is achieved by first applying the offsets determined by comparing the super-wavecal and the contemporaneous wavecals for the G190H data from Table \ref{tbl-3} and then registering wavelength scales for the G270H data to those of the G190H data by applying the offsets from Table \ref{tbl-4}.  All offsets are applied in an additive sense.  For Q0107$-$025A and Q0107$-$0232, there were two visits each of FOS data taken with the G190H grating. The two visits for each object were combined by applying the offsets from Table \ref{tbl-3} before the combination, to correct for epoch-to-epoch changes in the zero points of the wavelength scale. The spectra were combined in the same way as described above. After combination, there resulted five calibrated FOS spectra: G190H spectra for all three quasars and G270H spectra for Q0107$-$025A and B (see Figure \ref{allspec}).

 It is not possible to make a formal error budget, since the components of the error have not all been measured or constrained so they cannot, for example, be combined in quadrature.  The limit on wavelength accuracy of the zero point for both G190H and G270H data, based on the FOS calibration process, is 0.20 pixels, and the limit on the relative wavelength scales is 0.025 pixels. The offsets applied based on the reconciliation of superwavecals, which set the zeropoint, and wavecals that were contemporaneous with the observations had an rms of 0.083 pixels. The registration of the G270H data to the G190H data has an uncertainty that is higher, 0.33 pixels for Q0107$-$025A and 0.55 pixels for Q0107$-$025B. For GIM, all we can say is that the effect is not detected, and the rms bound on its effect on the wavelength scale is 0.77 pixels. Up to this point, a fair estimate of the zero point error would be 0.083 pixels, and a fair estimate of the relative error among lines measured in either grating would be 0.5 pixels. With a resolution element of approximately 3.9 pixels for each grating, only the strongest absorption lines would have their redshift uncertainty limited by the calibration and reduction process.

\section{REDUCTION AND CALIBRATION OF THE GHRS DATA}

Quasars Q0107$-$025A and B were both observed with {\it HST}'s GHRS with the G140L grating (2.0 arcsec aperture and ACCUM mode) in two different epochs, 1994 and 1996 (see Table~\ref{tbl-2}).  The data sets were retrieved from the Canadian Astronomy Data Centre (CADC) and were pipeline reduced and calibrated. Using IDL, the flux and error array for each subexposure of each data set were interpolated onto a linearized wavelength scale.  All data points with negative error values were identified and masked.  All of the subexposures were then summed, weighted by inverse variance, to produce the final spectra (see Figure~\ref{allspec}).

The Q0107$-$025A spectrum is the result of combining four GHRS G140L data sets observed on 1996 December 2, 1996 December 4, 1996 November 26, and 1994 November 15 (all post-COSTAR).  All four were originally observed by Weymann et al. (1998), the first three as part of program 6260 and the fourth as part of program 5172.  Exposure times for the data sets were 10445, 7507, 5222, and 9792 s, and each data set is comprised of 20, 24, 12, and 12 subexposures, respectively. The Q0107$-$025B spectrum is the result of combining two GHRS G140L data sets observed on 1994 September 22 and 1996 December 5 (both post-COSTAR).  Both were originally observed by Weymann et al., the first as part of program 5172 and the second as part of program 6260.  Exposure times for the data sets were 9792 and 10445 s, and the data sets were comprised of 20 and 24 subexposures, respectively.

The G140L data were reduced onto a standard wavelength scale by the pipeline created by CADC. In principle, however, this could still result in a slight offset with respect to the FOS wavelength calibration.  It is possible to register the GHRS and FOS wavelength scales empirically by using strong Galactic lines or seeking Ly$\alpha$-Ly$\beta$ pairs in which the Ly$\alpha$ line is in the FOS wavelength range and the Ly$\beta$ line is in the GHRS wavelength range.  However, in practice, the GHRS spectra are of low enough S/N that they are of limited use in the analysis, so this registration is not performed. 

\section{LINE MEASUREMENT AND IDENTIFICATION}

Continuous flux statistics are required to use the full information content of the spectra, but the traditional approach of fitting and counting lines is favored at low redshift, when the mean opacity is low, and in situations of low S/N. The condition for favoring continuous flux measurements is given by $(S/N)^2 > (n/f) (P_n/P_f)$, where $n$ is the mean line density, $f$ is the mean flux transmission, and $P_n$ and $P_f$ are the line and flux power spectrum amplitudes respectively (Hui 2003). The mean transmission and flux power spectrum amplitude at a fixed column density both decline toward lower redshifts, but the latter has not been measured at the redshift of the spectra presented here. If it is assumed that the product $n(P_n/P_f)$ stays constant, the criterion at $z \sim 0.7$ becomes $(S/N) \gtrsim 6$. Near the midpoint of the G190H data for the fainter quasar, S/N per pixel is about 9. Given the difficulties of fitting and normalizing the continuum for a flux analysis and the uncertainty in the evolution of $P_n$ and $P_f$, the simpler approach of line measurement is preferred in this paper.

\subsection{The Line Profiles}

The line-spread function for all instrumental setups, HRS G140L, FOS G190H and FOS G270H, are well characterized by a Gaussian profile.  For the FOS, the FWHM of the profile for the 1.0 aperture (diameter $=$ 0.86 arcsec) for a point source at 3400 \AA\ is 0.96 diodes (Keyes et al. 1995). This corresponds to 1.39 \AA\ for the G190H grating and 1.97 \AA\ for the G270H grating (using the dispersion values, G190H$=$1.45 \AA\ diode$^{-1}$ and G270H$=$2.05 \AA\ diode$^{-1}$, also given in Keyes et al.). To verify these numbers, the published electronic profiles of unresolved spectral lines are obtained for each of the gratings for the FOS red side.  These profiles are then fitted using the ANIMALS software (described in \S 5.2), and the fitted FWHM for each configuration is $1.41\pm0.03$ \AA\ ($\chi^{2}_{\nu} =0.66$) for the G190H grating and $1.99\pm0.05$ \AA\ ($\chi^{2}_{\nu}=0.61$) for the G270H grating, which is consistent with the published results. Although, in principle, the line-spread function should have Lorentzian wings, as well as a Gaussian core, in practice we find simple Gaussians to be a good fit to both the data and the published profile of an unresolved line.  For the GHRS, the line-spread function is a Gaussian distribution with FWHM $= 0.80$ \AA\ \citep{gil94,heap95}.

\subsection{ANIMALS}

The algorithms used to fit absorption features affect the resulting line lists in ways that can be subtle and complex. The velocity resolution of the {\it HST} observations is greater than the largest Doppler parameters measured at high spectral resolution for absorbers with these column densities (Penton, Stocke \& Shull 2002). In principle, this means that the absorption lines are all unresolved and should be well fitted by Gaussian profiles at the instrumental resolution. However, some low-redshift absorption features are genuinely broad even when observed at high resolution (Penton et al. 2004). Also, the effects of noise and continuum variations can lead the software to overfit complex regions of absorption, with a resulting overestimate in the line correlation amplitude. As a result, in this paper we favor an approach in which the Gaussian width is retained as a free parameter. Experience shows that this leads to a reliable identification of opacity ``peaks'' in the spectrum. 

We have developed software to fit the continuum flux of the quasar spectra and to fit Gaussian profiles to the regions of absorption. This code --- named ANIMALS, for ANalysis of the Intergalactic Medium via Absorption Lines Software --- is described in its original incarnation in \citet{cep98} and was designed to handle typical observed quasar spectra. Subsequently, in order to work with spectra extracted from hydrodynamic simulations, it was further refined and developed to be completely automated and the continuum fitting was performed using a slightly modified algorithm to accommodate the much shorter spectra \citep{cep02}.  Also, the algorithm to select and fit Gaussian profiles to absorption regions was expanded to handle data with a wider range in S/N and line density by allowing a greater number of components to be fitted in the final phase of the process and by allowing the FWHM of the profile to vary to accommodate lines having a width larger than instrumental resolution. ANIMALS comprises the interactive continuum-fitting procedure from \citet{cep98}, which was designed for observational data, and the more flexible line selection and fitting algorithm from \citet{cep02}. 

The software is used to find and fit regions of absorption with Gaussian profiles.  The wavelength of the line center, the amplitude, and the FWHM of each profile are allowed to vary.  The final ``best'' fit is chosen as described in \citet{cep02} as the fit with the lowest $\chi^{2}_{\nu}$, where $\chi^{2}_{\nu} <=100$ (in practice $\chi^{2}_{\nu} <3$) and the errors in the fitted equivalent width and the FWHM are smaller than the measurements themselves.  No restriction is made on the range of equivalent widths or on how close two lines may be fitted to one another, although criteria and limits can easily be applied if needed. Allowing the FWHM to vary results occasionally in lines narrower than the instrumental resolution, but this is always within the range anticipated due to the limits of the S/N. Visual inspection of the distributions of the $\chi^{2}_{\nu}$, equivalent widths, FWHM, and velocity separation of neighboring lines, as well as inspection of the fit overplotted on the spectra, confirms that the results of the fitting procedures fall within acceptable ranges.

\subsection{Line Selection}

The input parameters of the ANIMALS software are selected to include all regions of localized variation that might reasonably be attributed to \ion{H}{1} opacity. There is no completely reliable prescription for selecting absorption features, since the weak and broad absorption features can be mistaken for continuum variations, and this problem becomes worse when the S/N or resolution is low. With generous inputs to ANIMALS for defining an absorption region, twin significance parameters are used to select lines: one that relates to goodness of fit, and one that relates to the equivalent width of the feature. The goal of the line-selection process is to maximize the yield of real opacity features and to minimize the contamination from spurious variations, which might be caused either by real structure in the continuum or by difficulty in ascertaining the true value of the continuum. 

The strength of an absorption feature, or observed equivalent width $W_{obs}$, is fundamental to its ``significance,'' which we define in two ways. First, we define the detection significance as $S\sigma_{det}=W_{obs}/\sigma_{det}$, where $\sigma_{det}$ is related to the flux error (computationally, the convolution of the instrumental profile with the $1 \sigma$ flux error array) and compares the strength of a line to the detection limit of the data. Secondly, we define the fitted significance as $S\sigma_{fit}=W_{obs}/\sigma_{fit}$, where $\sigma_{fit}$ is the error in the equivalent width fitted by the software and is a measure of line reliability.  In practice, $S\sigma_{det}$ can be used to set a uniform detection threshold for absorbers and is directly related to the S/N, where $S\sigma_{fit}$ is an indicator of how well the equivalent width is determined: lines with large $S\sigma_{fit}$ might be found in noisy regions of the spectrum or in the wings of broader absorption features.  

We compare independent epochs of observation for Q0107$-$025A and Q0107$-$0232 to determine the criteria for selecting real absorption features. ANIMALS was used to fit two epochs of spectra for each object and the combined spectrum for each object. Features selected by the software were divided into three categories: (1) features that appeared in each epoch and in the combined spectra, (2) features that appeared in one epoch or the other and in the combined spectra, and (3) features that appeared in either of the two epochs but not in the combined spectra. Category 1 yields a list of features that are very likely to be real, category 2 yields a set of features that are consistent with being real, and category 3 yields a set of features that are very likely to be spurious. In addition, there were features that appeared in the combined spectrum for each object that did not appear in either of the independent epoch observations. These accounted for less than 10\% of the sample, and because they are mostly likely the weakest lines and cannot contribute to the discrimination between real and spurious lines, we do not present them here.

\begin{figure}[hp]
\vskip 2truecm
\centering
\mbox{
\includegraphics[scale=.8, angle=0]{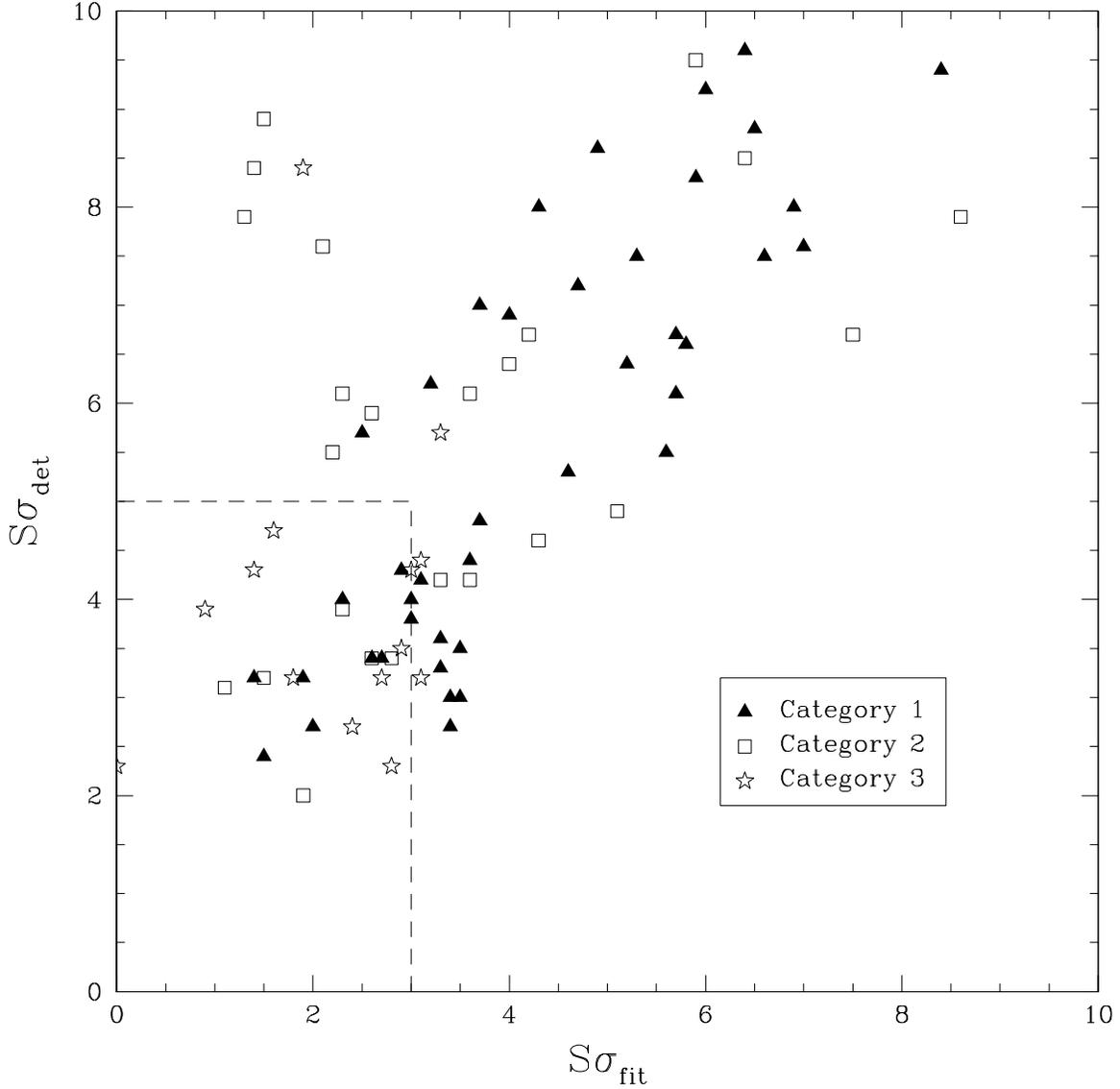}
}
\caption{
Detected significance, $S\sigma_{det}=W/\sigma_{det}$, vs. fitted significance, $S\sigma_{fit}=W/\sigma_{fit}$ for features detected with ANIMALS in the two independent epochs of observations for Q0107$-$025A and Q0107$-$0232.  The categories are as described in the text.  The triangles represent features that appear in each epoch and in the combined spectra, the squares represent features that appeared in one epoch or the other and in the combined spectra, and the stars represent features that appeared in either of the two epochs but not in the combined spectra.  The dashed line shows the criteria imposed to optimize the elimination of spurious features while retaining features that are most likely real.
}
\label{signif}
\end{figure}

Figure~\ref{signif} shows the fitted and detected significance of all features selected by ANIMALS in the G190H spectra for both epochs of Q0107$-$025A and Q0107$-$0232. Different symbols are used for the categories described above. There is a good correlation between fitted and detected significance overall, as expected. However, the correlation is not perfect; a feature can have high detected significance and low fitted significance if it is strong but poorly fitted by an unresolved line profile, and it can have high fitted significance and low detected significance if it is shallow but broad. 

The dashed lines in Figure~\ref{signif} show the thresholds that were used to define features that are included in the line lists and used in the subsequent analysis. Any features with $\sigma_{fit} < 3$ and $\sigma_{det} < 5$ are excluded. Among features that are very likely to be real (category 1), this threshold only excludes 8\% on average. On the other hand, among features that are very likely to be spurious (category 3), only 4\% on average are found with $\sigma_{fit} > 3$ and $\sigma_{det} > 5$. The same plot can be used to look at the effect of varying the thresholds. If a more conservative threshold of $\sigma_{fit} < 4$ and $\sigma_{det} < 7$ is used, real features are excluded at a faster rate than spurious features. Thus, in this work we assume that features with $\sigma_{fit} > 3$ and $\sigma_{det} > 5$ represent legitimate absorption features, and we infer from this ``reality check'' that the completeness is above 90\% and the spurious detection rate is under 5\%. The line lists for each object and each grating are presented in Tables~\ref{lineA140}-\ref{lineC190}. Expanded plots of each normalized spectrum are presented in Figures~3-5, and the lines fitted to each absorption region are overplotted, using the numbering scheme from   Tables~\ref{lineA140}-\ref{lineC190}.

\subsection{Line Identification}

Having located significant regions of absorption or opacity in the spectra of the three quasars, the next step is to identify the atomic transition responsible for each feature or line. The overall goal is a study of the IGM as traced by the Ly$\alpha$ transition. However, the UV spectra also contain transitions that act as contaminants in this experiment: highly ionized metal lines from gas in the halo of the Milky Way, metal lines associated with intervening absorbers (previously known or newly discovered), and higher order Lyman transitions from the IGM. Each of these must be dealt with in turn. 

In practice, the line identification process is complex and nonunique. Given good-quality UV spectra, the observed line density is high. Even though lines can be centroided accurately, there are many possible high-excitation metal-line transitions, giving a significant probability of chance identifications when wavelength matching is the only criterion. Line strength can also be used to evaluate an identification, predicted according to oscillator strength, but the metallicity and excitation conditions of enriched gas along particular sight lines in ISM or IGM gas are very uncertain, so anticipated line strength cannot always be used in a prescriptive way. In general, we follow the methods established by the {\it HST} Absorption Line Key Project \citep{bah96}, as summarized and applied in our earlier work on absorbers in the direction of the Virgo Cluster \citep{imp99}.

\subsubsection{Galactic Lines}

The first step in identifying metal lines is to search for transitions that arise from absorption in the halo of the Milky Way galaxy, so their redshift is essentially zero. Our comparison list of metal lines for this search is taken from \citet{imp99}. Any line that matches a transition in the Galactic line comparison list within $\pm 0.37$~\AA\, or 2.5 times the mean error in the determination of line central wavelengths is considered a possible identification. The result of this process yielded five, one, and nine possible Galactic line identifications in the G140L, G190H, and G270H gratings, respectively.  As a control experiment to test for chance coincidences, the wavelengths of the Galactic line comparison list were shifted by 10 \AA\ blueward and redward. The mean number of line identifications was one, zero, and one in the G140L, G190H, and G270H gratings. On purely statistical grounds, therefore, we expect that most of the possible G270H identifications are real, a few of the G140L identifications are real, and the single G190H identification may or may not be real.

In Tables~\ref{lineA140}-\ref{lineC190} the suggested identifications are based purely on the matches that fall within the 2.5$\sigma$ search window, with two exceptions. First, in the spectrum for component C taken with the G190H grating, the \ion{Al}{2} $\lambda$1670 line meets the criterion for identification as a Galactic metal line (see discussion in \S 3.2.4).  Visual inspection of the spectra for the A and B components shows a strong line at this wavelength, but the residuals of 0.49 and 0.40 \AA, respectively, exceed the match criterion. We claim these as valid identifications because these lines are strong and unblended and are thus most likely to be Galactic lines. Second, in the G270H spectrum for component B, the strong \ion{Mg}{2} doublet has a large systematic residual of 0.39 \AA\ such that five additional \ion{Fe}{2} lines are not identified. Visual inspection of the spectrum, comparison with the G270H spectrum for component A, and the fact that these lines are in the low line density region redward of Ly$\alpha$ argue that the \ion{Fe}{2} lines are also valid identifications. All identifications of the Galactic lines are listed in Tables~\ref{lineA140}-\ref{lineC190}.

\subsubsection{Metal-Line Systems}

The Ly$\alpha$ forest will, in general, be ``polluted'' by metals associated with intervening high column density absorbers. These absorbers are often discovered by searching for strong \ion{Mg}{2} and \ion{C}{4} doublets redward of the Ly$\alpha$ emission line. The rationale for excluding these metal-line systems from a sample of absorbers designed to study the IGM is the fact that absorbers selected by the \ion{Mg}{2} or \ion{C}{4} doublets are high-density systems statistically associated with the halos of galaxies.  This metal-enriched halo gas is distinct from the far less enriched gas of the IGM.  The latter may reveal metals when observed with sufficient resolution and sensitivity, but those weak metal lines do not represent ``contamination'' in this particular experiment. 

In searching for additional lines associated with known metal systems, we are guided by the earlier work of our group. \citet{din95,din97} used G140L and G190H data to identify three intervening metal-line systems, which we use as a starting point to identify the absorbers and re-measure the redshifts for these systems.  As with the Galactic lines, a search window of 2.5 $\sigma$ relative to the mean line centroid error is used for a match. We re-measure the redshifts for these systems, and in some cases identify additional lines associated with them.  Table~\ref{linemetal} is a compilation of these absorption systems and it illustrates common absorption across the three lines of sight. The Table also lists the recomputed redshift and number of lines used in each spectrum to calculate the redshift, which is an average weighted by the strength of the line.  We used the procedure for line identification from \S 3.2 of \citet{imp99}, including the rules for how to prioritize a particular identification when multiple identifications are made.  This results in some slight differences in the identifications we make compared to \citet{din97}.

The higher S/N of this combined data set offers the possibility that new metal-line systems might be identified. Redward of Ly$\alpha$ emission, we see no sign of new \ion{C}{4} or \ion{Mg}{2} doublets that might place higher excitations in the Ly$\alpha$ forest. Blueward of Ly$\alpha$ emission, there is the possibility of discovering low-redshift \ion{C}{4} doublets and associated metal species. However, the yield on a blind search and the reliability of possible identifications are expected to be low for the reasons discussed earlier in the description of the Galactic line search. Therefore, we did not carry out a search for new metal-line systems. 

\subsubsection{Higher Order Lyman Series}

Over the full wavelength range of the data presented here, higher order Lyman transitions are mixed in with the Ly$\alpha$ lines of interest. One strategy for keeping the identifications clean is simply to avoid doing any analysis that extends into the Ly$\beta$ forest. However, the fact that Q0107$-$0232 differs substantially in redshift from the other two quasars motivates the use of the full wavelength coverage. Therefore, we apply a statistical method to estimate the contamination of the data with higher order lines, recognizing that the reliability of the individual identifications may be low. 

We perform a check on the level of contamination by unidentified \lyb\ lines by comparing the computed $dN/dz$ in the \lyb\ forest region with the $dN/dz$ from the \lya-only forest region. Since the cosmic evolution of \lya\ is negligible across this wavelength or redshift range, there should be no trend if, on average, unrecognized \lyb\ lines are not present. For all components, the values of $dN/dz$ agree to within the 1.5 $\sigma$ statistical counting errors. This result is anticipated, since none of the spectra are deep enough to readily detect higher order lines in the Lyman series. 

\section{COHERENCE MEASUREMENT}\label{ironzero} 

Line matching across paired lines of sight is a simple way to detect common structures or coherence on scales typical of the transverse separation. The null hypothesis in all these tests is that the lines are randomly distributed in redshift (zero autocorrelation) and that coincidences are random across two or more sight lines (zero cross-correlation). If absorbing structures have a scale length much larger than the transverse separation of two sight lines, the coincidence rate might still be low if the structures are very elongated or filamentary. Line-matching statistics are not well suited to determining the geometry of absorbing structures, but they can be used to sensitively test for departures from randomness. Coincidences across three sight lines strongly point to a planar structure if the random probability of a triple coincidence is low. Hydrodynamic simulations of the cosmic web lead to the expectation of a mixture of filaments and sheets with a complex spatial geometry. 

\subsection{The Samples}

The complete sample of \lya\ features is defined as all the \lya\ and unidentified features blueward of the peak of \lya\ emission for each quasar. For the analysis, the primary sample consists of all \lya\ and unidentified features in the wavelength range common to all three lines of sight, which is 1620.60$-$2089.85 \AA. To account for the proximity effect, two lines are excluded for component A and one line for component C because they fall within 1200 \kms\ of the \lya\ emission peak. Since component C has a substantially lower redshift than either of the other quasars, a secondary sample is formed from the \lya\ path length that overlaps between components A and B, which spans $\lambda\lambda$ 1620.60-2368.33 \AA. This range includes a 1200 \kms\ allowance for the proximity effect, but no lines are excluded from the sample as a result of this.  The upper bound on this range stretches into the G270H data, and in the overlap region between G190H and G270H the identified lines are essentially identical, with one line found only in the G190H spectra of component A being close to the detection limit.  Since the errors in equivalent width for the G190H data are smaller than for the G270H data, and subsequently the detection and fitted significances are higher, we choose to use the measured lines from the G190H data and add only the lines from the G270H data that extend in wavelength past the end of the G190H data. As mentioned previously, the GHRS data for components A and B are of too poor quality to warrant including them in the line matching. The actual lines comprising each sample are footnoted in Tables~\ref{lineA140}-\ref{lineC190}.

\subsection{Matching Methods}

We form pairs of \lya\ absorbers using two different methods among the three quasar sight lines. A ``nearest neighbor'' pair is formed for each \lya\ absorber by locating the absorber closest in velocity in the adjacent sight line. This means that there are as many absorber pairs as there are individual \lya\ absorbers, and although there will be some double counting of pairs, this method has the advantage of matching multiple components of a complex region of absorption with a single feature in the adjacent sight line. A ``symmetric'' pair requires that the match is reversible: the nearest counterpart in velocity  going from sight line A to B is the same as the nearest counterpart in velocity going from sight line B to A. By definition, symmetric matches form a subset of nearest neighbor matches.  Both methods were investigated because they have different implications for the ability to detect spatial coherence. Nearest neighbor matching ensures that all absorbers in each spectrum have a partner, giving maximum statistics but at the expense of double-counting some absorber pairs. Symmetric matching ensures that coincidences are not driven by the sight line with superior depth of data, but the statistics are poorer because a number of absorbers will be orphaned in the matching process. 

We form triples of \lya\ absorbers using the two methods as for the pairs, but extended for use with all three lines of sight. A nearest neighbor triple is formed for each \lya\ absorber by locating the absorber closest in velocity in an adjacent line of sight, and then the absorber closest in velocity in the third line of sight is located for this second absorber, forming a triple. Again, there are as many absorber triples as there are individual \lya\ absorbers. A symmetric triple is identified if the nearest counterpart in velocity forms a closed loop going from sight line A to B, then B to C, then C to A (i.e. as with symmetric pairs, the matching is commutative).  These symmetric triples, by definition, form a subset of the nearest neighbor triples.

The description of each of these methods is purely algorithmic, with no justification given so far that matches will represent intersections of each sight line with a single, contiguous cosmic structure. The largest velocity difference anticipated for a true physical match is a few hundred km s$^{-1}$, since that is the shear that might be imprinted by the peculiar velocity of cosmic structures spanning $\sim 1$ Mpc. In practice, the two techniques yielded similar results. Symmetric matching is more reliable because a line match represents a physically coherent structure and the procedure avoids double counting. We present pairs defined this way in Tables \ref{xppairsAB} -- \ref{xppairsBC} and triples in Table \ref{xptriples}. The mean and rms velocity differences for absorber pairs are $14 \pm 560$ km~s$^{-1}$ for AB (15 lines), $-48 \pm 534$ km~s$^{-1}$ for AC (19 lines), and $29 \pm 409$ km~s$^{-1}$ for BC (18 lines). The analogous numbers for the 12 symmetric triple matches are $102 \pm 314$ km~s$^{-1}$ for AB, $56 \pm 372$ km~s$^{-1}$ for AC, and $-158 \pm 555$ km~s$^{-1}$ for BC. For all important statistical tests, we perform calculations and present results using both types of matching technique.

These matching procedures give an independent cross-check on the zero-point wavelength error discussed in \S 3.2. One potential cross-check involved Galactic metal lines and was described in \S 5.4.1. Unfortunately, the small number of species in each sight line and the ambiguity in their identifications precluded them from being used to draw any conclusion on zero-point shifts. In \S 3.2.4, we noted that target placement in the aperture was the largest contributor to zero-point error. For A and B, acquired with the same guide star, the centering uncertainty corresponded to 90 km~s$^{-1}$. For C relative to A or B, the uncertainty might have been as high as 300-400 km~s$^{-1}$, since C was acquired with a different guide star. Mean velocity offsets for symmetric matches from the paired and triple sight lines quoted above are all much smaller than the rms velocity differences. This test, using the IGM itself, shows that there are no large, unrecognized systematic errors in the velocity zero point. 

\subsection{Monte Carlo Simulations}

To detect correlations in the \lya\ forest, it is necessary to generate a randomly distributed parent population of absorbers.  Rather than generate synthetic spectra, we generate lists of absorption lines and subject them to a suitable selection function.  We then form pair and triple line matches from this simulated population of lines. For each realization of the data, we select a number of lines per line of sight from a Poisson distribution, with the observed number as the mean of that distribution.  The central wavelength of an absorption line is randomly placed in redshift space, and line strength is drawn from a power-law distribution in equivalent width with fitted coefficients as measured for the Quasar Absorption Line Key Project by  \citet{wey98}. 

The minimum allowed equivalent width is set by the sensitivity of the observed data to line strength. For the observed data, this can be characterized by the detection limit, which is the 1 $\sigma$ error in the flux as a function of wavelength convolved with the instrumental profile.  For the Monte Carlo simulation, we approximate this limit on the detectable equivalent width for each line of sight by computing the average of the detection limits at the wavelength of each observed absorber.  Three times this value is used as the flat detection threshold for the simulated lines of sight.

In forming the observed line lists we use a joint criterion to define a ``real'' or ``reliable'' line.  For the simulation there is no error in a fit with which to form the fitted significance, $S_{fit}$. However, the detection significance $S_{det}$ is a true reflection of the $1 \sigma$ limiting equivalent width, and from Figure \ref{signif} it can be seen that if we choose $S_{det}=3$, we include most of the real lines (category 1). Therefore, we apply this value to decide to keep or reject the line. To simulate the paired matches, we create 10,000 realizations of each of the three lines of sight. To simulate the triple matches, we create an additional 1000 realizations of each of the three lines of sight, from which the nearest neighbor and symmetric triples are formed.

\subsection{Absorber Matches in Paired Sight Lines}

Comparing the lists of symmetrically paired lines to the lists presented by \citet{you01}, we ``rediscover'' all of Young's pairs, except in one case in which a feature at 2018.5 \AA\ in the B line of sight is culled from the final line list because it does not meet the significance criterion. With deeper data we have increased the number of pairs by a factor of 2. We present the counts of \lya\ nearest neighbor absorber pairs for each pair of quasar sight lines in Figure \ref{nnmonte}. The expected mean number of absorber pairs for a random distribution in each sight line is shown by the thin solid line through the data. The three upper curves show the random pair counts in each bin that include 90\%, 95\%, and 99\% of the 10,000 Monte Carlo realizations. Thus, these curves are confidence intervals on departures from random behavior.

No strong clustering signal is seen in Figure~\ref{nnmonte}. There is a slight excess of absorber pairs at small $|\Delta v|$ in A-B relative to random, but only for B-C does it rise above the 95\% confidence interval in any bin. Transverse scales of 1$-$2 $h_{70}^{-1}$ Mpc correspond to groups or poor clusters, so line-of-sight power would show up as excess pairs of velocity scales of 100$-$500 \kms, but much of this power would not be resolved with data of this spectral resolution.  For the transverse measurement, all three pairs of sight lines show an excess of small velocity separation absorber matches, but in each case the signal is of marginal significance. To gain more sensitivity, we combine the three sight line pairings in Figure 7, giving both the symmetric and nearest neighbor absorber matching algorithms to show that they agree. The signals are combined by addition, and the errors are combined in quadrature.  

If the absorbers were in coherent two-dimensional structures that are larger than the largest transverse separation of the three quasars, then it would be inappropriate to combine an absorber pair count from the three baselines of the triangle, since they would be different measurements of the same structure. Since cosmic structures sampled in neutral hydrogen are likely to be inhomogeneous, the absorber correlation would be strong but not perfect. However, it is justified to combine the signals as independent either if the absorbers are not generally in large coherent structures, or if they are in large coherent structures but the structures are filamentary and do not generally cover more than one sight line. In Figure~\ref{allpairs}, which can only be sensibly interpreted if all coherence is on scales less than 1$-$2 $h_{70}^{-1}$ Mpc, an excess of absorber matches is detected above the 99\% confidence level on 100$-$200 km s$^{-1}$ scales, typical of low-density regions or poor groups of galaxies. 

Genuine absorber coherence shows up as an excess of paired absorbers with small velocity differences, so we reduce the ``noise'' of random matches by using different criteria for the velocity difference than we require for a match. Table \ref{prob} lists the probability that the number of observed line pairs is random for three different matching criteria in $|\Delta v|$ and for all three quasar pairings. The probability is the number of times that equal to or greater than the number of observed absorber pairs was found in the random experiment, divided by 10,000. The detection of nonrandom behavior or coherence is significant at the 96\% confidence level across the A and B sight lines, and is significant at the 99\% level across the B and C sight lines. The absorber pairs across the sight lines with the largest transverse separation, A and C, show no excess above random. 

The marginal detection of absorber coherence across two quasar-quasar spans, despite improved data and superior absorber statistics compared to those of \citet{you01}, suggests that the coherence signal lies preferentially with the stronger absorbers. To test this hypothesis, we must include absorber equivalent width as a parameter in the analysis, since the more abundant weak lines will produce a larger number of random matches at small velocity splittings. Figure~\ref{nnslice} presents a plot of the average rest equivalent width for a \lya\ pair versus its velocity separation for both the observed absorber pairs ({\it left panels}) and absorber pairs drawn from the random Monte Carlo experiment ({\it right panels}). There is a visual suggestion that the strongest lines have preferentially smaller velocity separations.

We test the hypothesis that the stronger absorbers have smaller velocity matches than random using the nearest neighbor matches, which have slightly greater statistical power. Figure~\ref{nnstrong} is a histogram of the distributions of velocity separations of the strongest absorbers as measured by their rest equivalent widths, $W_{rest}\ge 0.6$ \AA. We test the hypothesis in a second way in Figure~\ref{nnclose}, with a histogram of the equivalent widths of all absorbers with $|\Delta v| < 400$ \kms, splittings that might be caused by physical coherence. The overplotted comparisons in both figures are the distributions from the Monte Carlo experiment. We test for deviations of the data from a random expectation with a one-sided K-S test, which has good power and sensitivity for the small data samples used here. 

The results show that symmetric strong absorber matches are not drawn from a distribution of velocity differences appropriate to random line placement at a high confidence level. The K-S test yields probabilities of 0.003\% (AB), 0.3\% (AC), 0.2\% (BC), and $<10^{-5}$\% (all) that the observed matches are drawn from a random distribution, in the sense that symmetric matches have smaller values of $|\Delta v|$. The average velocity difference is $\sim 200$ km~s$^{-1}$ across a 1$-$2 $h_{70}$ Mpc transverse separation (Figure 9), as is appropriate for sampling a low velocity dispersion cosmic environment. The second test uses a cut on $|\Delta v|$ to select the absorbers most likely to be part of coherent structures and asks whether they tend to be stronger or weaker absorbers on average (Figure 10). The K-S test yields probabilities of 1.5\% (AB), 0.2\% (AC), 0.6\% (BC), and 0.02\% (all) that the observed matches are drawn from the equivalent width distribution that would arise from random absorber placement. Taken together, the results imply that higher column density absorbers are responsible for the relatively weak coherence signal.

With experiments limited by small number statistics, it is advisable to address significance in a different way, by asking whether the observed absorber matches could arise as a statistical fluctuation. Table 19 uses 10,000 Monte Carlo simulations of the absorbers to ask how many times random absorber placement results in a number of absorber matches that equals or exceeds the observed number. The resulting probabilities are presented for both methods of line matching, for all quasar pairings, and for three different cuts in $|\Delta v|$. Only A to C, the largest transverse separation quasar pair, fails to yield a significant result in any case.

\subsection{Absorber Matches across the Triple Sight Lines}

The triple coincidences give qualitatively new information, because any triple absorber matches that would not be expected by chance must be due to a coherent structure that spans all three sight lines of the quasar asterism. This would indicate a planar cosmic structure. The most direct way to interpret the triple matches is to ask how likely it is that the observed number would occur by chance, given the method of matching. Figure \ref{xptriple12} shows the number of symmetric triple absorber matches seen in 1000 Monte Carlo realizations of the experiment. The mean number found in the control experiment was 5, and the observed number of 12 did not occur once in 1000 realizations. The triple coincident absorbers are therefore significant at a 99.9\% confidence level or greater. The best physical explanation for this result is multiple planar structures along the line of sight to all three quasars.

To attempt to go beyond the counting experiment risks diminishing returns from small number statistics, but we run one further test using a cut of $|\Delta v| < 400$ \kms\, since this corresponds to a velocity match criterion that sharpens the detection of coincident absorber pairs. Figure \ref{xptriple400} shows the strength of triple-matches with $|\Delta v| < 400$ \kms\ relative to a random control experiment. With Kolmogorov-Smirnov (K-S) significance of 93\% (symmetric matching) or 99\% (nearest neighbor matching), the triple matched absorbers are stronger, on average. The interpretation is that planar intervening structures are best traced by the higher column density absorbing gas. 

\subsection{Extended Wavelength Coverage of Q0107$-$025A, B}

We increase the absorber statistics by extending the wavelength coverage of components A and B. This secondary sample of Ly$\alpha$ absorbers is described in \S6.1. The number of absorbers increases from 32 to 50 for component A and from 20 to 36 for component B.  The number of nearest neighbor pairs increases from 52 to 86, and the number of symmetric matches increases from 15 to 25. Figure \ref{ABpair} shows the distribution of the symmetric and nearest neighbor pairs with velocity separation.  As before, variation around the random pair counts at the 90\%, 95\%, and 99\% confidence intervals was determined using a Monte Carlo simulation of 10,000 realizations, where the number of absorbers per sight line in each realization is a Poisson deviate of the observed number of absorbers. With the increased wavelength coverage of this sample, the significance of the clustering in both pairing methods is at the 95\% confidence level. 

\section{SUMMARY} 

We present new observations of the quasar triplet Q0107$-$025A, Q0107$-$025B, and Q0107$-$0232 (C) and analyze double and triple absorber coincidences as tracers of the coherence of the IGM on scales of 1$-$2 $h_{70}^{-1}$ Mpc over the redshift range $0.4 < z_a < 0.9$. Unfortunately, the GHRS data with G140L are of insufficient quality to extend the experiment for sight lines A and B to zero redshift. The sum of the {\it HST} FOS data over several epochs yields G190H spectra with $10 < S/N < 35$ over the wavelength range 1570$-$3280 \AA, making this the highest quality data available for any measure of Ly$\alpha$ coherence in this redshift range, at least until the anticipated installation of COS. Redder data with G270L for sight lines A and B have been used to find new metal line systems and to check for metal line contamination in the Ly$\alpha$ ``forest.'' This asterism is unprecedented among quasars with $B < 18.5$.

Since the measurement of IGM coherence relies on absorbers matched across the sight lines in velocity space, particular care was taken with relative and absolute calibration of the wavelength scales of the spectra. We requested contemporaneous wavecals especially for this purpose. Random or systmatic errors due to geomagnetically induced image motion are shown to be less than 20\% of the resolution. Uncertain target placement in the 0.9 arcsecond aperture can in principle cause a large velocity offset but the absorber matches show that there is no unrecognized error in the velocity scales above a level of 50 km s$^{-1}$. 

The data is in the regime of moderate S/N and low \ion{H}{1} opacity, where the method of absorption line fitting and identification is preferred to a continuous flux statistic that might, in principle, use more of the data. At this resolution, all the absorbers are unresolved. Automated software is used to select, fit, and identify absorbers. The significance of a line is based on a combination of its degree of departure from the level of the local continuum and the quality of the fit to a Gaussian profile. Detection limits were tuned to give a maximum yield of real features and a minimum contamination by spurious features, based on a comparison of spectra of the same objects taken at different epochs. The final line lists have a completeness of 90\% and a contamination rate from spurious features of under 5\%.

A velocity match window corresponding to 2.5 times the error in absorber centroiding was used to identify absorbers. Galactic lines were identified in each spectrum and excluded from the rest of the analysis. Metal lines from intervening absorbers were searched for in the region redward of the Ly$\alpha$ emission line. Higher excitation species of these systems were searched for in the Ly$\alpha$ forest, but no blind search for metals was done, since the reliability of any identifications would have been low. The line density across the Ly$\alpha$ and Ly$\beta$ forests shows that higher order Lyman series do not contribute significantly to the data. Absorber matches across sight lines were defined either by being nearest neighbors in velocity, one for each absorber, or as symmetric matches, in which an absorber uniquely and reversibly matches across sight lines.

The significance of absorber matches, both doubly and triply, was measured with respect to a Monte Carlo simulation where the null hypothesis was that the absorbers are randomly distributed in velocity. This control experiment was conducted by drawing absorbers from a realistic power-law distribution of rest equivalent widths, adding noise, and ``selecting'' lines according to the S/N and limiting equivalent width of the actual data. In general, there was little difference in the effectiveness of the two algorithms for matching absorbers. 

The main scientific conclusions of this work are as follows:

1. Four intervening metal-line systems are seen in these spectra, one of which is previously unpublished. Three of the four are detected in the A and B pair, and one is seen in all three sight lines. The rarity of metals ensures that these joint detections would not occur by chance, so there is evidence that the metal lines represent enriched gas that is distributed on cluster scales.

2. The transverse absorber pairing statistics show no strong clustering on scales of 1$-$2 $h_{70}^{-1}$ Mpc, with only the B-C pairing showing an excess of absorber pairs at a significance level slightly above 95\%. With all sight lines combined, there is evidence of non-random pairings at the 99\% confidence level on line-of-sight velocity scales of 100$-$200 kms$^{-1}$. This result is consistent with a fraction of the IGM gas being associated with groups or poor clusters.

3. The marginal detection of absorber coherence across the sight lines, after substantially improving the data and absorber statistics relative to prior studies, suggests that the weak clustering rests with higher column density absorbers. This supposition is confirmed by showing that the absorber matches involve preferentially stronger lines at a 98.5\% confidence level or better for all sight line pairings.

4. For the A and B sight lines alone, the extended wavelength coverage of overlap allows for increased absorber statistics relative to pairings with the C sight line. Including this extra region increases the absorber samples by about 30\%. The result increases the significance of the A-to-B coherence signal from $<$80\% to $>$95\%. This clustering is far smaller than that of luminous galaxies on these physical scales.

5. The statistics of triple coincidences give a good test for the underlying geometry of the absorbing gas. There are 12 observed triple absorber matches; the mean number in a random control experiment was 5, and 12 or more did not occur at all in 1000 realizations. The absorbers that span all three sight lines have a higher column density than average, with 99\% confidence. This is evidence that, even though it is inhomogeneous, the higher column density gas of the IGM has a planar or sheet-like geometry. 

\acknowledgements
We gratefully acknowledge the support and perseverance of the scientific staff of the STScI and the ST-ECF, who answered endless technical questions about the FOS and GHRS instruments on the {\it HST}. The earlier work and insights of Nadine Dinshaw and Patrick Young are also acknowledged.

\clearpage
\begin{deluxetable}{crrrrrrr}
\tablecaption{Program Quasars \label{tbl-1}}
\tablewidth{0pt}
\tablehead{
\colhead{Quasar} & 
\colhead{$\alpha$} & 
\colhead{$\delta$} & 
\colhead{Mag.} & 
\colhead{$z_{em}$} & 
\colhead{$\Delta\theta$} & 
\colhead{$\Delta\theta$} & 
\colhead{$\Delta\theta$}
\\
\colhead{} & 
\colhead{(J2000.0)} & 
\colhead{(J2000.0)} & 
\colhead{} & 
\colhead{} & 
\colhead{$\rightarrow$ 025A} & 
\colhead{$\rightarrow$ 025B} & 
\colhead{$\rightarrow$ 0232} \\
\colhead{(1)} &
\colhead{(2)} &
\colhead{(3)} &
\colhead{(4)} &
\colhead{(5)} &
\colhead{(6)} &
\colhead{(7)} &
\colhead{(8)} \\
}
\startdata
Q0107$-$025A & $01^{\rm h}$ $10^{\rm m}$ 13\fs10 & $-$02\arcdeg 
19\arcmin 52\farcs0 & 18.1 & 0.960 & 0 & 1\farcm29 & 2\farcm93\\
Q0107$-$025B & $01^{\rm h}$ $10^{\rm m}$ 16\fs20 & $-$02\arcdeg 
18\arcmin 50\farcs0 & 17.4 & 0.956 & 1\farcm29 & 0 & 1\farcm92\\
Q0107$-$0232 & $01^{\rm h}$ $10^{\rm m}$ 14\fs51 & $-$02\arcdeg 
16\arcmin 57\farcs5 & 18.4 & 0.726 & 2\farcm93 & 1\farcm92 & 0\\
\enddata
\tablecomments{
For the three program quasars in col. (1), the J2000.0 right ascension and declination from \citet{bar01} and \citet{hew95} are in cols. (2) and (3). The photographic $J$ magnitude (which is close to $B$) for Q0107$-$025A and Q0107$-$0232 and the $V$ magnitude for Q0107$-$025B are in col. (4).  The emission redshift for the quasars from \citet{gos97} and \citet{hew95} are in col. (5).  The angular separation in arcminutes of each quasar from the object listed in the column heading is in cols. (6)$-$(8).
}
\end{deluxetable}

\clearpage
\begin{deluxetable}{rrrrrrrrr}
\tabletypesize{\footnotesize}
\tablecolumns{9}
\tablewidth{0pt}
\tablecaption{Journal of {\it HST} Observations \label{tbl-2}}
\tablehead{
\colhead{Proposal} & 
\colhead{Epoch} & 
\colhead{Instrument} & 
\colhead{Grating} & 
\colhead{Exposure} & 
\colhead{Wavelength} & 
\colhead{Mean} & 
\colhead{Resolution} & 
\colhead{Dispersion}
\\
\colhead{ID} & 
\colhead{} & 
\colhead{} & 
\colhead{} & 
\colhead{Time} & 
\colhead{Range} & 
\colhead{S/N} & 
\colhead{} & 
\colhead{} 
\\
\colhead{} & 
\colhead{} & 
\colhead{} & 
\colhead{} & 
\colhead{(s)} & 
\colhead{(\AA)} & 
\colhead{} & 
\colhead{(\AA)} & 
\colhead{(\AA\ pixel$^{-1}$)} \\ 
\colhead{(1)} & 
\colhead{(2)} & 
\colhead{(3)} & 
\colhead{(4)} & 
\colhead{(5)} & 
\colhead{(6)} & 
\colhead{(7)} & 
\colhead{(8)} & 
\colhead{(9)} \\
}
\startdata
\cutinhead{Q0107$-$025A}
 5172 & 1994 & GHRS   & G140L & 9792  & 1212$-$1498  & $\sim3$ &  0.8 & 0.143 \\
 6260 & 1996 & GHRS   & G140L & 23174 & 1212$-$1498  & $\sim4$ &  0.8 & 0.143 \\
 5320 & 1994 & FOS/RD & G190H & 13320 & 1572$-$2311  & 15 & 1.39 & 0.36\\
 6592 & 1997 & FOS/RD & G190H & 13550 & 1572$-$2311  & 25 & 1.39 & 0.36\\
 6100 & 1996 & FOS/RD & G270H & 8760  & 2222$-$3277  & 23 & 1.97 & 0.51\\
\cutinhead{Q0107$-$025B}
 5172 & 1994 & GHRS   & G140L & 9792  & 1212$-$1498  & $\sim5$ &  0.8 & 0.143\\
 6260 & 1996 & GHRS   & G140L & 10445 & 1212$-$1498  & $\sim5$ &  0.8 & 0.143\\
 5320 & 1994 & FOS/RD & G190H & 6480  & 1572$-$2311  & 13 & 1.39 & 0.36\\
 6100 & 1996 & FOS/RD & G270H & 6430  & 2222$-$3277  & 34 & 1.97  & 0.51\\
 7752 & 2000 & STIS MAMA & G230L & 19245  & 1568$-$3184  & \nodata & \nodata & \nodata \\
\cutinhead{Q0107$-$0232}
 6100 & 1996 & FOS/RD & G190H & 16980 & 1572$-$2311  & 13 & 1.39 & 0.36\\
 6592 & 1997 & FOS/RD & G190H & 15910 & 1572$-$2311  & 10 & 1.39 & 0.36\\
\enddata
\tablecomments{For each of the three quasars observed, the proposal identification number and the year the observations were taken are in cols. (1) and (2).  The instrumental configuration of the observation is described in cols. (3) and (4), with the total exposure time in col. (5).  All FOS observations were made with the 0\farcs86 circular aperture, and all GHRS observations were made with the 2\farcs0 circular aperture.  The wavelength coverage obtained is in col. (6); the mean S/N ratio, the instrumental resolution (FWHM), and the spectral dispersion for each spectrum are given in cols. (7)$-$(9).}
\end{deluxetable}

\clearpage
\begin{deluxetable}{crrrr}
\tablecaption{Comparison between the Super-wavecal and Contemporaneous Wavecal
for G190H Observations\label{tbl-3}}
\tablewidth{300pt}
\tablehead{
\colhead{Quasar} &  
\colhead{Epoch} &
\colhead{Proposal ID} & 
\colhead{Offset} &
\colhead{Offset}
\\
\colhead{} &  
\colhead{} &
\colhead{} & 
\colhead{(\AA)} & 
\colhead{(\kms)} 
\\
\colhead{(1)} &  
\colhead{(2)} &  
\colhead{(3)} &  
\colhead{(4)} &  
\colhead{(5)} \\
}
\startdata
Q0107$-$025A & 1994 & 5320 & 0.80 $\pm$\ 0.02 & 124 $\pm$\ 3\\
              & 1997 & 6592 & 0.91 $\pm$\ 0.04 & 141 $\pm$\ 6\\
Q0107$-$025B & 1994 & 5320 & 0.83 $\pm$\ 0.03 & 129 $\pm$\ 4\\
Q0107$-$0232 & 1996 & 6100 & 0.76 $\pm$\ 0.03 & 117 $\pm$\ 5\\
              & 1997 & 6592 & 0.06 $\pm$\ 0.03 & ~~8 $\pm$\ 5\\
\enddata
\tablecomments{
For each epoch of observations specified in cols. (1)$-$(3), the offset between the super-wavecal and the contemporaneous wavecal is computed and listed in cols. (4) and (5). These measured offsets are added to the wavelength scale for each epoch of observation before combining the epochs.}
\end{deluxetable}

\clearpage
\begin{deluxetable}{ccclcl}
\tablecaption{Wavelength Zero-Point Registration Techniques \label{tbl-4}}
\tablewidth{0pt}
\tablecolumns{6}
\tablehead{
\colhead{Quasar} &
\colhead{Epoch} &
\colhead{Proposal ID} &
\colhead{Technique} &
\colhead{Offset} &
\colhead{Offset}
\\
\colhead{} &
\colhead{} &
\colhead{} &
\colhead{} &
\colhead{(\AA)} &
\colhead{(\kms)}
\\
\colhead{(1)} &
\colhead{(2)} &
\colhead{(3)} &
\colhead{(4)} &
\colhead{(5)} &
\colhead{(6)} \\
}
\startdata
Q0107$-$025A  &   1996 &  6100  &      Cross-correlation &     $1.60 \pm 0.17$ & $148 \pm 16$ \\
 	      &        &	&      Strongest lines    &     $1.68 \pm 0.07$ & $157 \pm 7 $\\
	      &        &	&      Galactic lines    &     $1.81 \pm 0.35$ & $169 \pm 33$ \\
Q0107$-$025B  &   1996 &  6100  &      Cross-correlation &     $1.46 \pm 0.28$ & $136 \pm 26$ \\
	      &	       &	&      Strongest lines    &     $1.57 \pm 0.07$ & $146 \pm 7 $\\
	      &        &	&      Galactic lines    &     $1.49 \pm 0.13$ & $138 \pm 12$ \\
%
%
\enddata
\tablecomments{For the two observations described in cols. (1)$-$(3), the zeropoint offset in the wavelength scale between the G190H and G270H observations is computed using the technique specified in col. (4):  cross-correlating the data in the overlap region, measuring the wavelength centers of the strongest absorption lines in the overlap region, and using Galactic lines over the entire wavelength range of the spectra.  The results are listed in cols. (5) and (6). The offsets from the cross-correlation technique are applied to the data in the sense that they are added to the G270H data in order to register them to the G190H wavelength scale.}
\end{deluxetable}

\clearpage
\begin{deluxetable}{crrrrrr}
\tablecaption{Wavelength Zero-Point Variations on the Timescale of an Orbit \label{tbl-5}}
\tablewidth{0pt}
\tablehead{
\colhead{Quasar} &  
\colhead{Epoch} &
\colhead{Grating} &
\colhead{Proposal ID} & 
\colhead{Exposure} & 
\colhead{Offset} &
\colhead{Offset} \\
\colhead{} &
\colhead{} &
\colhead{} &
\colhead{} &
\colhead{Time} &
\colhead{} & 
\colhead{} \\
\colhead{} &  
\colhead{} &
\colhead{} &
\colhead{} &
\colhead{(s)} &
\colhead{(\AA)} &
\colhead{(\kms)}
\\
\colhead{(1)} &
\colhead{(2)} &
\colhead{(3)} &
\colhead{(4)} &
\colhead{(5)} &
\colhead{(6)} &
\colhead{(7)} \\
}
\startdata
Q0107$-$025A & 1994 & G190H & 5320 & 522 & $-$0.04 $\pm$\ 0.20 & ~$-$6 $\pm$\ 31\\
              & 1997 & G190H & 6592 & 600 & $-$0.05 $\pm$\ 0.20 & ~$-$8 $\pm$\ 31\\
Q0107$-$025B & 1994 & G190H & 5320 & 381 & $-$0.02 $\pm$\ 0.22 & ~$-$3 $\pm$\ 34\\
Q0107$-$0232 & 1996 & G190H & 6100 & 863 &   ~0.02 $\pm$\ 0.24 &   ~3  $\pm$\ 36\\
              & 1997 & G190H & 6592 & 720 & $-$0.16 $\pm$\ 0.24 & $-$25 $\pm$\ 38\\ 	
Q0107$-$025A & 1996 & G270H & 6100 & 360 & $-$0.07 $\pm$\ 0.15 & $-$11 $\pm$\ 16\\
Q0107$-$025B & 1996 & G270H & 6100 & 240 & $-$0.15 $\pm$\ 0.43 & $-$23 $\pm$\ 47\\
\enddata
\tablecomments{For the observations described in cols. (1)$-$(5), the variation of the zero point in the wavelength scale on the timescale of an orbit is calculated and listed in cols. (6) and (7).  These offsets are calculated for diagnostic purposes and were not actually applied to the wavelength scale.}
\end{deluxetable}

\clearpage
\begin{deluxetable}{crrrrrrr}
\tablecaption{Wavelength Zero-Point Variations on the Timescale of a Visit \label{tbl-6}}
\tablewidth{0pt}
\tablehead{
\colhead{Quasar} &  
\colhead{Epoch} &
\colhead{Grating} &
\colhead{Proposal ID} & 
\colhead{Exposure} & 
\colhead{Offset} &
\colhead{Offset} &\\
\colhead{} &
\colhead{} &
\colhead{} &
\colhead{} &
\colhead{Time} &
\colhead{} & 
\colhead{} &\\
\colhead{} &  
\colhead{} &
\colhead{} &
\colhead{} &
\colhead{(s)} &
\colhead{(\AA)} &
\colhead{(\kms)}
\\
\colhead{(1)} &  
\colhead{(2)} &  
\colhead{(3)} &  
\colhead{(4)} &  
\colhead{(5)} &  
\colhead{(6)} &  
\colhead{(7)} \\
}
\startdata
Q0107$-$025A & 1994 & G190H & 5320 & 522 &   ~0.11 $\pm$\ 0.24 &   ~17 $\pm$\ 37\\
              & 1997 & G190H & 6592 & 600 &   ~0.03 $\pm$\ 0.13 &   ~~5 $\pm$\ 20\\
Q0107$-$025B & 1994 & G190H & 5320 & 381 &   ~0.08 $\pm$\ 0.20 &   ~12 $\pm$\ 31\\
Q0107$-$0232 & 1996 & G190H & 6100 & 863 &   ~0.12 $\pm$\ 0.26 &   ~19 $\pm$\ 40\\
              & 1997 & G190H & 6592 & 720 & $-$0.06 $\pm$\ 0.17 & ~$-$9 $\pm$\ 26\\ 	
Q0107$-$025A & 1996 & G270H & 6100 & 360 & $-$0.01 $\pm$\ 0.02 & ~$-$2 $\pm$\ ~3\\
Q0107$-$025B & 1996 & G270H & 6100 & 240 &   ~0.00 $\pm$\ 0.03 &   ~~0 $\pm$\ ~5\\
\enddata
\tablecomments{For the observations described in cols. (1)$-$(5), the variation of the zero point in the wavelength scale on the timescale of a visit is calculated and listed in cols. (6) and (7).  These offsets are calculated for diagnostic purposes and were not actually applied to the wavelength scale.}
\end{deluxetable}

\begin{deluxetable}{rrrrrrrrlr}
\tabletypesize{\footnotesize}
\tablecaption{Absorption Lines for Q0107$-$025A (HRS G140L)\label{lineA140}}
\tablewidth{0pt}
\tablehead{
\colhead{Line} &
\colhead{$\lambda_{c}$} &
\colhead{$W_{obs}$} &
\colhead{FWHM} &
\colhead{$S\sigma_{fit}$} &
\colhead{$S\sigma_{det}$} &
\colhead{$\chi_{\nu}^{2}$}  &
\colhead{$z_{abs}$} &
\colhead{Identification\tablenotemark{a}} &
\colhead{$\Delta\lambda$} \\
\colhead{} &
\colhead{(\AA)} &
\colhead{(\AA)} &
\colhead{(\AA)} &
\colhead{} &
\colhead{} &
\colhead{} &
\colhead{} &
\colhead{} &
\colhead{(\AA)} \\
\colhead{(1)} &
\colhead{(2)} &
\colhead{(3)} &
\colhead{(4)} &
\colhead{(5)} &
\colhead{(6)} &
\colhead{(7)} &
\colhead{(8)} &
\colhead{(9)} &
\colhead{(10)} \\
}
\startdata
   1 & $1226.53\pm 0.33$ & $ 1.26\pm 0.54$ & $ 1.80\pm 0.97$ &  2.30 &  6.12 &  1.08 &   &   &  \\ 
   2 & $1258.54\pm 0.19$ & $ 1.09\pm 0.28$ & $ 1.73\pm 0.48$ &  4.00 & 14.30 &  0.80 &   &   &  \\ 
   3 & $1260.30\pm 0.16$ & $ 0.73\pm 0.25$ & $ 1.23\pm 0.35$ &  2.90 &  9.61 &  0.80 & 0.0000  & \ion{Si}{2} 1260 & $-$0.12 \\ 
     & $1260.30\pm 0.16$ & $ 0.73\pm 0.25$ & $ 1.23\pm 0.35$ &  2.90 &  9.61 &  0.80 & 0.2286  & Ly$\beta$ & 0.10 \\ 
   4 & $1265.50\pm 0.09$ & $ 0.43\pm 0.11$ & $ 0.80\pm 0.23$ &  4.10 &  5.79 &  1.21 &   &   &  \\ 
   5 & $1279.39\pm 0.15$ & $ 0.53\pm 0.14$ & $ 1.22\pm 0.41$ &  3.70 &  6.94 &  0.96 &   &   &  \\ 
   6 & $1301.41\pm 0.13$ & $ 0.32\pm 0.10$ & $ 0.92\pm 0.34$ &  3.30 &  4.87 &  0.48 &   &   &  \\ 
   7 & $1310.02\pm 1.78$ & $ 0.67\pm 1.57$ & $ 4.69\,\pm$\nodata\tablenotemark{b} &  0.40 &  9.47 &  2.52 &   &   &  \\ 
   8 & $1334.53\pm 0.15$ & $ 0.68\pm 0.13$ & $ 1.63\pm 0.38$ &  5.10 &  9.53 &  0.57 & 0.0000  & \ion{C}{2} 1334 & 0.00 \\ 
   9 & $1355.44\pm 0.18$ & $ 0.94\pm 0.20$ & $ 2.02\pm 0.56$ &  4.70 & 11.88 &  1.25 &   &   &  \\ 
  10 & $1375.08\pm 0.27$ & $ 0.36\pm 0.15$ & $ 1.47\pm 0.77$ &  2.40 &  5.38 &  0.66 &   &   &  \\ 
  11 & $1393.58\pm 0.12$ & $ 0.34\pm 0.11$ & $ 0.80\pm 0.32$ &  3.10 &  4.11 &  1.20 & 0.0000 & \ion{Si}{4} 1393 & $-$0.18 \\ 
  12 & $1491.20\pm 0.30$ & $ 1.39\pm 0.42$ & $ 2.26\pm 0.86$ &  3.30 &  8.60 &  0.53 &   &   &  \\ 
  13 & $1493.50\pm 0.15$ & $ 0.82\pm 0.25$ & $ 1.03\pm 0.37$ &  3.30 &  4.85 &  0.22 & 0.2286  & Ly$\alpha$  & $-$0.07 \\ 
  14 & $1500.44\pm 0.32$ & $ 1.51\pm 0.51$ & $ 2.45\pm 1.07$ &  3.00 & 10.37 &  0.78 &   &   &  \\ 
  15 & $1504.52\pm 0.09$ & $ 0.64\pm 0.16$ & $ 0.80\pm 0.24$ &  4.00 &  4.99 &  1.06 &   &   &  \\ 
\enddata
\tablenotetext{a}{Because of the low S/N of these data, no features 
from this table were included in the samples for the analysis.}
\tablenotetext{b}{This feature has an atypically poor fit due to the fact that
it is broad and shallow.  It may be a blend of unresolved weak features or
a region not well fitted by the continuum.}
\tablecomments{
For each absorption feature listed in col. (1), col. (2) gives the central 
wavelength in angstroms, col. (3) gives the equivalent width in angstroms, and
col. (4) gives the full width at half-maximum of the fitted Gaussian.
Col. (5) gives the significance of the line defined as $W_{obs}/\sigma_{fit}$, 
where $\sigma_{fit}$ is the error in the equivalent width, and col. (6) gives 
the 
significance of the line defined as $W_{obs}/\sigma_{det}$, where $\sigma_{det}$
is the detection limit of the data at $\lambda_{c}$ (see text for further 
details).
The reduced $\chi^{2}$ is listed in col. (7), and if this value is identical
to that of an adjacent line, it indicates that the lines were fitted 
simultaneously. Cols. (8)$-$(10) contain the redshift system associated with
the identification and the residual in the sense $\lambda_{c} - 
\lambda_{pred}$, where $\lambda_{pred} = \lambda_0 (1+z_{abs})$.
}
\end{deluxetable}

\begin{deluxetable}{rrrrrrrrlr}
\tabletypesize{\footnotesize}
\tablecaption{Absorption Lines for Q0107$-$025A (FOS G190H)\label{lineA190}}
\tablewidth{0pt}
\tablehead{
\colhead{Line} &
\colhead{$\lambda_{c}$} &
\colhead{$W_{obs}$} &
\colhead{{FWHM}} &
\colhead{$S\sigma_{fit}$} &
\colhead{$S\sigma_{det}$} &
\colhead{$\chi_{\nu}^{2}$} &
\colhead{$z_{abs}$} &
\colhead{Identification\tablenotemark{a}} &
\colhead{$\Delta\lambda$} \\
\colhead{} &
\colhead{(\AA)} &
\colhead{(\AA)} &
\colhead{(\AA)} &
\colhead{} &
\colhead{} &
\colhead{} &
\colhead{} &
\colhead{} &
\colhead{(\AA)} \\
\colhead{(1)} &
\colhead{(2)} &
\colhead{(3)} &
\colhead{(4)} &
\colhead{(5)} &
\colhead{(6)} &
\colhead{(7)} &
\colhead{(8)} &
\colhead{(9)} &
\colhead{(10)} \\
}
\startdata
   1 & $1637.49\pm 0.25$ & $ 0.55\pm 0.18$ & $ 1.58\pm 0.62$ &  3.10 &  4.27 &  0.78 & 0.2272 & \ion{C}{2} 1334 & $-$0.25 \\ 
   2 & $1671.28\pm 0.13$ & $ 0.88\pm 0.11$ & $ 2.09\pm 0.31$ &  8.00 & 11.30 &  0.61 & 0.0000 & \ion{Al}{2} 1670 & 0.49 \\ 
     & $1671.28\pm 0.13$ & $ 0.88\pm 0.11$ & $ 2.09\pm 0.31$ &  8.00 & 11.30 &  0.61 & 0.7188 & \lyc & $-$0.32 \\ 
   3 & $1683.17\pm 0.09$ & $ 0.81\pm 0.09$ & $ 1.54\pm 0.20$ &  8.80 & 10.74 &  0.93 &  &  &  \\ 
   4 & $1701.30\pm 0.16$ & $ 0.61\pm 0.09$ & $ 2.15\pm 0.39$ &  6.40 &  9.64 &  0.84 & 0.3995 & Ly$\alpha$ & $-$0.03 \\ 
   5 & $1721.69\pm 0.13$ & $ 0.38\pm 0.08$ & $ 1.24\pm 0.31$ &  4.60 &  5.41 &  1.12 &  &  &  \\ 
   6 & $1727.48\pm 0.38$ & $ 0.46\pm 0.12$ & $ 2.87\pm 0.92$ &  3.70 &  7.07 &  0.60 &  &  &  \\ 
   7 & $1738.09\pm 0.08$ & $ 0.79\pm 0.07$ & $ 1.70\pm 0.18$ & 10.80 & 13.84 &  1.56 & 0.7865  & \lyc & 0.65 \\ 
     & $1738.09\pm 0.08$ & $ 0.79\pm 0.07$ & $ 1.70\pm 0.18$ & 10.80 & 13.84 &  1.56 & 0.8766  & \lyg & $-$0.07 \\ 
   8 & $1746.32\pm 0.10$ & $ 0.16\pm 0.05$ & $ 0.73\pm 0.25$ &  3.50 &  2.95 &  0.08 & 0.8766  & \lyf & $-$0.32 \\ 
   9 & $1759.87\pm 0.16$ & $ 0.39\pm 0.07$ & $ 1.78\pm 0.39$ &  5.30 &  7.40 &  0.53 & 0.8766  & \lye & $-$0.01 \\ 
  10 & $1763.23\pm 0.09$ & $ 0.83\pm 0.09$ & $ 1.87\pm 0.23$ &  9.70 & 13.52 &  0.53 & 0.7188  & \lyb & 0.22 \\ 
  11 & $1773.33\pm 0.13$ & $ 0.42\pm 0.06$ & $ 1.76\pm 0.32$ &  6.50 &  8.79 &  0.50 & 0.7286  & \lyb & 0.27 \\ 
  12 & $1782.15\pm 0.08$ & $ 0.60\pm 0.06$ & $ 1.72\pm 0.21$ &  9.80 & 12.82 &  0.39 & 0.8766  & \lyd & $-$0.13 \\ 
  13 & $1808.65\pm 0.15$ & $ 0.52\pm 0.07$ & $ 2.32\pm 0.36$ &  7.60 & 11.46 &  0.82 &  &  &  \\ 
  14 & $1824.96\pm 0.04$ & $ 1.82\pm 0.06$ & $ 2.42\pm 0.10$ & 29.50 & 40.36 &  1.58 & 0.8766  & \lyc & $-$0.11 \\ 
     & $1824.96\pm 0.04$ & $ 1.82\pm 0.06$ & $ 2.42\pm 0.10$ & 29.50 & 40.36 &  1.58 & 0.5012 & \lya\ only & 0.00 \\ 
  15 & $1832.63\pm 0.11$ & $ 0.57\pm 0.06$ & $ 1.99\pm 0.26$ &  9.00 & 12.81 &  0.48 & 0.7865  & \lyb & 0.18 \\ 
  16 & $1853.54\pm 0.22$ & $ 0.21\pm 0.06$ & $ 1.68\pm 0.55$ &  3.70 &  4.78 &  0.19 &  &  &  \\ 
  17 & $1867.12\pm 0.03$ & $ 2.60\pm 0.06$ & $ 2.56\pm 0.07$ & 42.10 & 54.57 &  0.52 &  &  &  \\ 
  18 & $1877.42\pm 0.14$ & $ 0.14\pm 0.04$ & $ 0.97\pm 0.34$ &  3.30 &  3.27 &  0.56 &  &  &  \\ 
  19 & $1893.49\pm 0.07$ & $ 0.77\pm 0.07$ & $ 1.69\pm 0.19$ & 10.50 & 18.23 &  0.44 &  &  &  \\ 
  20 & $1899.97\pm 0.41$ & $ 0.86\pm 0.20$ & $ 3.83\pm 0.97$ &  4.30 & 19.97 &  0.94 & 0.2272 & \ion{C}{4} 1548 & 0.02  \\ 
  21 & $1903.04\pm 0.21$ & $ 0.44\pm 0.17$ & $ 1.98\pm 0.47$ &  2.60 & 10.34 &  0.94 & 0.2272 & \ion{C}{4} 1550 & $-$0.06 \\ 
  22 & $1911.51\pm 0.12$ & $ 0.25\pm 0.04$ & $ 1.41\pm 0.29$ &  5.70 &  6.72 &  0.51 &  &  &  \\ 
  23 & $1921.19\pm 0.11$ & $ 0.67\pm 0.06$ & $ 2.51\pm 0.27$ & 11.30 & 18.68 &  0.62 &  &  &  \\ 
  24 & $1924.78\pm 0.05$ & $ 0.80\pm 0.05$ & $ 1.70\pm 0.11$ & 17.50 & 23.11 &  0.62 & 0.8766 & \lyb & $-$0.09 \\ 
  25 & $1928.64\pm 0.19$ & $ 0.25\pm 0.05$ & $ 1.89\pm 0.50$ &  4.70 &  7.24 &  1.09 &  &  &  \\ 
  26 & $1947.43\pm 0.07$ & $ 0.10\pm 0.03$ & $ 0.59\pm 0.27$ &  3.40 &  2.75 &  0.53 &  &  &  \\ 
  27 & $1952.16\pm 0.10$ & $ 0.11\pm 0.03$ & $ 0.68\pm 0.24$ &  3.40 &  3.07 &  0.43 &  &  &  \\ 
  28 & $1955.37\pm 0.10$ & $ 0.30\pm 0.04$ & $ 1.41\pm 0.24$ &  6.90 &  8.01 &  0.43 &  &  &  \\ 
  29 & $1976.50\pm 0.18$ & $ 0.32\pm 0.05$ & $ 2.16\pm 0.42$ &  6.00 &  9.23 &  0.52 &  &  &  \\ 
  30 & $1985.65\pm 0.13$ & $ 0.29\pm 0.06$ & $ 1.38\pm 0.31$ &  4.90 &  8.64 &  0.61 &  &  &  \\ 
  31 & $1987.40\pm 0.14$ & $ 0.23\pm 0.06$ & $ 1.26\pm 0.34$ &  4.00 &  6.92 &  0.61 &  &  &  \\ 
  32 & $1994.20\pm 0.07$ & $ 0.47\pm 0.05$ & $ 1.49\pm 0.18$ & 10.10 & 14.83 &  0.53 &  0.2881 & \ion{C}{4} 1548 & $-$0.04  \\ 
  33 & $1998.14\pm 0.11$ & $ 0.21\pm 0.04$ & $ 1.33\pm 0.27$ &  5.80 &  6.64 &  1.58 &  0.2881 & \ion{C}{4} 1550 & 0.59 \\ 
  34 & $2007.02\pm 0.07$ & $ 0.32\pm 0.03$ & $ 1.26\pm 0.16$ &  9.00 & 14.42 &  1.49 &  &  &  \\ 
  35 & $2015.66\pm 0.07$ & $ 1.34\pm 0.05$ & $ 4.29\pm 0.18$ & 28.00 & 83.34 &  1.19 &  &  &  \\ 
  36 & $2022.43\pm 0.11$ & $ 0.29\pm 0.09$ & $ 1.64\pm 0.29$ &  3.10 & 21.33 &  1.15 &  &  &  \\ 
  37 & $2025.66\pm 0.17$ & $ 1.13\pm 0.22$ & $ 3.99\pm 0.86$ &  5.20 & 82.36 &  1.15 &  &  &  \\ 
  38 & $2028.98\pm 0.25$ & $ 0.30\pm 0.14$ & $ 2.18\pm 0.51$ &  2.20 & 20.34 &  1.15 &  &  &  \\ 
  39 & $2034.10\pm 0.35$ & $ 0.10\pm 0.04$ & $ 1.92\pm 0.94$ &  2.50 &  5.88 &  0.67 &  &  &  \\ 
  40 & $2036.60\pm 0.20$ & $ 0.12\pm 0.04$ & $ 1.38\pm 0.49$ &  3.20 &  6.43 &  0.67 &  &  &  \\ 
  41 & $2041.44\pm 0.04$ & $ 0.41\pm 0.03$ & $ 1.31\pm 0.11$ & 14.40 & 19.16 &  0.47 &  &  &  \\ 
  42 & $2054.27\pm 0.05$ & $ 0.42\pm 0.03$ & $ 1.50\pm 0.13$ & 14.20 & 16.26 &  0.55 &  &  &  \\ 
  43 & $2059.89\pm 0.10$ & $ 0.22\pm 0.03$ & $ 1.38\pm 0.23$ &  7.00 &  7.89 &  0.41 &  &  &  \\ 
  44 & $2073.69\pm 0.12$ & $ 0.19\pm 0.03$ & $ 1.39\pm 0.29$ &  5.70 &  6.36 &  0.58 &  &  &  \\ 
  45 & $2082.10\pm 0.11$ & $ 0.18\pm 0.03$ & $ 1.25\pm 0.26$ &  5.60 &  5.78 &  0.30 &  &  &  \\ 
  46 & $2089.48\pm 0.03$ & $ 1.59\pm 0.04$ & $ 2.31\pm 0.07$ & 41.60 & 48.99 &  1.31 &  0.7188 & \lya & $-$0.01 \\ 
  47 & $2101.39\pm 0.02$ & $ 1.24\pm 0.03$ & $ 1.79\pm 0.05$ & 39.10 & 38.72 &  0.78 &  0.7286 & \lya & $-$0.02 \\ 
  48 & $2122.79\pm 0.13$ & $ 0.17\pm 0.03$ & $ 1.43\pm 0.32$ &  5.20 &  6.32 &  0.73 &  &  &  \\ 
  49 & $2131.56\pm 0.19$ & $ 0.12\pm 0.03$ & $ 1.39\pm 0.47$ &  3.60 &  4.39 &  1.17 &  &  &  \\ 
  50 & $2146.95\pm 0.16$ & $ 0.10\pm 0.03$ & $ 1.07\pm 0.38$ &  3.30 &  3.59 &  0.36 &  &  &  \\ 
  51 & $2166.83\pm 0.15$ & $ 0.22\pm 0.04$ & $ 1.88\pm 0.37$ &  5.90 &  8.30 &  0.46 & 0.3995 & \ion{C}{4} 1548 & 0.12 \\ 
  52 & $2171.74\pm 0.03$ & $ 1.18\pm 0.03$ & $ 1.94\pm 0.06$ & 35.20 & 44.57 &  2.17 & 0.7865 & \lya & $-$0.05 \\ 
  53 & $2187.61\pm 0.16$ & $ 0.29\pm 0.04$ & $ 2.25\pm 0.43$ &  6.50 & 11.17 &  0.91 &  &  &  \\ 
  54 & $2196.86\pm 0.10$ & $ 0.19\pm 0.03$ & $ 1.29\pm 0.23$ &  6.60 &  7.52 &  0.62 &  &  &  \\ 
  55 & $2240.78\pm 0.07$ & $ 0.27\pm 0.03$ & $ 1.41\pm 0.18$ &  9.20 & 10.90 &  0.59 &  &  &  \\ 
  56 & $2246.33\pm 0.04$ & $ 0.47\pm 0.03$ & $ 1.34\pm 0.09$ & 17.20 & 19.01 &  1.16 &  &  &  \\ 
  57 & $2251.63\pm 0.07$ & $ 0.23\pm 0.03$ & $ 1.27\pm 0.18$ &  8.40 &  9.40 &  0.69 & 0.8766 & \ion{N}{1} 1200  & $-$0.12 \\ 
  58 & $2262.96\pm 0.35$ & $ 0.33\pm 0.06$ & $ 3.89\pm 0.95$ &  5.10 & 13.79 &  1.14 &  &  &  \\ 
  59 & $2271.83\pm 0.06$ & $ 0.39\pm 0.03$ & $ 1.52\pm 0.13$ & 13.30 & 15.38 &  0.65 &  &  &  \\ 
  60 & $2281.30\pm 0.02$ & $ 1.18\pm 0.03$ & $ 1.71\pm 0.04$ & 44.80 & 51.35 &  0.71 & 0.8766 & \lya & $-$0.03 \\ 
  61 & $2288.84\pm 0.12$ & $ 0.08\pm 0.02$ & $ 0.88\pm 0.29$ &  3.50 &  3.54 &  0.16 &  &  &  \\ 
  62 & $2297.18\pm 0.04$ & $ 0.39\pm 0.03$ & $ 1.32\pm 0.10$ & 15.10 & 16.89 &  0.15 &  &  &  \\ 
  63 & $2308.66\pm 0.05$ & $ 0.36\pm 0.03$ & $ 1.16\pm 0.12$ & 11.70 & 12.49 &  0.44 &  &  &  \\ 
\enddata
\tablenotetext{a}{
The primary sample from the analysis comprises \lya\ and unidentified
features with $1620.60 < \lambda_{c} < 2089.85$. This corresponds to \lya\
and unidentified lines up through line 46 in this table.  All \lya\ and 
unidentified features in this table are included in the secondary sample, 
which has $1620.60 < \lambda_{c} < 2368.33$ as described in the analysis.
}
\tablecomments{
For each absorption feature listed in col. (1), col. (2) gives the central
wavelength in angstroms, col. (3) gives the equivalent width in angstroms, and
col. (4) gives the full width at half-maximum of the fitted Gaussian.
Col. (5) gives the significance of the line defined as $W_{obs}/\sigma_{fit}$,
where $\sigma_{fit}$ is the error in the equivalent width, and col. (6) is the
significance of the line defined as $W_{obs}/\sigma_{det}$, where $\sigma_{det}$
is the detection limit of the data at $\lambda_{c}$ (see text for further
details).
The reduced $\chi^{2}$ is listed in col. (7), and if this value is identical 
to that of an adjacent line, it indicates that the lines were fitted
simultaneously. Cols. (8)$-$(10) contain the redshift system associated with
the identification and the residual in the sense $\lambda_{c} -
\lambda_{pred}$, where $\lambda_{pred} = \lambda_0 (1+z_{abs})$.
}
\end{deluxetable}

\begin{deluxetable}{rrrrrrrrlr}
\tabletypesize{\footnotesize}
\tablecaption{Absorption Lines for Q0107$-$025A (FOS G270H)\label{lineA270}}
\tablewidth{0pt}
\tablehead{
\colhead{Line} &
\colhead{$\lambda_{c}$} &
\colhead{$W_{obs}$} &
\colhead{{FWHM}} &
\colhead{$S\sigma_{fit}$} &
\colhead{$S\sigma_{det}$} &
\colhead{$\chi_{\nu}^{2}$} &
\colhead{$z_{abs}$} &
\colhead{Identification\tablenotemark{a}} &
\colhead{$\Delta\lambda$} \\
\colhead{} &
\colhead{(\AA)} &
\colhead{(\AA)} &
\colhead{(\AA)} &
\colhead{} &
\colhead{} &
\colhead{} &
\colhead{} &
\colhead{} &
\colhead{(\AA)} \\
\colhead{(1)} &
\colhead{(2)} &
\colhead{(3)} &
\colhead{(4)} &
\colhead{(5)} &
\colhead{(6)} &
\colhead{(7)} &
\colhead{(8)} &
\colhead{(9)} &
\colhead{(10)} \\
}
\startdata
   1 & $2247.19\pm 0.20$ & $ 0.39\pm 0.08$ & $ 1.88\pm 0.47$ &  4.60 &  5.31 &  0.11 &  &  &  \\ 
   2 & $2251.98\pm 0.22$ & $ 0.36\pm 0.08$ & $ 1.92\pm 0.53$ &  4.20 &  5.01 &  0.21 & 0.8769 & \ion{N}{1} 1200  & $-$0.13 \\ 
   3 & $2272.78\pm 0.21$ & $ 0.37\pm 0.08$ & $ 1.95\pm 0.51$ &  4.50 &  5.25 &  0.86 &  &  &  \\ 
   4 & $2281.72\pm 0.06$ & $ 1.39\pm 0.08$ & $ 2.21\pm 0.15$ & 16.90 & 18.87 &  0.40 & 0.8769 & \lya & 0.03 \\ 
   5 & $2297.30\pm 0.17$ & $ 0.35\pm 0.07$ & $ 1.79\pm 0.41$ &  5.10 &  5.66 &  0.45 &  &  &  \\ 
   6 & $2308.96\pm 0.18$ & $ 0.34\pm 0.07$ & $ 1.86\pm 0.43$ &  5.00 &  5.86 &  0.19 &  &  &  \\ 
   7 & $2318.24\pm 0.06$ & $ 1.21\pm 0.07$ & $ 2.34\pm 0.15$ & 17.60 & 21.69 &  1.20 &  &  &  \\ 
   8 & $2342.76\pm 0.06$ & $ 1.58\pm 0.06$ & $ 3.45\pm 0.15$ & 26.30 & 39.88 &  1.04 &  &  &  \\ 
   9 & $2361.46\pm 0.06$ & $ 0.49\pm 0.03$ & $ 1.72\pm 0.13$ & 15.10 & 21.48 &  0.80 &  &  &  \\ 
  10 & $2370.48\pm 0.09$ & $ 0.30\pm 0.03$ & $ 1.90\pm 0.22$ &  9.90 & 17.45 &  0.90 &  &  &  \\ 
  11 & $2374.96\pm 0.10$ & $ 0.17\pm 0.02$ & $ 1.38\pm 0.23$ &  7.00 & 11.61 &  1.25 &  &  &  \\ 
  12 & $2379.06\pm 0.03$ & $ 0.83\pm 0.02$ & $ 2.04\pm 0.07$ & 33.40 & 66.83 &  2.59 &  &  &  \\ 
  13 & $2382.87\pm 0.05$ & $ 0.50\pm 0.16$ & $ 1.59\pm 0.20$ &  3.20 & 40.65 &  2.20 & 0.0000 & \ion{Fe}{2} 2382 & 0.10 \\ 
  14 & $2384.81\pm 0.86$ & $ 0.31\pm 0.17$ & $ 3.30\pm 1.49$ &  1.80 & 23.24 &  2.20 &  &  &  \\ 
  15 & $2389.70\pm 0.06$ & $ 0.49\pm 0.03$ & $ 2.11\pm 0.16$ & 15.90 & 30.08 &  0.80 &  &  &  \\ 
  16 & $2396.46\pm 0.24$ & $ 0.13\pm 0.03$ & $ 1.97\pm 0.57$ &  4.10 &  6.70 &  0.90 & 0.7194 & \ion{Si}{4} 1393 &  0.03 \\ 
  17 & $2411.81\pm 0.19$ & $ 0.16\pm 0.04$ & $ 1.68\pm 0.45$ &  4.30 &  5.39 &  0.35 & 0.7194 & \ion{Si}{4} 1402 & $-$0.11 \\ 
  18 & $2586.76\pm 0.22$ & $ 0.66\pm 0.09$ & $ 3.58\pm 0.55$ &  7.70 &  6.94 &  0.57 & 0.0000 & \ion{Fe}{2} 2586 & 0.11 \\ 
  19 & $2600.38\pm 0.09$ & $ 0.51\pm 0.06$ & $ 1.73\pm 0.23$ &  9.10 &  5.16 &  1.42 & 0.0000 & \ion{Fe}{2} 2600 & 0.21 \\ 
  20 & $2796.37\pm 0.07$ & $ 0.77\pm 0.06$ & $ 1.95\pm 0.18$ & 12.90 &  6.88 &  0.25 & 0.0000 & \ion{Mg}{2} 2796 & 0.02 \\ 
  21 & $2803.58\pm 0.07$ & $ 0.65\pm 0.06$ & $ 1.79\pm 0.18$ & 11.50 &  5.76 &  0.69 & 0.0000 & \ion{Mg}{2} 2803 & 0.05 \\ 
\enddata
\tablenotetext{a}{
The secondary sample from the analysis comprises \lya\ and unidentified
features with $1620.60 < \lambda_{c} < 2368.33$. This corresponds to 
the lines specified in the footnote of Table~\ref{lineA190}, excluding
features up through line 6 in this table because the same features are
better measured in Table~ref{lineA190} and including lines 7, 8, and 9
from this table.
}
\tablecomments{
For each absorption feature listed in col. (1), col. (2) gives the central
wavelength in angstroms, col. (3) gives the equivalent width in angstroms, and
col. (4) gives the full width at half-maximum of the fitted Gaussian.
Col. (5) gives the significance of the line defined as $W_{obs}/\sigma_{fit}$,
where $\sigma_{fit}$ is the error in the equivalent width, and col. (6) gives the
significance of the line defined as $W_{obs}/\sigma_{det}$, where $\sigma_{det}$
is the detection limit of the data at $\lambda_{c}$ (see text for further
details).
The reduced $\chi^{2}$ is listed in col. (7), and if this value is identical to
that of an adjacent line, it indicates that the lines were fitted
simultaneously. Cols. (8)$-$(10) contain the redshift system associated with
the identification and the residual in the sense $\lambda_{c} -
\lambda_{pred}$, where $\lambda_{pred} = \lambda_0 (1+z_{abs})$.
}
\end{deluxetable}

\begin{deluxetable}{rrrrrrrrlr}
\tabletypesize{\footnotesize}
\tablecaption{Absorption Lines for Q0107$-$025B (HRS G140L)\label{lineB140}}
\tablewidth{0pt}
\tablehead{
\colhead{Line} &
\colhead{$\lambda_{c}$} &
\colhead{$W_{obs}$} &
\colhead{{FWHM}} &
\colhead{$S\sigma_{fit}$} &
\colhead{$S\sigma_{det}$} &
\colhead{$\chi_{\nu}^{2}$} &
\colhead{$z_{abs}$} &
\colhead{Identification\tablenotemark{a}} &
\colhead{$\Delta\lambda$} \\
\colhead{} &
\colhead{(\AA)} &
\colhead{(\AA)} &
\colhead{(\AA)} &
\colhead{} &
\colhead{} &
\colhead{} &
\colhead{} &
\colhead{} &
\colhead{(\AA)} \\
\colhead{(1)} &
\colhead{(2)} &
\colhead{(3)} &
\colhead{(4)} &
\colhead{(5)} &
\colhead{(6)} &
\colhead{(7)} &
\colhead{(8)} &
\colhead{(9)} &
\colhead{(10)} \\
}
\startdata
   1 & $1259.66\pm 0.17$ & $ 0.74\pm 0.15$ & $ 1.77\pm 0.43$ &  4.90 &  9.11 &  0.75 & 0.0000 & \ion{Si}{2} 1260 & $-$0.76 \\ 
   2 & $1265.82\pm 0.22$ & $ 0.66\pm 0.19$ & $ 1.68\pm 0.61$ &  3.40 &  8.37 &  1.02 & 0.2268 & \ion{O}{6} 1031 & $-$0.15 \\ 
   3 & $1273.56\pm 0.13$ & $ 0.30\pm 0.10$ & $ 0.86\pm 0.34$ &  3.10 &  4.13 &  0.75 & 0.2268 & \ion{O}{6} 1037 & 0.61 \\ 
   4 & $1279.40\pm 0.15$ & $ 0.33\pm 0.10$ & $ 1.10\pm 0.41$ &  3.30 &  5.30 &  0.63 &  &  &  \\ 
   5 & $1301.33\pm 0.11$ & $ 0.60\pm 0.11$ & $ 1.45\pm 0.33$ &  5.60 & 12.52 &  2.04 & 0.0000 & \ion{O}{1} 1302 & $-$0.84 \\ 
   6 & $1311.89\pm 0.15$ & $ 0.26\pm 0.08$ & $ 1.07\pm 0.42$ &  3.20 &  5.27 &  0.68 &  &  &  \\ 
   7 & $1322.84\pm 0.10$ & $ 0.22\pm 0.06$ & $ 0.76\pm 0.26$ &  3.60 &  4.49 &  0.71 &  &  &  \\ 
   8 & $1328.64\pm 0.08$ & $ 0.50\pm 0.08$ & $ 1.18\pm 0.22$ &  6.60 &  9.96 &  1.40 &  &  &  \\ 
   9 & $1334.06\pm 0.10$ & $ 0.82\pm 0.09$ & $ 1.86\pm 0.25$ &  8.90 & 16.80 &  0.55 &  0.0000 & \ion{C}{2} 1334 & $-$0.47 \\ 
  10 & $1355.46\pm 0.12$ & $ 0.51\pm 0.10$ & $ 1.44\pm 0.34$ &  5.40 & 10.03 &  0.98 &  &  &  \\ 
  11 & $1360.66\pm 0.19$ & $ 1.05\pm 0.17$ & $ 2.79\pm 0.56$ &  6.30 & 19.78 &  0.87 &  &  &  \\ 
  12 & $1366.35\pm 0.14$ & $ 0.27\pm 0.08$ & $ 1.09\pm 0.36$ &  3.60 &  5.38 &  0.63 &  &  &  \\ 
  13 & $1380.88\pm 0.12$ & $ 0.36\pm 0.08$ & $ 1.16\pm 0.30$ &  4.70 &  7.10 &  1.15 &  &  &  \\ 
  14 & $1392.94\pm 0.19$ & $ 0.38\pm 0.11$ & $ 1.43\pm 0.50$ &  3.60 &  7.03 &  0.94 & 0.0000 & \ion{Si}{4} 1393 & $-$0.82 \\ 
  15 & $1403.00\pm 0.09$ & $ 0.67\pm 0.09$ & $ 1.39\pm 0.24$ &  7.00 & 11.37 &  0.60 &  &  &  \\ 
  16 & $1434.69\pm 0.10$ & $ 0.42\pm 0.10$ & $ 0.96\pm 0.27$ &  4.30 &  6.11 &  0.83 &  &  &  \\ 
  17 & $1458.31\pm 0.07$ & $ 0.54\pm 0.11$ & $ 0.78\pm 0.19$ &  5.10 &  5.91 &  0.70 &  &  &  \\ 
  18 & $1461.64\pm 0.08$ & $ 1.15\pm 0.14$ & $ 1.39\pm 0.20$ &  8.20 & 12.85 &  0.86 &  &  &  \\ 
  19 & $1491.46\pm 0.43$ & $ 1.24\pm 0.45$ & $ 3.44\pm 1.60$ &  2.80 & 13.90 &  0.55 & 0.2268 & Ly$\alpha$  & 0.08 \\ 
\enddata
\tablenotetext{a}{Because of the low S/N of these data, no features 
from this table were included in the samples for the analysis.}
\tablecomments{
For each absorption feature listed in col. (1), col. (2) gives the central
wavelength in angstroms, col. (3) gives the equivalent width in angstroms, and
col. (4) gives the full width at half-maximum of the fitted Gaussian.
Col. (5) gives the significance of the line defined as $W_{obs}/\sigma_{fit}$,
where $\sigma_{fit}$ is the error in the equivalent width, and col. (6) gives the
significance of the line defined as $W_{obs}/\sigma_{det}$, where $\sigma_{det}$
is the detection limit of the data at $\lambda_{c}$ (see text for further
details).
The reduced $\chi^{2}$ is listed in col. (7), and if this value
is identical to that of an adjacent line, it indicates that the lines were fitted
simultaneously. Cols. (8)$-$(10) contain the redshift system associated with
the identification and the residual in the sense $\lambda_{c} -
\lambda_{pred}$, where $\lambda_{pred} = \lambda_0 (1+z_{abs})$.
}
\end{deluxetable}

\begin{deluxetable}{rrrrrrrrlr}
\tabletypesize{\footnotesize}
\tablecaption{Absorption Lines for Q0107$-$025B (FOS G190H)\label{lineB190}}
\tablewidth{0pt}
\tablehead{
\colhead{Line} &
\colhead{$\lambda_{c}$} &
\colhead{$W_{obs}$} &
\colhead{{FWHM}} &
\colhead{$S\sigma_{fit}$} &
\colhead{$S\sigma_{det}$} &
\colhead{$\chi_{\nu}^{2}$} &
\colhead{$z_{abs}$} &
\colhead{Identification\tablenotemark{a}} &
\colhead{$\Delta\lambda$} \\
\colhead{} &
\colhead{(\AA)} &
\colhead{(\AA)} &
\colhead{(\AA)} &
\colhead{} &
\colhead{} &
\colhead{} &
\colhead{} &
\colhead{} &
\colhead{(\AA)} \\
\colhead{(1)} &
\colhead{(2)} &
\colhead{(3)} &
\colhead{(4)} &
\colhead{(5)} &
\colhead{(6)} &
\colhead{(7)} &
\colhead{(8)} &
\colhead{(9)} &
\colhead{(10)} \\
}
\startdata
   1 & $1637.82\pm 0.08$ & $ 0.51\pm 0.14$ & $ 0.55\pm 0.20$ &  3.60 &  2.34 &  1.52 & 0.2273 & \ion{C}{2} 1334 & 0.05 \\ 
   2 & $1671.19\pm 0.13$ & $ 0.82\pm 0.15$ & $ 1.55\pm 0.33$ &  5.60 &  6.53 &  0.85 & 0.0000 & \ion{Al}{2} 1670 & 0.40 \\
     & $1671.19\pm 0.13$ & $ 0.82\pm 0.15$ & $ 1.55\pm 0.33$ &  5.60 &  6.53 &  0.85 & 0.7183 & \lyc  & 0.07 \\
   3 & $1680.16\pm 0.15$ & $ 0.37\pm 0.12$ & $ 1.00\pm 0.39$ &  3.20 &  3.23 &  1.29 &  &  &  \\ 
   4 & $1701.06\pm 0.07$ & $ 1.28\pm 0.12$ & $ 1.69\pm 0.19$ & 10.80 & 12.42 &  2.07 & 0.3993 & Ly$\alpha$ & $-$0.03 \\ 
   5 & $1717.51\pm 0.25$ & $ 0.62\pm 0.16$ & $ 1.99\pm 0.60$ &  3.90 &  5.57 &  0.44 &  &  &  \\ 
   6 & $1738.05\pm 0.16$ & $ 0.32\pm 0.10$ & $ 1.11\pm 0.39$ &  3.30 &  3.57 &  0.24 & 0.7874 & \lyc  & $-$0.27 \\ 
   7 & $1746.62\pm 0.06$ & $ 0.40\pm 0.07$ & $ 0.74\pm 0.16$ &  5.60 &  4.49 &  1.38 &  &  &  \\ 
   8 & $1759.89\pm 0.17$ & $ 0.55\pm 0.13$ & $ 1.56\pm 0.41$ &  4.20 &  6.77 &  0.71 &  &  &  \\ 
   9 & $1762.53\pm 0.10$ & $ 1.44\pm 0.16$ & $ 2.03\pm 0.28$ &  9.10 & 14.08 &  0.71 & 0.7183 & \lyb & 0.04 \\ 
  10 & $1784.15\pm 0.77$ & $ 0.79\pm 0.34$ & $ 4.13\pm 1.99$ &  2.30 & 10.67 &  0.36 &  &  &  \\ 
  11 & $1823.58\pm 0.09$ & $ 1.50\pm 0.11$ & $ 2.66\pm 0.23$ & 13.90 & 20.90 &  0.86 &  &  &  \\ 
  12 & $1844.60\pm 0.07$ & $ 0.68\pm 0.07$ & $ 1.31\pm 0.17$ &  9.20 &  9.61 &  0.25 &  &  &  \\ 
  13 & $1852.68\pm 0.12$ & $ 0.42\pm 0.08$ & $ 1.28\pm 0.29$ &  5.10 &  6.02 &  0.51 &  &  &  \\ 
  14 & $1855.65\pm 0.10$ & $ 1.19\pm 0.11$ & $ 2.32\pm 0.25$ & 11.10 & 16.79 &  0.51 & 0.8094 & \lyb & $-$0.29 \\ 
  15 & $1866.73\pm 0.10$ & $ 1.05\pm 0.11$ & $ 2.09\pm 0.25$ &  9.70 & 13.14 &  1.51 &  &  &  \\ 
  16 & $1919.41\pm 0.34$ & $ 0.60\pm 0.12$ & $ 3.56\pm 0.83$ &  5.10 & 10.05 &  0.36 &  &  &  \\ 
  17 & $1924.62\pm 0.08$ & $ 0.93\pm 0.08$ & $ 2.04\pm 0.21$ & 11.60 & 16.25 &  0.36 & 0.8763 & \lyb & 0.06 \\ 
  18 & $1950.21\pm 0.21$ & $ 0.34\pm 0.08$ & $ 1.78\pm 0.51$ &  4.00 &  5.63 &  0.37 & 0.3993 & \ion{Si}{4} 1393 & $-$0.08 \\ 
  19 & $1956.40\pm 0.17$ & $ 0.51\pm 0.10$ & $ 1.96\pm 0.45$ &  5.40 &  7.97 &  1.12 &  &  &  \\ 
  20 & $1971.91\pm 0.15$ & $ 0.42\pm 0.08$ & $ 1.72\pm 0.38$ &  5.40 &  7.24 &  0.35 &  &  &  \\ 
  21 & $1997.16\pm 0.34$ & $ 0.37\pm 0.10$ & $ 2.72\pm 0.93$ &  3.60 &  7.15 &  0.62 &  &  &  \\ 
  22 & $2044.48\pm 0.15$ & $ 0.10\pm 0.03$ & $ 0.40\pm 0.63$ &  3.80 &  2.16 &  0.25 &  &  &  \\ 
  23 & $2048.84\pm 0.23$ & $ 0.17\pm 0.06$ & $ 1.42\pm 0.55$ &  3.10 &  3.76 &  0.26 &  &  &  \\ 
  24 & $2054.50\pm 0.12$ & $ 0.32\pm 0.05$ & $ 1.44\pm 0.29$ &  6.00 &  6.88 &  1.37 &  &  &  \\ 
  25 & $2064.73\pm 0.38$ & $ 0.38\pm 0.11$ & $ 3.07\pm 1.11$ &  3.60 &  7.83 &  0.86 &  &  &  \\ 
  26 & $2083.57\pm 0.10$ & $ 0.60\pm 0.06$ & $ 2.02\pm 0.25$ &  9.60 & 13.31 &  0.72 &  &  &  \\ 
  27 & $2088.83\pm 0.03$ & $ 1.42\pm 0.05$ & $ 1.81\pm 0.08$ & 28.20 & 31.87 &  0.72 & 0.7183 & \lya & $-$0.06 \\ 
  28 & $2097.15\pm 1.13$ & $ 0.35\pm 0.19$ & $ 5.76\pm 4.18$ &  1.80 &  8.11 &  0.47 &  &  &  \\ 
  29 & $2101.69\pm 0.15$ & $ 0.20\pm 0.05$ & $ 1.23\pm 0.37$ &  3.90 &  4.47 &  0.84 &  &  &  \\ 
  30 & $2125.45\pm 0.22$ & $ 0.22\pm 0.06$ & $ 1.66\pm 0.57$ &  3.50 &  4.81 &  0.76 &  &  &  \\ 
  31 & $2166.32\pm 0.10$ & $ 0.30\pm 0.05$ & $ 1.22\pm 0.23$ &  6.20 &  6.80 &  0.69 & 0.3993 & \ion{C}{4} 1548  & 0.08 \\ 
  32 & $2170.02\pm 0.16$ & $ 0.24\pm 0.06$ & $ 1.42\pm 0.39$ &  4.30 &  5.52 &  1.34 & 0.3993 & \ion{C}{4} 1550  & 0.18 \\ 
  33 & $2172.83\pm 0.07$ & $ 0.75\pm 0.06$ & $ 1.89\pm 0.19$ & 12.10 & 17.19 &  1.34 & 0.7874 & \lya & $-$0.06 \\ 
  34 & $2185.55\pm 0.17$ & $ 0.32\pm 0.06$ & $ 1.88\pm 0.42$ &  5.30 &  7.33 &  0.45 &  &  &  \\ 
  35 & $2199.64\pm 0.03$ & $ 1.59\pm 0.05$ & $ 2.00\pm 0.08$ & 30.70 & 37.62 &  0.74 & 0.8094 & \lya & 0.01 \\ 
  36 & $2210.67\pm 0.13$ & $ 0.23\pm 0.05$ & $ 1.25\pm 0.31$ &  4.80 &  5.26 &  0.46 &  &  &  \\ 
  37 & $2215.33\pm 0.49$ & $ 0.24\pm 0.08$ & $ 3.10\pm 1.18$ &  3.10 &  5.76 &  0.48 &  &  &  \\ 
  38 & $2226.93\pm 0.09$ & $ 0.42\pm 0.05$ & $ 1.47\pm 0.21$ &  8.20 &  9.99 &  0.43 &  &  &  \\ 
  39 & $2230.35\pm 0.11$ & $ 0.34\pm 0.05$ & $ 1.48\pm 0.26$ &  6.60 &  8.17 &  0.54 &  &  &  \\ 
  40 & $2246.48\pm 0.26$ & $ 0.26\pm 0.06$ & $ 2.14\pm 0.64$ &  4.00 &  6.37 &  0.83 &  &  &  \\ 
  41 & $2280.94\pm 0.03$ & $ 1.08\pm 0.04$ & $ 1.71\pm 0.08$ & 24.60 & 29.45 &  0.35 & 0.8763 & \lya & $-$0.02 \\ 
  42 & $2297.60\pm 0.08$ & $ 0.25\pm 0.04$ & $ 1.09\pm 0.20$ &  6.40 &  6.67 &  1.87 &  &  &  \\ 
\enddata
\tablenotetext{a}{
The primary sample from the analysis comprises \lya\ and unidentified
features with $1620.60 < \lambda_{c} < 2089.85$. This corresponds to \lya\
and unidentified lines up through line 27 in this table.  All \lya\ and 
unidentified features in this table are included in the secondary sample, 
which has $1620.60 < \lambda_{c} < 2368.33$ as described in the analysis.
}
\tablecomments{
For each absorption feature listed in col. (1), col. (2) gives the central
wavelength in angstroms, col. (3) gives the equivalent width in angstroms, and
col. (4) gives the full width at half-maximum of the fitted Gaussian.
Col. (5) gives the significance of the line defined as $W_{obs}/\sigma_{fit}$,
where $\sigma_{fit}$ is the error in the equivalent width, and col. (6) gives the
significance of the line defined as $W_{obs}/\sigma_{det}$, where $\sigma_{det}$
is the detection limit of the data at $\lambda_{c}$ (see text for further
details).
The reduced $\chi^{2}$ is listed in col. (7), and if this value
is identical to that of an adjacent line, it indicates that the lines were fitted
simultaneously. Cols. (8)$-$(10) contain the redshift system associated with
the identification and the residual in the sense $\lambda_{c} -
\lambda_{pred}$, where $\lambda_{pred} = \lambda_0 (1+z_{abs})$.
}
\end{deluxetable}

\begin{deluxetable}{rrrrrrrrlr}
\tabletypesize{\footnotesize}
\tablecaption{Absorption Lines for Q0107$-$025B (FOS G270H)\label{lineB270}}
\tablewidth{0pt}
\tablehead{
\colhead{Line} &
\colhead{$\lambda_{c}$} &
\colhead{$W_{obs}$} &
\colhead{{FWHM}} &
\colhead{$S\sigma_{fit}$} &
\colhead{$S\sigma_{det}$} &
\colhead{$\chi_{\nu}^{2}$} &
\colhead{$z_{abs}$} &
\colhead{Identification\tablenotemark{a}} &
\colhead{$\Delta\lambda$} \\
\colhead{} &
\colhead{(\AA)} &
\colhead{(\AA)} &
\colhead{(\AA)} &
\colhead{} &
\colhead{} &
\colhead{} &
\colhead{} &
\colhead{} &
\colhead{(\AA)} \\
\colhead{(1)} &
\colhead{(2)} &
\colhead{(3)} &
\colhead{(4)} &
\colhead{(5)} &
\colhead{(6)} &
\colhead{(7)} &
\colhead{(8)} &
\colhead{(9)} &
\colhead{(10)} \\
}
\startdata
   1 & $2246.72\pm 0.23$ & $ 0.22\pm 0.06$ & $ 1.86\pm 0.56$ &  3.90 &  4.66 &  0.68 &  &  &  \\ 
   2 & $2281.30\pm 0.05$ & $ 1.33\pm 0.06$ & $ 2.38\pm 0.12$ & 23.70 & 28.49 &  0.28 & 0.8766 & \lya  & 0.00  \\ 
   3 & $2298.37\pm 0.20$ & $ 0.19\pm 0.05$ & $ 1.69\pm 0.47$ &  4.20 &  4.70 &  0.40 &  &  &  \\ 
   4 & $2319.05\pm 0.04$ & $ 1.01\pm 0.04$ & $ 2.05\pm 0.10$ & 23.20 & 26.27 &  0.72 &  &  &  \\ 
   5 & $2342.12\pm 0.18$ & $ 0.28\pm 0.08$ & $ 1.67\pm 0.38$ &  3.40 &  8.43 &  0.53 &  &  &  \\ 
   6 & $2344.52\pm 0.22$ & $ 0.48\pm 0.09$ & $ 2.54\pm 0.50$ &  5.40 & 15.07 &  0.53 & 0.0000 & \ion{Fe}{2} 2344 & 0.31 \\ 
   7 & $2349.38\pm 0.19$ & $ 0.19\pm 0.04$ & $ 1.97\pm 0.48$ &  4.90 &  6.02 &  2.59 &  &  & \\ 
   8 & $2374.84\pm 0.08$ & $ 0.20\pm 0.02$ & $ 1.39\pm 0.20$ &  8.30 &  8.39 &  0.92 & 0.0000 & \ion{Fe}{2} 2374 & 0.38 \\ 
   9 & $2383.23\pm 0.05$ & $ 0.67\pm 0.03$ & $ 2.14\pm 0.11$ & 21.90 & 25.09 &  0.83 & 0.0000 & \ion{Fe}{2} 2382 & 0.46 \\ 
  10 & $2510.88\pm 0.28$ & $ 0.15\pm 0.05$ & $ 1.97\pm 0.69$ &  3.40 &  4.11 &  0.66 &  &  &  \\ 
  11 & $2587.18\pm 0.11$ & $ 0.41\pm 0.05$ & $ 2.06\pm 0.27$ &  8.90 & 11.01 &  0.45 & 0.0000 & \ion{Fe}{2} 2586 & 0.53 \\ 
  12 & $2600.63\pm 0.07$ & $ 0.53\pm 0.04$ & $ 1.89\pm 0.17$ & 13.10 & 14.69 &  0.60 & 0.0000 & \ion{Fe}{2} 2600 & 0.46 \\ 
  13 & $2796.64\pm 0.04$ & $ 0.91\pm 0.04$ & $ 1.99\pm 0.10$ & 22.50 & 24.49 &  0.55 & 0.0000 & \ion{Mg}{2} 2796 & 0.29 \\ 
  14 & $2803.92\pm 0.05$ & $ 0.72\pm 0.04$ & $ 1.78\pm 0.11$ & 18.60 & 19.65 &  0.27 & 0.0000 & \ion{Mg}{2} 2803 & 0.39 \\ 
  15 & $2853.53\pm 0.20$ & $ 0.19\pm 0.04$ & $ 1.86\pm 0.49$ &  4.50 &  5.12 &  0.69 & 0.0000 & \ion{Mg}{1} 2853 & 0.57 \\ 
\enddata
\tablenotetext{a}{
The secondary sample from the analysis comprises \lya\ and unidentified
features with $1620.60 < \lambda_{c} < 2368.33$. This corresponds to 
the lines specified in the footnote of Table~\ref{lineB190}, excluding
features up through line 3 in this table because the same features are better
measured in Table~\ref{lineB190} and including lines 4, 5, and 7 from
this table.
}
\tablecomments{
For each absorption feature listed in col. (1), col. (2) gives the central
wavelength in angstroms, col. (3) gives the equivalent width in angstroms, and
col. (4) gives the full width at half-maximum of the fitted Gaussian.
Col. (5) gives the significance of the line defined as $W_{obs}/\sigma_{fit}$,
where $\sigma_{fit}$ is the error in the equivalent width, and col. (6) gives the
significance of the line defined as $W_{obs}/\sigma_{det}$, where $\sigma_{det}$
is the detection limit of the data at $\lambda_{c}$ (see text for further
details).
The reduced $\chi^{2}$ is listed in col. (7), and if this value
is identical to that of an adjacent line, it indicates that the lines were fitted
simultaneously. Cols. (8)$-$(10) contain the redshift system associated with
the identification and the residual in the sense $\lambda_{c} -
\lambda_{pred}$, where $\lambda_{pred} = \lambda_0 (1+z_{abs})$.
}
\end{deluxetable}

\begin{deluxetable}{rrrrrrrrlr}
\tabletypesize{\footnotesize}
\tablecaption{Absorption Lines for Q0107$-$0232 (FOS G190H)\label{lineC190}}
\tablewidth{0pt}
\tablehead{
\colhead{Line} &
\colhead{$\lambda_{c}$} &
\colhead{$W_{obs}$} &
\colhead{{FWHM}} &
\colhead{$S\sigma_{fit}$} &
\colhead{$S\sigma_{det}$} &
\colhead{$\chi_{\nu}^{2}$} &
\colhead{$z_{abs}$} &
\colhead{Identification\tablenotemark{a}} &
\colhead{$\Delta\lambda$} \\
\colhead{} &
\colhead{(\AA)} &
\colhead{(\AA)} &
\colhead{(\AA)} &
\colhead{} &
\colhead{} &
\colhead{} &
\colhead{} &
\colhead{} &
\colhead{(\AA)} \\
\colhead{(1)} &
\colhead{(2)} &
\colhead{(3)} &
\colhead{(4)} &
\colhead{(5)} &
\colhead{(6)} &
\colhead{(7)} &
\colhead{(8)} &
\colhead{(9)} &
\colhead{(10)} \\
}
\startdata
   1 & $1656.72\pm 0.19$ & $ 1.02\pm 0.31$ & $ 1.39\pm 0.50$ &  3.30 &  4.24 &  1.12 &  &  &  \\ 
   2 & $1670.43\pm 0.14$ & $ 0.71\pm 0.17$ & $ 1.20\pm 0.33$ &  4.30 &  4.65 &  0.23 & 0.0000 & \ion{Al}{2} 1670  & $-$0.36 \\ 
   3 & $1677.88\pm 0.30$ & $ 1.64\pm 0.34$ & $ 3.30\pm 0.86$ &  4.80 & 12.34 &  0.93 &  &  &  \\ 
   4 & $1702.02\pm 0.28$ & $ 1.34\pm 0.22$ & $ 3.68\pm 0.78$ &  6.00 & 14.63 &  1.24 & 0.4001 & \lya -only  & 0.00 \\ 
   5 & $1732.50\pm 0.34$ & $ 0.85\pm 0.23$ & $ 2.63\pm 0.89$ &  3.60 &  5.99 &  0.56 &  &  &  \\ 
   6 & $1736.11\pm 0.18$ & $ 1.65\pm 0.24$ & $ 2.72\pm 0.48$ &  7.00 & 11.29 &  0.88 &  &  &  \\ 
   7 & $1744.17\pm 0.55$ & $ 0.73\pm 0.32$ & $ 2.90\pm 1.57$ &  2.30 &  6.17 &  0.84 &  &  &  \\ 
   8 & $1746.91\pm 0.22$ & $ 0.65\pm 0.25$ & $ 1.58\pm 0.48$ &  2.60 &  6.06 &  0.84 &  &  &  \\ 
   9 & $1754.24\pm 0.16$ & $ 1.08\pm 0.14$ & $ 2.72\pm 0.43$ &  7.70 & 12.77 &  0.92 &  &  &  \\ 
  10 & $1761.39\pm 0.10$ & $ 0.53\pm 0.26$ & $ 1.16\pm 0.35$ &  2.10 &  7.71 &  0.47 &  &  &  \\ 
  11 & $1763.62\pm 0.68$ & $ 0.77\pm 0.34$ & $ 3.35\pm 1.70$ &  2.20 & 12.55 &  0.47 &  &  &  \\ 
  12 & $1781.29\pm 0.89$ & $ 0.54\pm 0.29$ & $ 3.36\pm 1.50$ &  1.90 & 13.30 &  0.30 &  &  &  \\ 
  13 & $1783.14\pm 0.13$ & $ 0.32\pm 0.25$ & $ 1.35\pm 0.52$ &  1.30 &  7.98 &  0.30 &  &  &  \\ 
  14 & $1796.23\pm 0.20$ & $ 0.26\pm 0.07$ & $ 1.50\pm 0.49$ &  3.60 &  4.12 &  0.36 &  &  &  \\ 
  15 & $1807.05\pm 0.37$ & $ 0.62\pm 0.15$ & $ 3.24\pm 0.95$ &  4.20 &  6.63 &  0.76 &  &  &  \\ 
  16 & $1822.39\pm 0.14$ & $ 0.51\pm 0.10$ & $ 1.53\pm 0.36$ &  5.10 &  4.94 &  1.50 &  &  &  \\ 
  17 & $1853.41\pm 0.09$ & $ 0.79\pm 0.09$ & $ 1.64\pm 0.22$ &  8.60 &  7.83 &  0.83 &  &  &  \\ 
  18 & $1857.74\pm 0.09$ & $ 0.72\pm 0.10$ & $ 1.49\pm 0.24$ &  7.50 &  6.65 &  0.78 &  &  &  \\ 
  19 & $1864.76\pm 0.12$ & $ 1.58\pm 0.14$ & $ 2.70\pm 0.28$ & 11.20 & 13.02 &  0.91 &  &  &  \\ 
  20 & $1878.49\pm 0.07$ & $ 1.76\pm 0.10$ & $ 2.36\pm 0.17$ & 16.90 & 16.81 &  0.56 &  &  &  \\ 
  21 & $1892.65\pm 0.07$ & $ 4.75\pm 0.16$ & $ 4.28\pm 0.18$ & 28.90 & 55.70 &  0.70 &  &  &  \\ 
  22 & $1918.61\pm 0.07$ & $ 1.17\pm 0.09$ & $ 1.86\pm 0.17$ & 12.60 & 16.09 &  0.75 &  &  &  \\ 
  23 & $1962.31\pm 0.06$ & $ 1.33\pm 0.09$ & $ 1.94\pm 0.16$ & 14.60 & 19.03 &  1.59 &  &  &  \\ 
  24 & $1971.14\pm 0.14$ & $ 0.92\pm 0.11$ & $ 2.50\pm 0.37$ &  8.20 & 12.83 &  0.62 &  &  &  \\ 
  25 & $2003.89\pm 0.08$ & $ 1.11\pm 0.10$ & $ 2.03\pm 0.21$ & 11.60 & 20.70 &  0.79 &  &  &  \\ 
  26 & $2027.25\pm 0.07$ & $ 0.61\pm 0.07$ & $ 1.20\pm 0.16$ &  8.70 & 11.46 &  0.89 &  &  &  \\ 
  27 & $2045.29\pm 0.57$ & $ 0.39\pm 0.26$ & $ 1.83\pm 1.00$ &  1.50 &  8.95 &  0.46 &  &  &  \\ 
  28 & $2046.95\pm 0.38$ & $ 0.36\pm 0.25$ & $ 1.48\pm 0.58$ &  1.40 &  8.51 &  0.46 &  &  &  \\ 
  29 & $2053.70\pm 0.04$ & $ 1.19\pm 0.05$ & $ 1.77\pm 0.09$ & 23.00 & 30.91 &  0.95 &  &  &  \\ 
  30 & $2067.00\pm 0.16$ & $ 0.35\pm 0.06$ & $ 2.12\pm 0.39$ &  6.30 & 12.03 &  1.11 &  &  &  \\ 
  31 & $2077.75\pm 0.06$ & $ 1.43\pm 0.10$ & $ 1.89\pm 0.12$ & 13.80 & 66.47 &  1.13 &  &  &  \\ 
  32 & $2079.65\pm 0.17$ & $ 0.39\pm 0.10$ & $ 1.58\pm 0.31$ &  3.90 & 19.73 &  1.13 &  &  &  \\ 
  33 & $2087.85\pm 0.02$ & $ 2.16\pm 0.03$ & $ 2.53\pm 0.04$ & 68.80 & 164.89 &  1.06 &  &  &  \\ 
  34 & $2092.54\pm 0.29$ & $ 0.05\pm 0.02$ & $ 1.31\pm 0.68$ &  2.20 &  5.67 &  0.23 &  &  &  \\ 
  35 & $2113.12\pm 0.12$ & $ 0.20\pm 0.03$ & $ 1.42\pm 0.28$ &  5.90 &  9.35 &  0.34 &  &  &  \\ 
  36 & $2137.71\pm 0.08$ & $ 0.18\pm 0.03$ & $ 1.04\pm 0.19$ &  6.40 &  8.48 &  0.21 &  &  & \\ 
  37 & $2263.61\pm 0.25$ & $ 0.27\pm 0.07$ & $ 2.15\pm 0.64$ &  4.00 &  6.23 &  0.47 &  &  &  \\ 
\enddata
\tablenotetext{a}{
The primary sample from the analysis comprises \lya\ and unidentified
features with $1620.60 < \lambda_{c} < 2089.85$. This corresponds to \lya\
and unidentified lines up through line 33 in this table.}
\tablecomments{
For each absorption feature listed in col. (1), col. (2) gives the central
wavelength in angstroms, col. (3) gives the equivalent width in Angstroms, and
col. (4) gives the full width at half-maximum of the fitted Gaussian.
Col. (5) gives the significance of the line defined as $W_{obs}/\sigma_{fit}$,
where $\sigma_{fit}$ is the error in the equivalent width, and col. (6) gives the
significance of the line defined as $W_{obs}/\sigma_{det}$, where $\sigma_{det}$
is the detection limit of the data at $\lambda_{c}$ (see text for further
details).
The reduced $\chi^{2}$ is listed in col. (7), and if this value
is identical to that of an adjacent line, it indicates that the lines were fitted
simultaneously. Cols. (8)$-$(10) contain the redshift system associated with
the identification and the residual in the sense $\lambda_{c} -
\lambda_{pred}$, where $\lambda_{pred} = \lambda_0 (1+z_{abs})$.
}
\end{deluxetable}

\begin{deluxetable}{lc crlc crlc crl}
\tablewidth{0pt}
\tablecolumns{10}
\rotate
\tabletypesize{\footnotesize}
\tablecaption{Absorbers Associated with Metal-Line Systems \label{linemetal}}
\tablewidth{0pt}
\tablehead{
\colhead{} &
\colhead{} &
\multicolumn{3}{c}{Q0107$-$025A} &
\colhead{} &
\multicolumn{3}{c}{Q0107$-$025B} &
\colhead{} &
\multicolumn{3}{c}{Q0107$-$0232} \\
\colhead{\rule[0mm]{0mm}{5.0mm}} &
\colhead{} &
\colhead{No.} &
\colhead{$z_{abs}$} &
\colhead{Species} &
\colhead{} &
\colhead{No.} &
\colhead{$z_{abs}$} &
\colhead{Species} &
\colhead{} &
\colhead{No.} &
\colhead{$z_{abs}$} &
\colhead{Species}  \\
\colhead{\rule[0mm]{0mm}{6mm}(1)} &
\colhead{} &
\colhead{(2)} &
\colhead{(3)} &
\colhead{(4)} &
\colhead{} &
\colhead{(5)} &
\colhead{(6)} &
\colhead{(7)} &
\colhead{} &
\colhead{(8)} &
\colhead{(9)} &
\colhead{(10)} \\
}
\startdata
System 1 & &  & & & & & & & & & &\\
~~~~~HRS/G140L && 2 &   0.2268 &  Ly$\alpha$, Ly$\beta$ &&   2 & 0.2268 & \lya, \ion{O}{6} 1031/1037 &&     & {\rm No spectrum} &  \\
~~~~~FOS/G190H && 3 &   0.2272 &  \ion{C}{2} 1334, \ion{C}{4} 1548/1550  &&   1 & 0.2273 & \ion{C}{2} 1334  &&   0 & \nodata &  \nodata \\
~~~~~FOS/G270H && 0 &   \nodata & \nodata &&   0 & \nodata & \nodata &&     & {\rm No spectrum} &  \\
System 2 & & & & & & & & & & & &\\
~~~~~HRS/G140L && 0 &   \nodata & \nodata &&   0 & \nodata & \nodata &&     & {\rm No spectrum} &  \\
~~~~~FOS/G190H && 2 &   0.2881 & \ion{C}{4} 1548/1550   &&   0 & \nodata & \nodata &&   0 & \nodata & \nodata  \\ 
~~~~~FOS/G270H && 0 &   \nodata & \nodata &&   0 & \nodata & \nodata &&     & {\rm No spectrum} &  \\
System 3 & & & & & & & & & & & &\\
~~~~~HRS/G140L && 0 &   \nodata & \nodata &&   0 & \nodata & \nodata  &&     & {\rm No spectrum} &  \\
~~~~~FOS/G190H && 2 &   0.3995 & Ly$\alpha$, \ion{C}{4} 1548 &&   4 & 0.3993 & \lya, \ion{C}{4} 1548/1550, \ion{Si}{4} 1393  &&   1 & 0.4001  & \lya \\
~~~~~FOS/G270H && 0 &   \nodata & \nodata &&   0 & \nodata & \nodata  &&     & {\rm No spectrum} &  \\
System 4 & & & & & & & & & & & &\\
~~~~~HRS/G140L && 0 &   \nodata & \nodata &&   0 & \nodata & \nodata  &&     & {\rm No spectrum} &  \\
~~~~~FOS/G190H && 2 &   0.8766 & \lya-\lyg, \ion{N}{1} 1200  &&   2 & 0.8763 & \lya, \lyb  &&    & {\rm $z_{em}$ too low} &  \\
~~~~~FOS/G270H && 2 &   0.8769 & \lya, \ion{N}{1} 1200  &&   1 & 0.8766 & \lya  &&     & {\rm No spectrum} &  \\
\enddata
\tablecomments{We look for lines associated with any of the four
systems with associated metal lines in each of the three instrumental setups shown in 
col. (1). 
The number of strong absorbers used to determine the redshift of the system is
shown in cols. (2), (5), and (8). 
The measured redshift of the absorption-line system is shown in cols. 
(3), (6), and (9),
where ``No spectrum'' indicates that no data were taken with that instrumental
configuration.
All species identified with each system are listed in cols (4), (7), and (10).
}
\end{deluxetable}

\begin{deluxetable}{cccrcccccrccr}
\rotate
\tabletypesize{\footnotesize}
\tablecaption{Symmetric Absorber Pairs for Q0107$-$025A and Q0107$-$025B\label{xppairsAB}}
\tablewidth{0pt}
\tablehead{
\multicolumn{5}{c}{Q0107$-$025A} &
\colhead{} &
\multicolumn{5}{c}{Q0107$-$025B} &
\colhead{} &
\colhead{A$-$B} \\
\cline{1-5} 
\cline{7-11} 
\cline{13-13} 
\colhead{\rule[0mm]{0mm}{5.0mm}Number} &
\colhead{$\lambda$} &
\colhead{$W_{obs}$} &
\colhead{$S\sigma_{det}$} &
\colhead{$z_{abs}$} &
\colhead{} &
\colhead{Number} &
\colhead{$\lambda$} &
\colhead{$W_{obs}$} &
\colhead{$S\sigma_{det}$} &
\colhead{$z_{abs}$} &
\colhead{} &
\colhead{$\Delta v$} \\
\colhead{} &
\colhead{(\AA)} &
\colhead{(\AA)} &
\colhead{} &
\colhead{} &
\colhead{} &
\colhead{} &
\colhead{(\AA)} &
\colhead{(\AA)} &
\colhead{} &
\colhead{} &
\colhead{} &
\colhead{(km s$^{-1}$)} \\
\colhead{(1)} &
\colhead{(2)} &
\colhead{(3)} &
\colhead{(4)} &
\colhead{(5)} &
\colhead{} &
\colhead{(6)} &
\colhead{(7)} &
\colhead{(8)} &
\colhead{(9)} &
\colhead{(10)} &
\colhead{} &
\colhead{(11)} 
}
\startdata
  3 &  1683.17 & $ 0.81 \pm 0.09 $ & 10.74 &  0.3846 &  &  3 &  1680.16 & $ 0.37 \pm 0.12 $ &  3.23 &  0.3821 &  &    537 \\ 
 4 &  1701.30 & $ 0.61 \pm 0.09 $ &  9.64 &  0.3995 &  &  4 &  1701.06 & $ 1.28 \pm 0.12 $ & 12.42 &  0.3993 &  &     42 \\ 
 5 &  1721.69 & $ 0.38 \pm 0.08 $ &  5.41 &  0.4162 &  &  5 &  1717.51 & $ 0.62 \pm 0.16 $ &  5.57 &  0.4128 &  &    729 \\ 
14 &  1824.96 & $ 1.82 \pm 0.06 $ & 40.36 &  0.5012 &  & 11 &  1823.58 & $ 1.50 \pm 0.11 $ & 20.90 &  0.5001 &  &    227 \\ 
16 &  1853.54 & $ 0.21 \pm 0.06 $ &  4.78 &  0.5247 &  & 13 &  1852.68 & $ 0.42 \pm 0.08 $ &  6.02 &  0.5240 &  &    139 \\ 
17 &  1867.12 & $ 2.60 \pm 0.06 $ & 54.57 &  0.5359 &  & 15 &  1866.73 & $ 1.05 \pm 0.11 $ & 13.14 &  0.5356 &  &     63 \\ 
23 &  1921.19 & $ 0.67 \pm 0.06 $ & 18.68 &  0.5804 &  & 16 &  1919.41 & $ 0.60 \pm 0.12 $ & 10.05 &  0.5789 &  &    278 \\ 
28 &  1955.37 & $ 0.30 \pm 0.04 $ &  8.01 &  0.6085 &  & 19 &  1956.40 & $ 0.51 \pm 0.10 $ &  7.97 &  0.6093 &  &   $-$158 \\ 
29 &  1976.50 & $ 0.32 \pm 0.05 $ &  9.23 &  0.6259 &  & 20 &  1971.91 & $ 0.42 \pm 0.08 $ &  7.24 &  0.6221 &  &    697 \\ 
31 &  1987.40 & $ 0.23 \pm 0.06 $ &  6.92 &  0.6348 &  & 21 &  1997.16 & $ 0.37 \pm 0.10 $ &  7.15 &  0.6428 &  &  $-$1469 \\ 
41 &  2041.44 & $ 0.41 \pm 0.03 $ & 19.16 &  0.6793 &  & 22 &  2044.48 & $ 0.10 \pm 0.03 $ &  2.16 &  0.6818 &  &   $-$446 \\ 
42 &  2054.27 & $ 0.42 \pm 0.03 $ & 16.26 &  0.6898 &  & 24 &  2054.50 & $ 0.32 \pm 0.05 $ &  6.88 &  0.6900 &  &    $-$34 \\ 
43 &  2059.89 & $ 0.22 \pm 0.03 $ &  7.89 &  0.6944 &  & 25 &  2064.73 & $ 0.38 \pm 0.11 $ &  7.83 &  0.6984 &  &   $-$704 \\ 
45 &  2082.10 & $ 0.18 \pm 0.03 $ &  5.78 &  0.7127 &  & 26 &  2083.57 & $ 0.60 \pm 0.06 $ & 13.31 &  0.7139 &  &   $-$212 \\ 
46 &  2089.48 & $ 1.59 \pm 0.04 $ & 48.99 &  0.7188 &  & 27 &  2088.83 & $ 1.42 \pm 0.05 $ & 31.87 &  0.7183 &  &     93 \\ 
\enddata
\tablecomments{
For each pair of absorption features, cols. (1) and (6) give the line numbers
from Tables~\ref{lineA190} and \ref{lineB190}, respectively; cols. (2) and (7)
give the central wavelength in angstroms; and cols. (3) and (8) give the equivalent 
widths in angstroms. Cols. (4) and (9) give the significance of the lines 
defined as $W_{obs}/\sigma_{det}$, where $\sigma_{det}$ is the detection limit 
of the data at $\lambda_{c}$ (see text for further details).  Cols. (5) and 
(10) contain the redshift of the absorption features, and col. (11) give the 
separation of the two features in velocity units, in the sense $\lambda_{A} 
- \lambda_{B}$.
}
\end{deluxetable}

\begin{deluxetable}{cccrcccccrccr}
\rotate
\tabletypesize{\footnotesize}
\tablecaption{Symmetric Absorber Pairs for Q0107$-$025A and Q0107$-$0232\label{xppairsAC}}
\tablewidth{0pt}
\tablehead{
\multicolumn{5}{c}{Q0107$-$025A} &
\colhead{} &
\multicolumn{5}{c}{Q0107$-$0232} &
\colhead{} &
\colhead{A$-$C} \\
\cline{1-5} 
\cline{7-11} 
\cline{13-13} 
\colhead{\rule[0mm]{0mm}{5.0mm}Number} &
\colhead{$\lambda$} &
\colhead{$W_{obs}$} &
\colhead{$S\sigma_{det}$} &
\colhead{$z_{abs}$} &
\colhead{} &
\colhead{Number} &
\colhead{$\lambda$} &
\colhead{$W_{obs}$} &
\colhead{$S\sigma_{det}$} &
\colhead{$z_{abs}$} &
\colhead{} &
\colhead{$\Delta v$} \\
\colhead{} &
\colhead{(\AA)} &
\colhead{(\AA)} &
\colhead{} &
\colhead{} &
\colhead{} &
\colhead{} &
\colhead{(\AA)} &
\colhead{(\AA)} &
\colhead{} &
\colhead{} &
\colhead{} &
\colhead{(km s$^{-1}$)} \\
\colhead{(1)} &
\colhead{(2)} &
\colhead{(3)} &
\colhead{(4)} &
\colhead{(5)} &
\colhead{} &
\colhead{(6)} &
\colhead{(7)} &
\colhead{(8)} &
\colhead{(9)} &
\colhead{(10)} &
\colhead{} &
\colhead{(11)} 
}
\startdata
  3 &  1683.17 & $ 0.81 \pm 0.09 $ & 10.74 &  0.3846 &  &  3 &  1677.88 & $ 1.64 \pm 0.34 $ & 12.34 &  0.3802 &  &    944 \\ 
 4 &  1701.30 & $ 0.61 \pm 0.09 $ &  9.64 &  0.3995 &  &  4 &  1702.02 & $ 1.34 \pm 0.22 $ & 14.63 &  0.4001 &  &   $-$127 \\ 
 6 &  1727.48 & $ 0.46 \pm 0.12 $ &  7.07 &  0.4210 &  &  5 &  1732.50 & $ 0.85 \pm 0.23 $ &  5.99 &  0.4251 &  &   $-$870 \\ 
13 &  1808.65 & $ 0.52 \pm 0.07 $ & 11.46 &  0.4878 &  & 15 &  1807.05 & $ 0.62 \pm 0.15 $ &  6.63 &  0.4865 &  &    265 \\ 
14 &  1824.96 & $ 1.82 \pm 0.06 $ & 40.36 &  0.5012 &  & 16 &  1822.39 & $ 0.51 \pm 0.10 $ &  4.94 &  0.4991 &  &    422 \\ 
16 &  1853.54 & $ 0.21 \pm 0.06 $ &  4.78 &  0.5247 &  & 17 &  1853.41 & $ 0.79 \pm 0.09 $ &  7.83 &  0.5246 &  &     21 \\ 
17 &  1867.12 & $ 2.60 \pm 0.06 $ & 54.57 &  0.5359 &  & 19 &  1864.76 & $ 1.58 \pm 0.14 $ & 13.02 &  0.5339 &  &    379 \\ 
18 &  1877.42 & $ 0.14 \pm 0.04 $ &  3.27 &  0.5444 &  & 20 &  1878.49 & $ 1.76 \pm 0.10 $ & 16.81 &  0.5452 &  &   $-$171 \\ 
19 &  1893.49 & $ 0.77 \pm 0.07 $ & 18.23 &  0.5576 &  & 21 &  1892.65 & $ 4.75 \pm 0.16 $ & 55.70 &  0.5569 &  &    133 \\ 
23 &  1921.19 & $ 0.67 \pm 0.06 $ & 18.68 &  0.5804 &  & 22 &  1918.61 & $ 1.17 \pm 0.09 $ & 16.09 &  0.5782 &  &    403 \\ 
28 &  1955.37 & $ 0.30 \pm 0.04 $ &  8.01 &  0.6085 &  & 23 &  1962.31 & $ 1.33 \pm 0.09 $ & 19.03 &  0.6142 &  &  $-$1062 \\ 
29 &  1976.50 & $ 0.32 \pm 0.05 $ &  9.23 &  0.6259 &  & 24 &  1971.14 & $ 0.92 \pm 0.11 $ & 12.83 &  0.6214 &  &    814 \\ 
34 &  2007.02 & $ 0.32 \pm 0.03 $ & 14.42 &  0.6510 &  & 25 &  2003.89 & $ 1.11 \pm 0.10 $ & 20.70 &  0.6484 &  &    468 \\ 
37 &  2025.66 & $ 1.13 \pm 0.22 $ & 82.36 &  0.6663 &  & 26 &  2027.25 & $ 0.61 \pm 0.07 $ & 11.46 &  0.6676 &  &   $-$235 \\ 
41 &  2041.44 & $ 0.41 \pm 0.03 $ & 19.16 &  0.6793 &  & 27 &  2045.29 & $ 0.39 \pm 0.26 $ &  8.95 &  0.6824 &  &   $-$565 \\ 
42 &  2054.27 & $ 0.42 \pm 0.03 $ & 16.26 &  0.6898 &  & 29 &  2053.70 & $ 1.19 \pm 0.05 $ & 30.91 &  0.6894 &  &     83 \\ 
44 &  2073.69 & $ 0.19 \pm 0.03 $ &  6.36 &  0.7058 &  & 31 &  2077.75 & $ 1.43 \pm 0.10 $ & 66.47 &  0.7091 &  &   $-$586 \\ 
45 &  2082.10 & $ 0.18 \pm 0.03 $ &  5.78 &  0.7127 &  & 32 &  2079.65 & $ 0.39 \pm 0.10 $ & 19.73 &  0.7107 &  &    353 \\ 
46 &  2089.48 & $ 1.59 \pm 0.04 $ & 48.99 &  0.7188 &  & 33 &  2087.85 & $ 2.16 \pm 0.03 $ & 164.89 &  0.7174 &  &    234 \\ 
\enddata
\tablecomments{
For each pair of absorption features, cols. (1) and (6) give the line numbers
from Tables~\ref{lineA190} and \ref{lineC190}, respectively; cols. (2) and (7)
give the central wavelength in angstroms; and cols. (3) and (8) give the equivalent 
widths in angstroms. Cols. (4) and (9) give the significance of the lines 
defined as $W_{obs}/\sigma_{det}$, where $\sigma_{det}$ is the detection limit 
of the data at $\lambda_{c}$ (see text for further details).  Cols. (5) and 
(10) contain the redshift of the absorption features, and col. (11) gives the 
separation of the two features in velocity units, in the sense $\lambda_{A} 
- \lambda_{C}$.
}
\end{deluxetable}

\begin{deluxetable}{cccrcccccrccr}
\rotate
\tabletypesize{\footnotesize}
\tablecaption{Symmetric Absorber Pairs for Q0107$-$025B and Q0107$-$0232\label{xppairsBC}}
\tablewidth{0pt}
\tablehead{
\multicolumn{5}{c}{Q0107$-$025B} &
\colhead{} &
\multicolumn{5}{c}{Q0107$-$0232} &
\colhead{} &
\colhead{B$-$C} \\
\cline{1-5} 
\cline{7-11} 
\cline{13-13} 
\colhead{\rule[0mm]{0mm}{5.0mm}Number} &
\colhead{$\lambda$} &
\colhead{$W_{obs}$} &
\colhead{$S\sigma_{det}$} &
\colhead{$z_{abs}$} &
\colhead{} &
\colhead{Number} &
\colhead{$\lambda$} &
\colhead{$W_{obs}$} &
\colhead{$S\sigma_{det}$} &
\colhead{$z_{abs}$} &
\colhead{} &
\colhead{$\Delta v$} \\
\colhead{} &
\colhead{(\AA)} &
\colhead{(\AA)} &
\colhead{} &
\colhead{} &
\colhead{} &
\colhead{} &
\colhead{(\AA)} &
\colhead{(\AA)} &
\colhead{} &
\colhead{} &
\colhead{} &
\colhead{(km s$^{-1}$)} \\
\colhead{(1)} &
\colhead{(2)} &
\colhead{(3)} &
\colhead{(4)} &
\colhead{(5)} &
\colhead{} &
\colhead{(6)} &
\colhead{(7)} &
\colhead{(8)} &
\colhead{(9)} &
\colhead{(10)} &
\colhead{} &
\colhead{(11)} 
}
\startdata
  3 &  1680.16 & $ 0.37 \pm 0.12 $ &  3.23 &  0.3821 &  &  3 &  1677.88 & $ 1.64 \pm 0.34 $ & 12.34 &  0.3802 &  &    407 \\ 
 4 &  1701.06 & $ 1.28 \pm 0.12 $ & 12.42 &  0.3993 &  &  4 &  1702.02 & $ 1.34 \pm 0.22 $ & 14.63 &  0.4001 &  &   $-$169 \\ 
 7 &  1746.62 & $ 0.40 \pm 0.07 $ &  4.49 &  0.4368 &  &  8 &  1746.91 & $ 0.65 \pm 0.25 $ &  6.06 &  0.4370 &  &    $-$50 \\ 
 8 &  1759.89 & $ 0.55 \pm 0.13 $ &  6.77 &  0.4477 &  & 10 &  1761.39 & $ 0.53 \pm 0.26 $ &  7.71 &  0.4489 &  &   $-$255 \\ 
10 &  1784.15 & $ 0.79 \pm 0.34 $ & 10.67 &  0.4676 &  & 13 &  1783.14 & $ 0.32 \pm 0.25 $ &  7.98 &  0.4668 &  &    170 \\ 
11 &  1823.58 & $ 1.50 \pm 0.11 $ & 20.90 &  0.5001 &  & 16 &  1822.39 & $ 0.51 \pm 0.10 $ &  4.94 &  0.4991 &  &    196 \\ 
13 &  1852.68 & $ 0.42 \pm 0.08 $ &  6.02 &  0.5240 &  & 17 &  1853.41 & $ 0.79 \pm 0.09 $ &  7.83 &  0.5246 &  &   $-$118 \\ 
15 &  1866.73 & $ 1.05 \pm 0.11 $ & 13.14 &  0.5356 &  & 19 &  1864.76 & $ 1.58 \pm 0.14 $ & 13.02 &  0.5339 &  &    317 \\ 
16 &  1919.41 & $ 0.60 \pm 0.12 $ & 10.05 &  0.5789 &  & 22 &  1918.61 & $ 1.17 \pm 0.09 $ & 16.09 &  0.5782 &  &    125 \\ 
19 &  1956.40 & $ 0.51 \pm 0.10 $ &  7.97 &  0.6093 &  & 23 &  1962.31 & $ 1.33 \pm 0.09 $ & 19.03 &  0.6142 &  &   $-$904 \\ 
20 &  1971.91 & $ 0.42 \pm 0.08 $ &  7.24 &  0.6221 &  & 24 &  1971.14 & $ 0.92 \pm 0.11 $ & 12.83 &  0.6214 &  &    117 \\ 
21 &  1997.16 & $ 0.37 \pm 0.10 $ &  7.15 &  0.6428 &  & 25 &  2003.89 & $ 1.11 \pm 0.10 $ & 20.70 &  0.6484 &  &  $-$1009 \\ 
22 &  2044.48 & $ 0.10 \pm 0.03 $ &  2.16 &  0.6818 &  & 27 &  2045.29 & $ 0.39 \pm 0.26 $ &  8.95 &  0.6824 &  &   $-$119 \\ 
23 &  2048.84 & $ 0.17 \pm 0.06 $ &  3.76 &  0.6854 &  & 28 &  2046.95 & $ 0.36 \pm 0.25 $ &  8.51 &  0.6838 &  &    277 \\ 
24 &  2054.50 & $ 0.32 \pm 0.05 $ &  6.88 &  0.6900 &  & 29 &  2053.70 & $ 1.19 \pm 0.05 $ & 30.91 &  0.6894 &  &    117 \\ 
25 &  2064.73 & $ 0.38 \pm 0.11 $ &  7.83 &  0.6984 &  & 30 &  2067.00 & $ 0.35 \pm 0.06 $ & 12.03 &  0.7003 &  &   $-$329 \\ 
26 &  2083.57 & $ 0.60 \pm 0.06 $ & 13.31 &  0.7139 &  & 32 &  2079.65 & $ 0.39 \pm 0.10 $ & 19.73 &  0.7107 &  &    565 \\ 
27 &  2088.83 & $ 1.42 \pm 0.05 $ & 31.87 &  0.7183 &  & 33 &  2087.85 & $ 2.16 \pm 0.03 $ & 164.89 &  0.7174 &  &    141 \\ 
\enddata
\tablecomments{
For each pair of absorption features, cols. (1) and (6) give the line numbers
from Tables~\ref{lineB190} and \ref{lineC190}, respectively; cols. (2) and (7)
give the central wavelength in angstroms; and cols. (3) and (8) give the equivalent 
widths in angstroms. Cols. (4) and (9) give the significance of the lines 
defined as $W_{obs}/\sigma_{det}$, where $\sigma_{det}$ is the detection limit 
of the data at $\lambda_{c}$ (see text for further details).  Cols. (5) and 
(10) contain the redshift of the absorption features, and col. (11) gives the 
separation of the two features in velocity units, in the sense $\lambda_{B} 
- \lambda_{C}$.
}
\end{deluxetable}

\thispagestyle{empty}
\begin{deluxetable}{cccrcccccrcccccrccrrr}
\rotate
\tabletypesize{\tiny}
\tablecaption{Symmetric Triple Absorber Matches\label{xptriples}}
\tablewidth{0pt}
\tablehead{
\multicolumn{5}{c}{Q0107$-$025A} &
\colhead{} &
\multicolumn{5}{c}{Q0107$-$025B} &
\colhead{} &
\multicolumn{5}{c}{Q0107$-$0232 (C)} &
\colhead{} &
\colhead{A$-$B} &
\colhead{B$-$C} &
\colhead{C$-$A} \\
\cline{1-5} 
\cline{7-11} 
\cline{13-17} 
\cline{19-19} 
\cline{20-20} 
\cline{21-21} \\ 
\colhead{\rule[0mm]{0mm}{5.0mm}Number} &
\colhead{$\lambda$} &
\colhead{$W_{obs}$} &
\colhead{$S\sigma_{det}$} &
\colhead{$z_{abs}$} &
\colhead{} &
\colhead{Number} &
\colhead{$\lambda$} &
\colhead{$W_{obs}$} &
\colhead{$S\sigma_{det}$} &
\colhead{$z_{abs}$} &
\colhead{} &
\colhead{Number} &
\colhead{$\lambda$} &
\colhead{$W_{obs}$} &
\colhead{$S\sigma_{det}$} &
\colhead{$z_{abs}$} &
\colhead{} &
\colhead{$\Delta v$} &
\colhead{$\Delta v$} &
\colhead{$\Delta v$} \\
\colhead{} &
\colhead{(\AA)} &
\colhead{(\AA)} &
\colhead{} &
\colhead{} &
\colhead{} &
\colhead{} &
\colhead{(\AA)} &
\colhead{(\AA)} &
\colhead{} &
\colhead{} &
\colhead{} &
\colhead{} &
\colhead{(\AA)} &
\colhead{(\AA)} &
\colhead{} &
\colhead{} &
\colhead{} &
\colhead{(km s$^{-1}$)} &
\colhead{(km s$^{-1}$)} &
\colhead{(km s$^{-1}$)} \\
\colhead{(1)} &
\colhead{(2)} &
\colhead{(3)} &
\colhead{(4)} &
\colhead{(5)} &
\colhead{} &
\colhead{(6)} &
\colhead{(7)} &
\colhead{(8)} &
\colhead{(9)} &
\colhead{(10)} &
\colhead{} &
\colhead{(11)} &
\colhead{(12)} &
\colhead{(13)} &
\colhead{(14)} &
\colhead{(15)} &
\colhead{} &
\colhead{(16)} &
\colhead{(17)} &
\colhead{(18)} \\
}
\startdata
  3 &  1683.17 & $ 0.81 \pm 0.09 $ & 10.74 &  0.3846 &  &  3 &  1680.16 & $ 0.37 \pm 0.12 $ &  3.23 &  0.3821 &  &  3 &  1677.88 & $ 1.64 \pm 0.34 $ & 12.34 &  0.3802 &  &    537 &    407 &   $-$944 \\ 
 4 &  1701.30 & $ 0.61 \pm 0.09 $ &  9.64 &  0.3995 &  &  4 &  1701.06 & $ 1.28 \pm 0.12 $ & 12.42 &  0.3993 &  &  4 &  1702.02 & $ 1.34 \pm 0.22 $ & 14.63 &  0.4001 &  &     42 &   $-$169 &   127 \\ 
14 &  1824.96 & $ 1.82 \pm 0.06 $ & 40.36 &  0.5012 &  & 11 &  1823.58 & $ 1.50 \pm 0.11 $ & 20.90 &  0.5001 &  & 16 &  1822.39 & $ 0.51 \pm 0.10 $ &  4.94 &  0.4991 &  &    227 &    196 &   $-$422 \\ 
16 &  1853.54 & $ 0.21 \pm 0.06 $ &  4.78 &  0.5247 &  & 13 &  1852.68 & $ 0.42 \pm 0.08 $ &  6.02 &  0.5240 &  & 17 &  1853.41 & $ 0.79 \pm 0.09 $ &  7.83 &  0.5246 &  &    139 &   $-$118 &    $-$21 \\ 
17 &  1867.12 & $ 2.60 \pm 0.06 $ & 54.57 &  0.5359 &  & 15 &  1866.73 & $ 1.05 \pm 0.11 $ & 13.14 &  0.5356 &  & 19 &  1864.76 & $ 1.58 \pm 0.14 $ & 13.02 &  0.5339 &  &     63 &    317 &   $-$379 \\ 
23 &  1921.19 & $ 0.67 \pm 0.06 $ & 18.68 &  0.5804 &  & 16 &  1919.41 & $ 0.60 \pm 0.12 $ & 10.05 &  0.5789 &  & 22 &  1918.61 & $ 1.17 \pm 0.09 $ & 16.09 &  0.5782 &  &    278 &    125 &   $-$403 \\ 
28 &  1955.37 & $ 0.30 \pm 0.04 $ &  8.01 &  0.6085 &  & 19 &  1956.40 & $ 0.51 \pm 0.10 $ &  7.97 &  0.6093 &  & 23 &  1962.31 & $ 1.33 \pm 0.09 $ & 19.03 &  0.6142 &  &   $-$158 &   $-$904 &  1062 \\ 
29 &  1976.50 & $ 0.32 \pm 0.05 $ &  9.23 &  0.6259 &  & 20 &  1971.91 & $ 0.42 \pm 0.08 $ &  7.24 &  0.6221 &  & 24 &  1971.14 & $ 0.92 \pm 0.11 $ & 12.83 &  0.6214 &  &    697 &    117 &   $-$814 \\ 
41 &  2041.44 & $ 0.41 \pm 0.03 $ & 19.16 &  0.6793 &  & 22 &  2044.48 & $ 0.10 \pm 0.03 $ &  2.16 &  0.6818 &  & 27 &  2045.29 & $ 0.39 \pm 0.26 $ &  8.95 &  0.6824 &  &   $-$446 &   $-$119 &   565 \\ 
42 &  2054.27 & $ 0.42 \pm 0.03 $ & 16.26 &  0.6898 &  & 24 &  2054.50 & $ 0.32 \pm 0.05 $ &  6.88 &  0.6900 &  & 29 &  2053.70 & $ 1.19 \pm 0.05 $ & 30.91 &  0.6894 &  &    $-$34 &    117 &    $-$83 \\ 
45 &  2082.10 & $ 0.18 \pm 0.03 $ &  5.78 &  0.7127 &  & 26 &  2083.57 & $ 0.60 \pm 0.06 $ & 13.31 &  0.7139 &  & 32 &  2079.65 & $ 0.39 \pm 0.10 $ & 19.73 &  0.7107 &  &   $-$212 &    565 &   $-$353 \\ 
46 &  2089.48 & $ 1.59 \pm 0.04 $ & 48.99 &  0.7188 &  & 27 &  2088.83 & $ 1.42 \pm 0.05 $ & 31.87 &  0.7183 &  & 33 &  2087.85 & $ 2.16 \pm 0.03 $ & 164.89 &  0.7174 &  &     93 &    141 &   $-$234 \\ 
\enddata
\tablecomments{
For each pair of absorption features, cols. (1), (6) and (11) give the line 
numbers from Tables~\ref{lineA190}, \ref{lineB190}, and \ref{lineC190}, 
respectively; cols. (2), (7), and (12) give the central wavelength in angstroms; 
and cols. (3), (8), and (13) give the equivalent 
widths in angstroms. Cols. (4), (9), and (14) give the significance of the lines 
defined as $W_{obs}/\sigma_{det}$, where $\sigma_{det}$ is the detection limit 
of the data at $\lambda_{c}$ (see text for further details).  Cols. (5), (10),
and (15) contain the redshift of the absorption features, and cols. (16), (17),
and (18) give the separation of the features in velocity units, in the sense 
$\lambda_{\rm A} - \lambda_{\rm B}$, $\lambda_{\rm B} - \lambda_{\rm C}$, and
$\lambda_{\rm C} - \lambda_{\rm A}$.
}
\end{deluxetable}

\begin{deluxetable}{ccccccccc}
\rotate
\tabletypesize{\footnotesize}
\tablecaption{Probability that Observed Absorber Matches Are Random\label{prob}}
\tablewidth{0pt}
\tablehead{
\multicolumn{1}{c}{} &
\colhead{} &
\multicolumn{3}{c}{Symmetric Matches} &
\colhead{} &
\multicolumn{3}{c}{Nearest Neighbor Matches} \\
\cline{3-5} 
\cline{7-9} 
\colhead{\rule[0mm]{0mm}{5.0mm}Paired Lines} &
\colhead{} &
\colhead{$\le$240} &
\colhead{$\le$420} &
\colhead{$\le$600} &
\colhead{} &
\colhead{$\le$240} &
\colhead{$\le$420} &
\colhead{$\le$600} \\
\colhead{of Sight} &
\colhead{} &
\colhead{(km\  s$^{-1}$)} &
\colhead{(km\  s$^{-1}$)} &
\colhead{(km\  s$^{-1}$)} &
\colhead{} &
\colhead{(km\  s$^{-1}$)} &
\colhead{(km\  s$^{-1}$)} &
\colhead{(km\  s$^{-1}$)} \\
\colhead{(1)} &
\colhead{} &
\colhead{(2)} &
\colhead{(3)} &
\colhead{(4)} &
\colhead{} &
\colhead{(5)} &
\colhead{(6)} &
\colhead{(7)} 
}
\startdata
A,B & & 0.046 & 0.148 & 0.140 & & 0.050 & 0.191 & 0.228  \\
A,C & & 0.405 & 0.329 & 0.176 & & 0.415 & 0.351 & 0.299  \\
B,C & & 0.010 & 0.002 & 0.004 & & 0.010 & 0.002 & 0.009  \\
\enddata
\tablecomments{
For each configuration of quasar lines of sight in col. (1), we compute the 
probability that the number of observed absorber matches (symmetric or nearest 
neighbor) would occur randomly.
This is done with a Monte Carlo simulation having 10,000 realizations for each
of the quasar pair configurations and 
counting the number of times the number of absorber matches in each Monte 
Carlo realization equals or exceeds the number of observed absorber matches.
}
\end{deluxetable}

\begin{figure}[hp]
\vskip -1.5truecm
\mbox{ \includegraphics[scale=.85, angle=0]{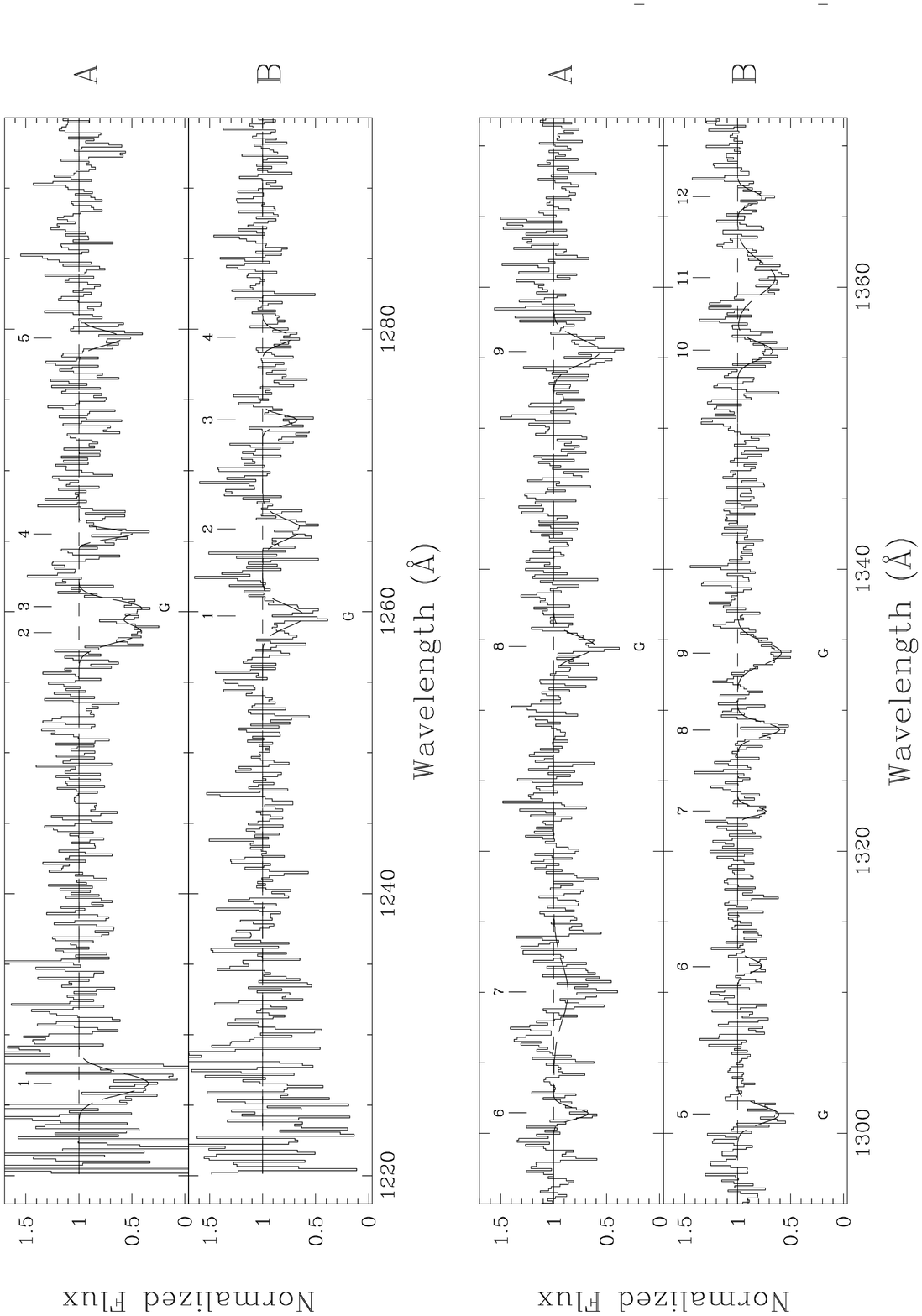}}
\label{expspec}
\end{figure}

\begin{figure}[hp]
\vskip -1.5truecm
\mbox{ \includegraphics[scale=.85, angle=0]{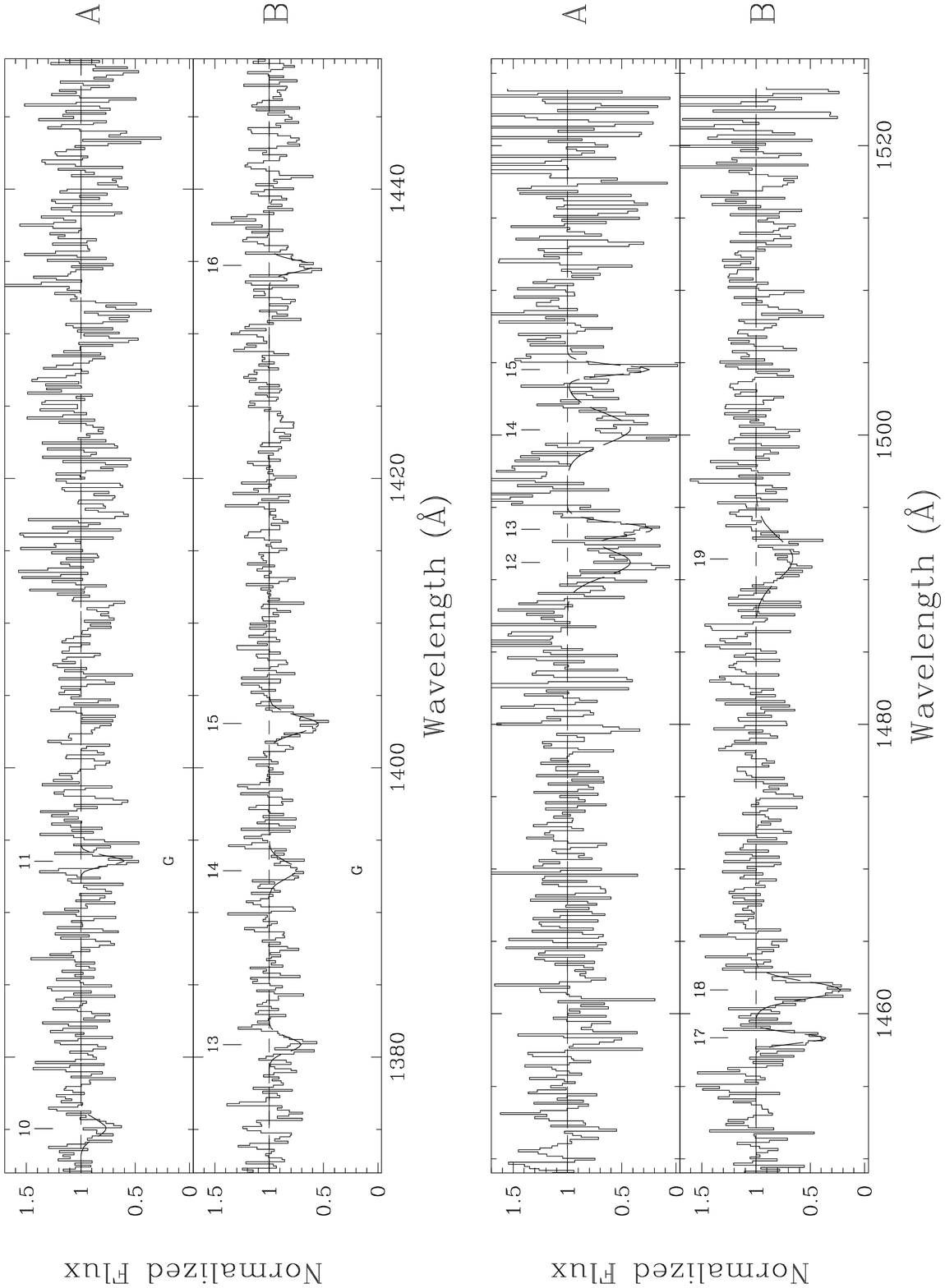}}
\caption{Expanded plot of the {\it HST} HRS G140L spectra for Q0107$-$025 A, B. The absorption features identified in Tables \ref{lineA140} and \ref{lineB140} are marked, and the fits are overplotted with a solid line. }
\end{figure}
\begin{figure}[hp]
\vskip -1.5truecm
\mbox{ \includegraphics[scale=.85, angle=0]{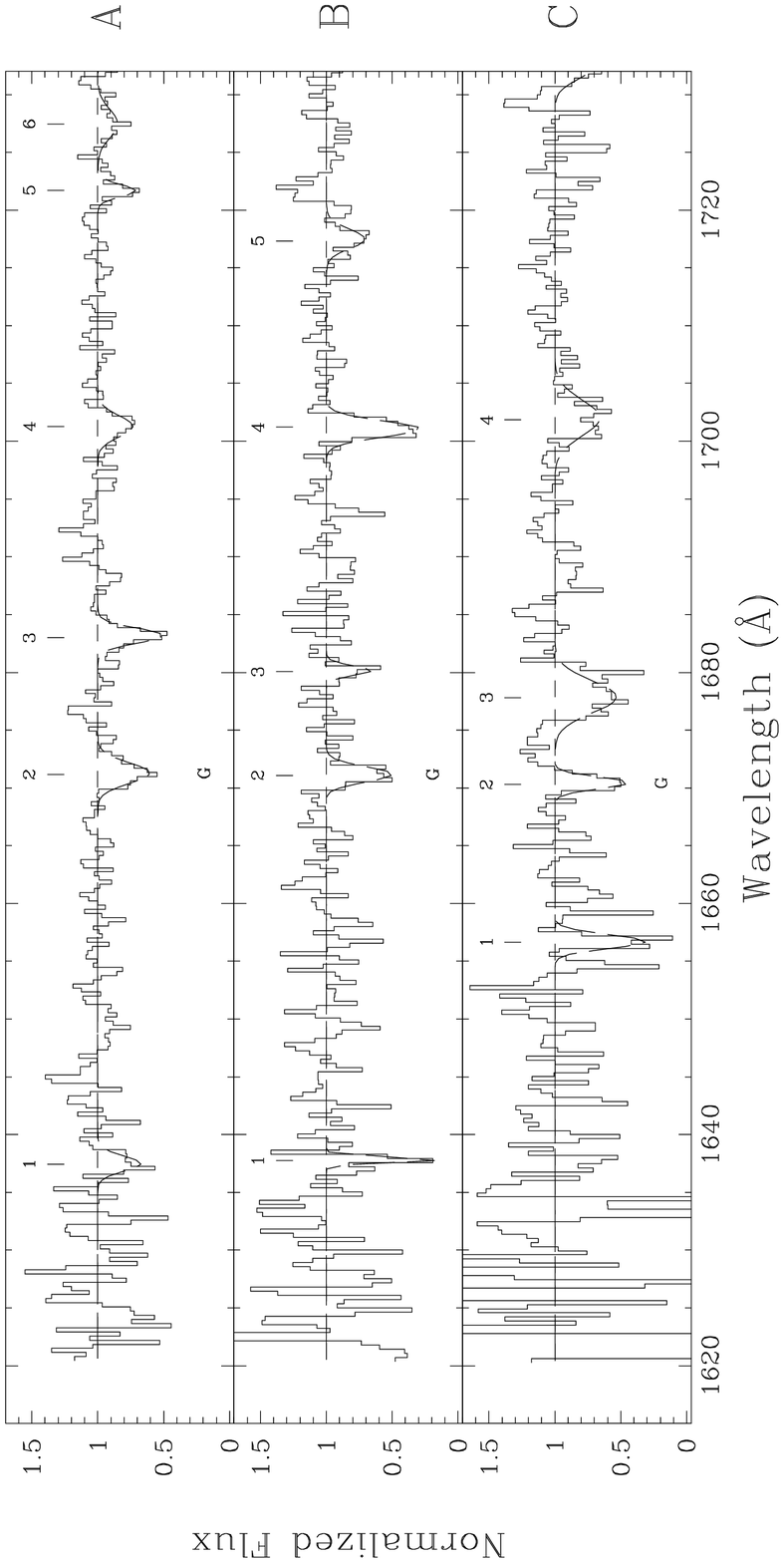}}
\end{figure}
\begin{figure}[hp]
\vskip -1.5truecm
\mbox{ \includegraphics[scale=.85, angle=0]{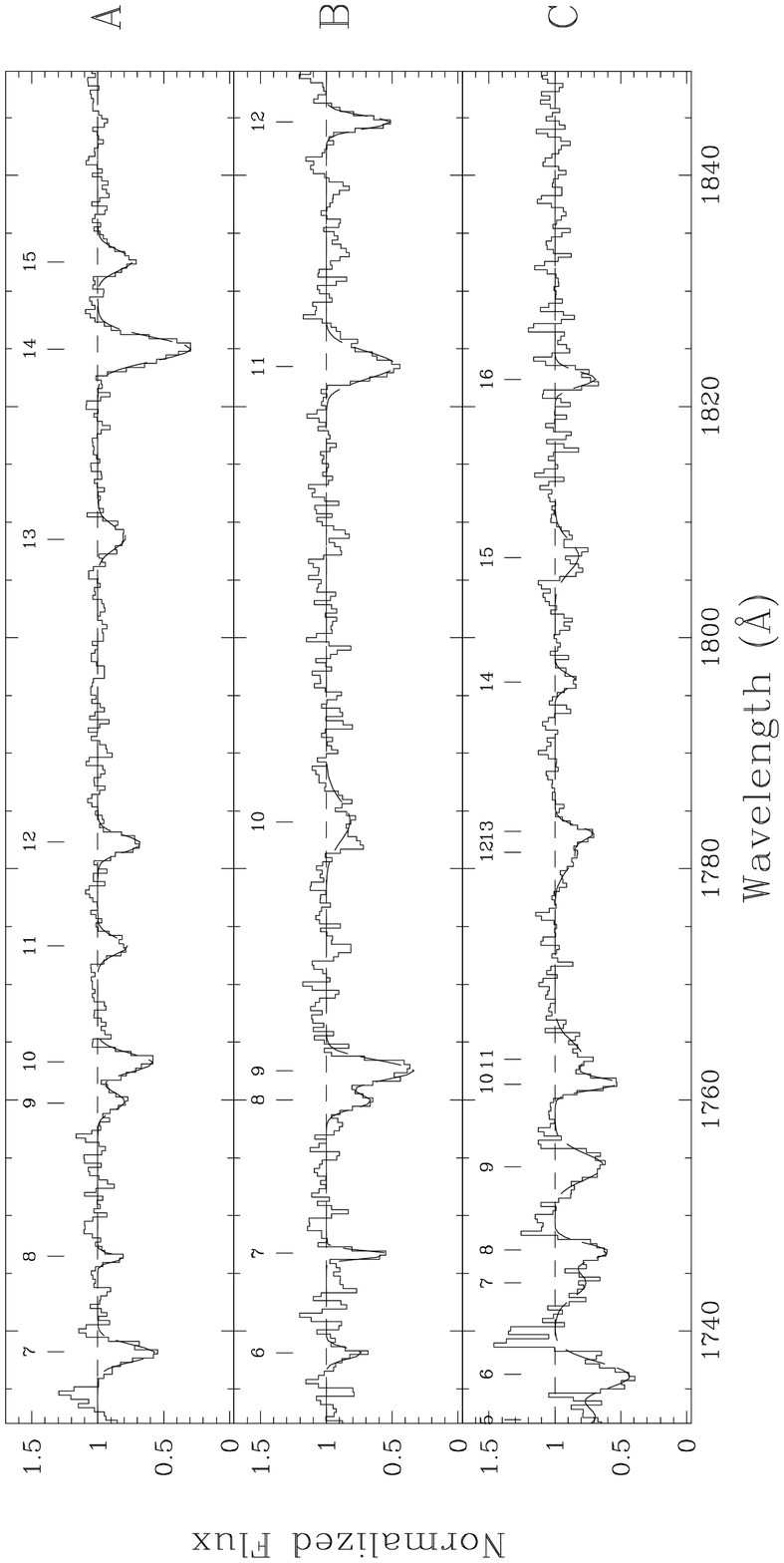}}
\end{figure}
\begin{figure}[hp]
\vskip -1.5truecm
\mbox{ \includegraphics[scale=.85, angle=0]{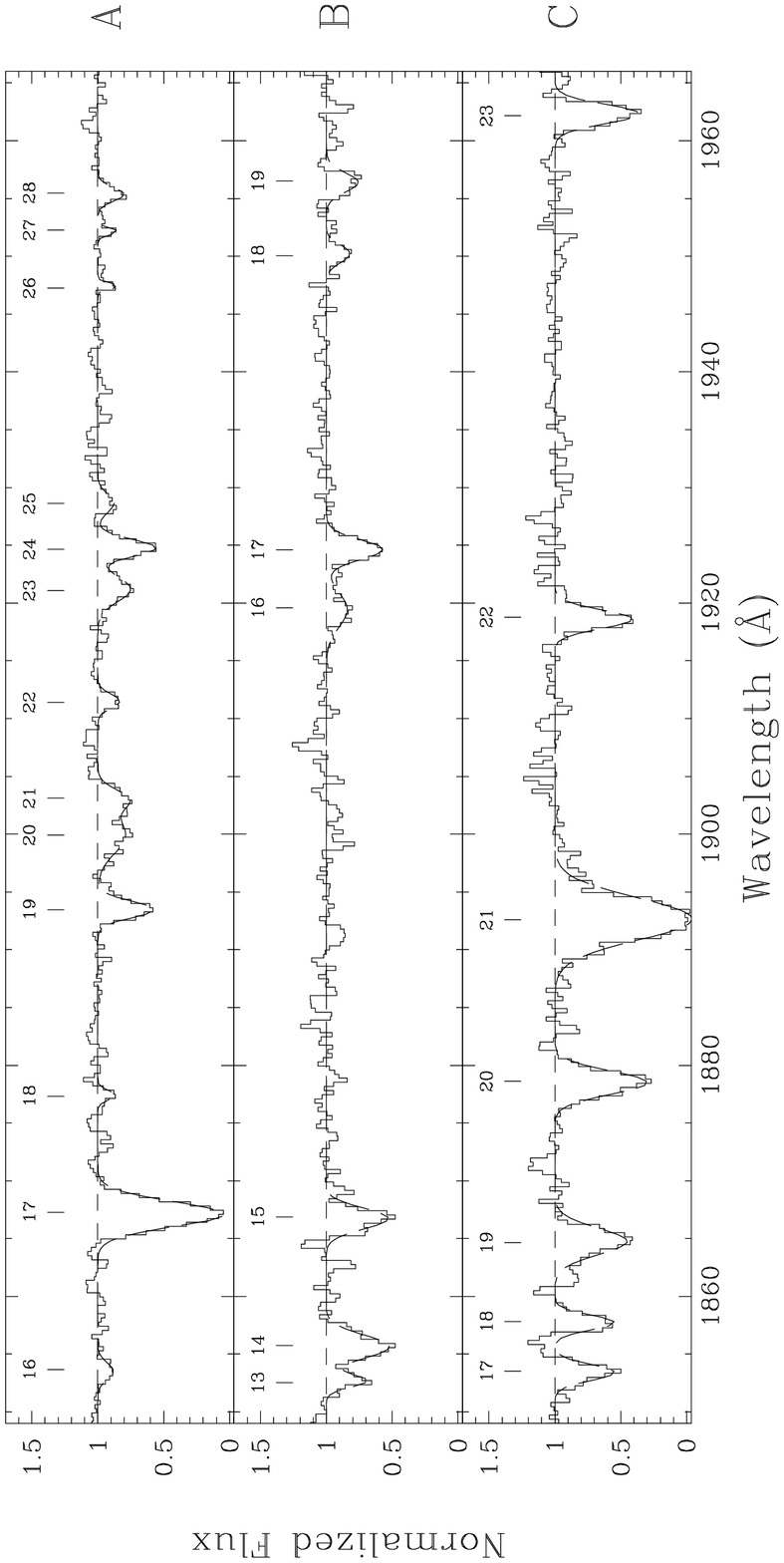}}
\end{figure}
\begin{figure}[hp]
\vskip -1.5truecm
\mbox{ \includegraphics[scale=.85, angle=0]{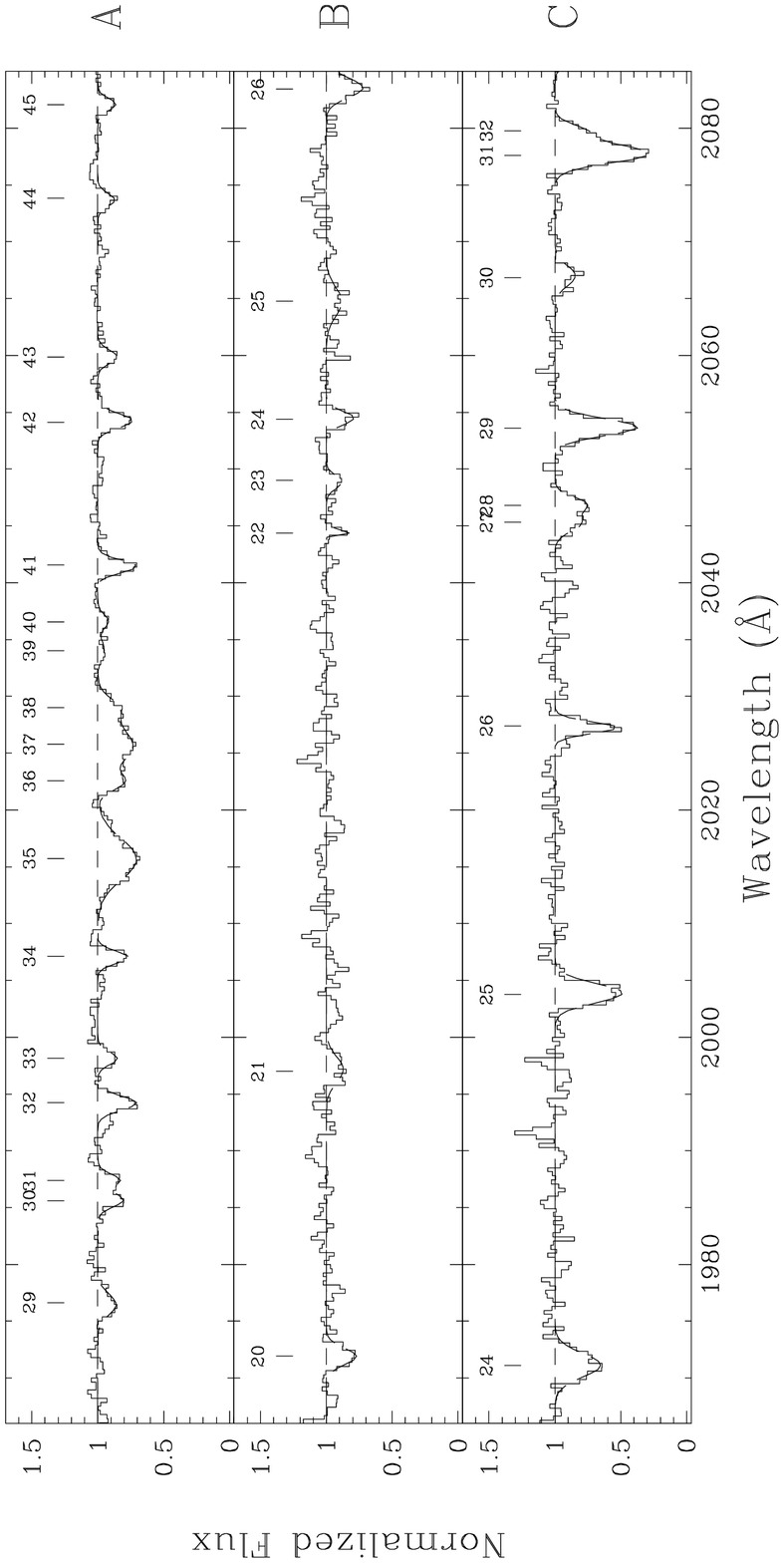}}
\end{figure}
\begin{figure}[hp]
\vskip -1.5truecm
\mbox{ \includegraphics[scale=.85, angle=0]{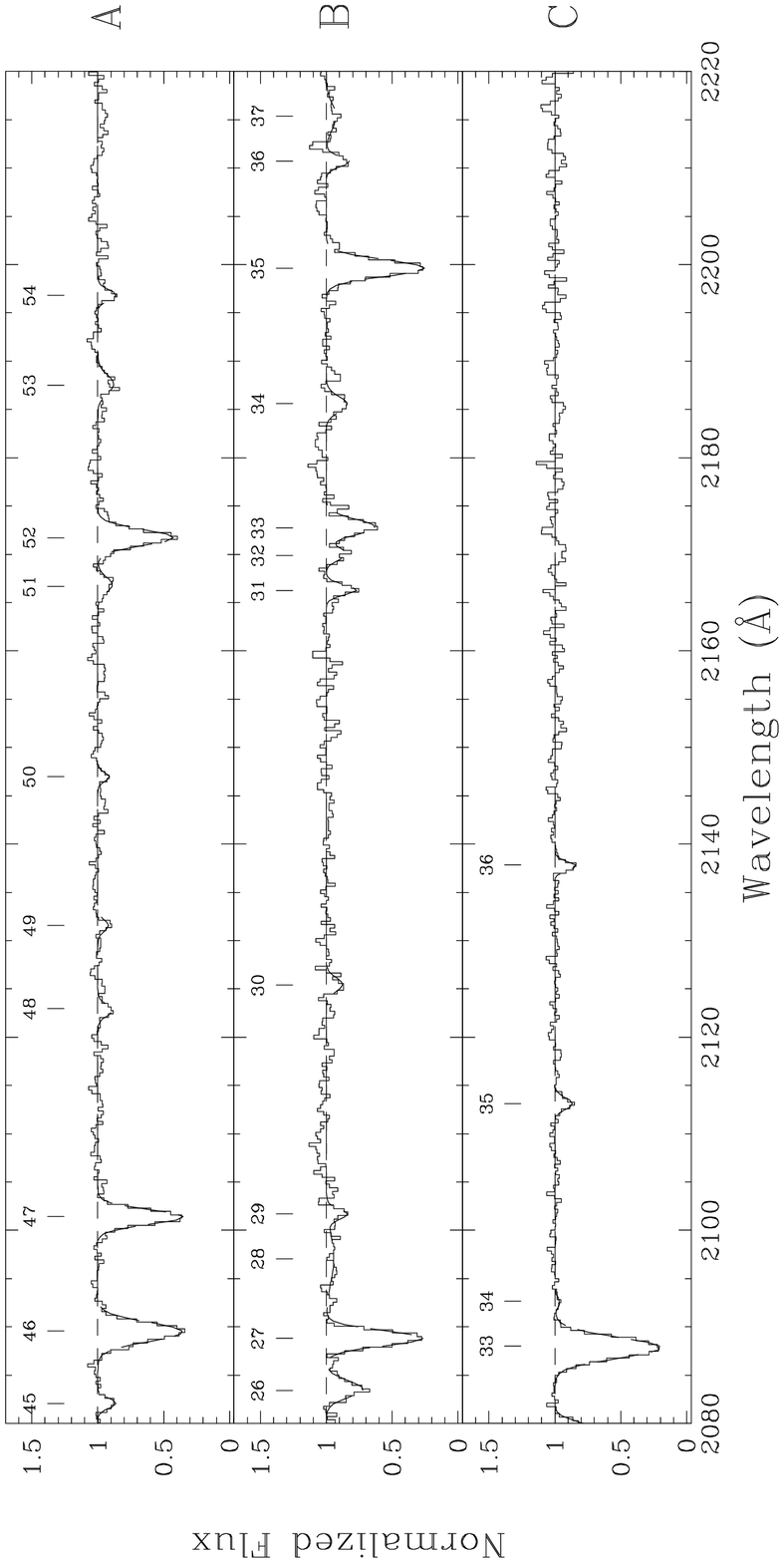}}
\end{figure}
\begin{figure}[hp]
\vskip -1.5truecm
\mbox{ \includegraphics[scale=.85, angle=0]{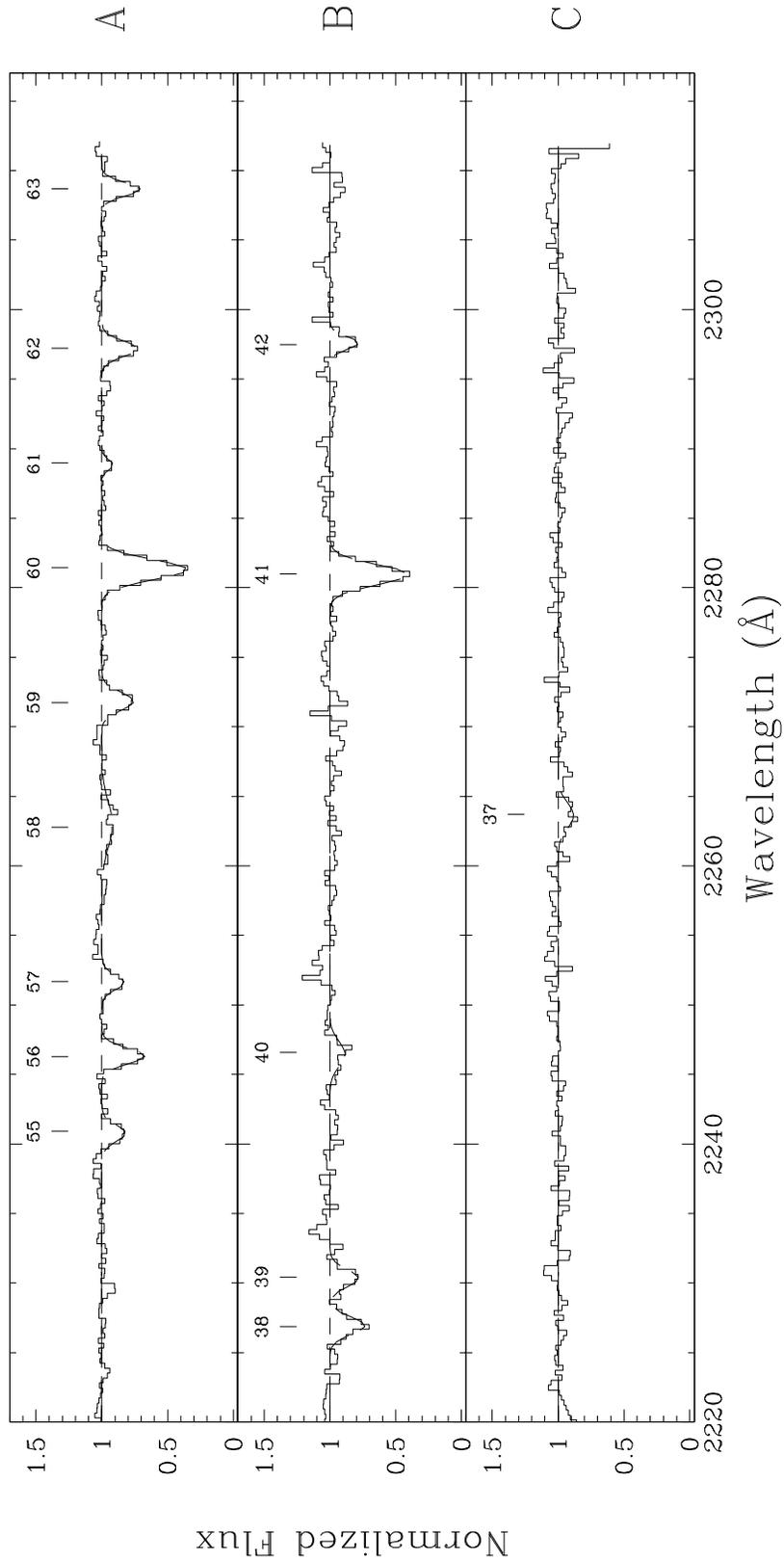}}
\caption{Expanded plot of the {\it HST} FOS G190H spectra for Q0107$-$025 A, B and Q0107$-$0232 (labelled ``C''). The absorption features identified in Tables \ref{lineA190}, \ref{lineB190}, and \ref{lineC190} are marked, and the fits are overplotted with a solid line.}
\end{figure}
\begin{figure}[hp]
\vskip -1.5truecm
\mbox{ \includegraphics[scale=.85, angle=0]{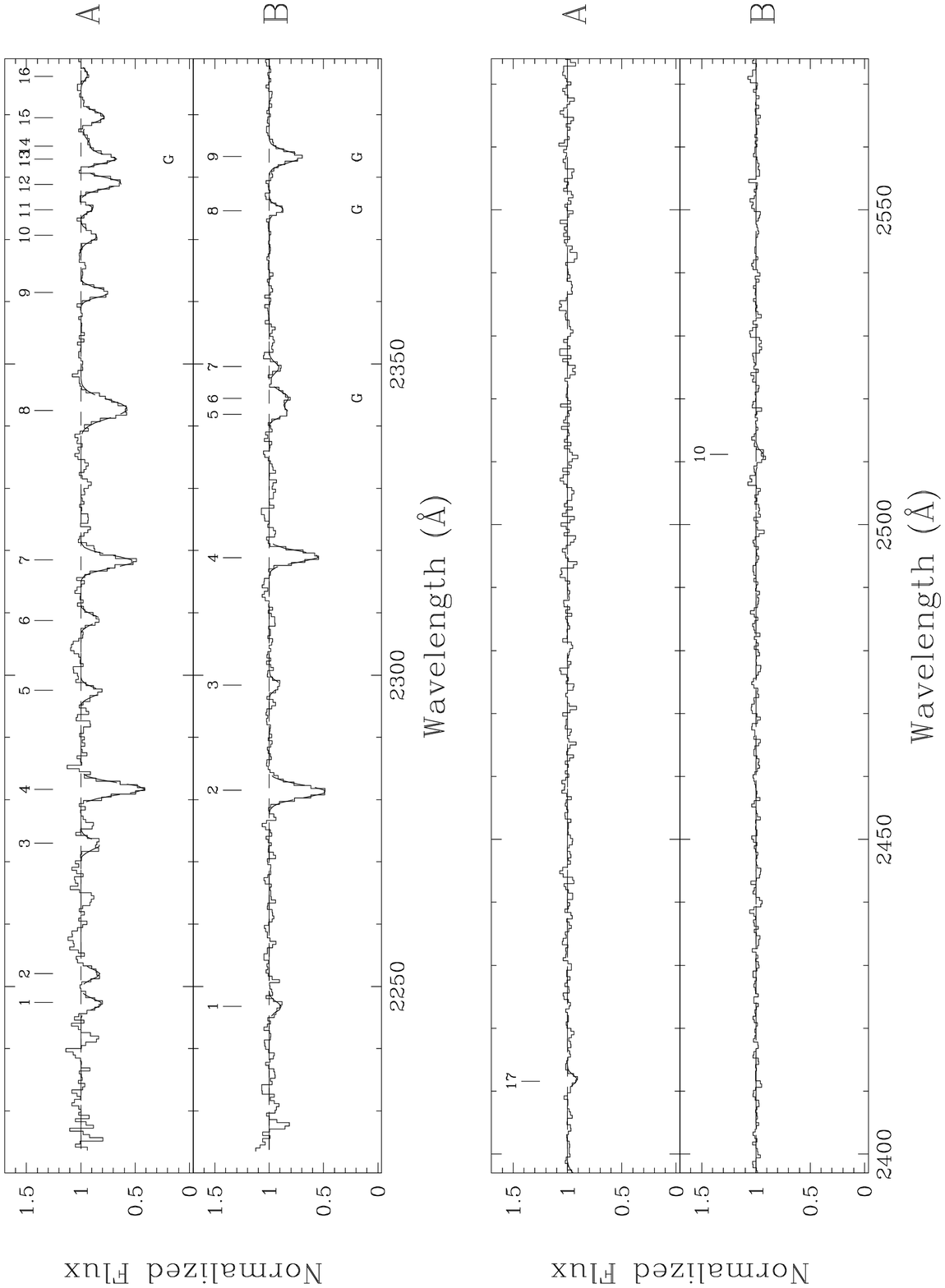}}
\end{figure}
\begin{figure}[hp]
\vskip -1.5truecm
\mbox{ \includegraphics[scale=.85, angle=0]{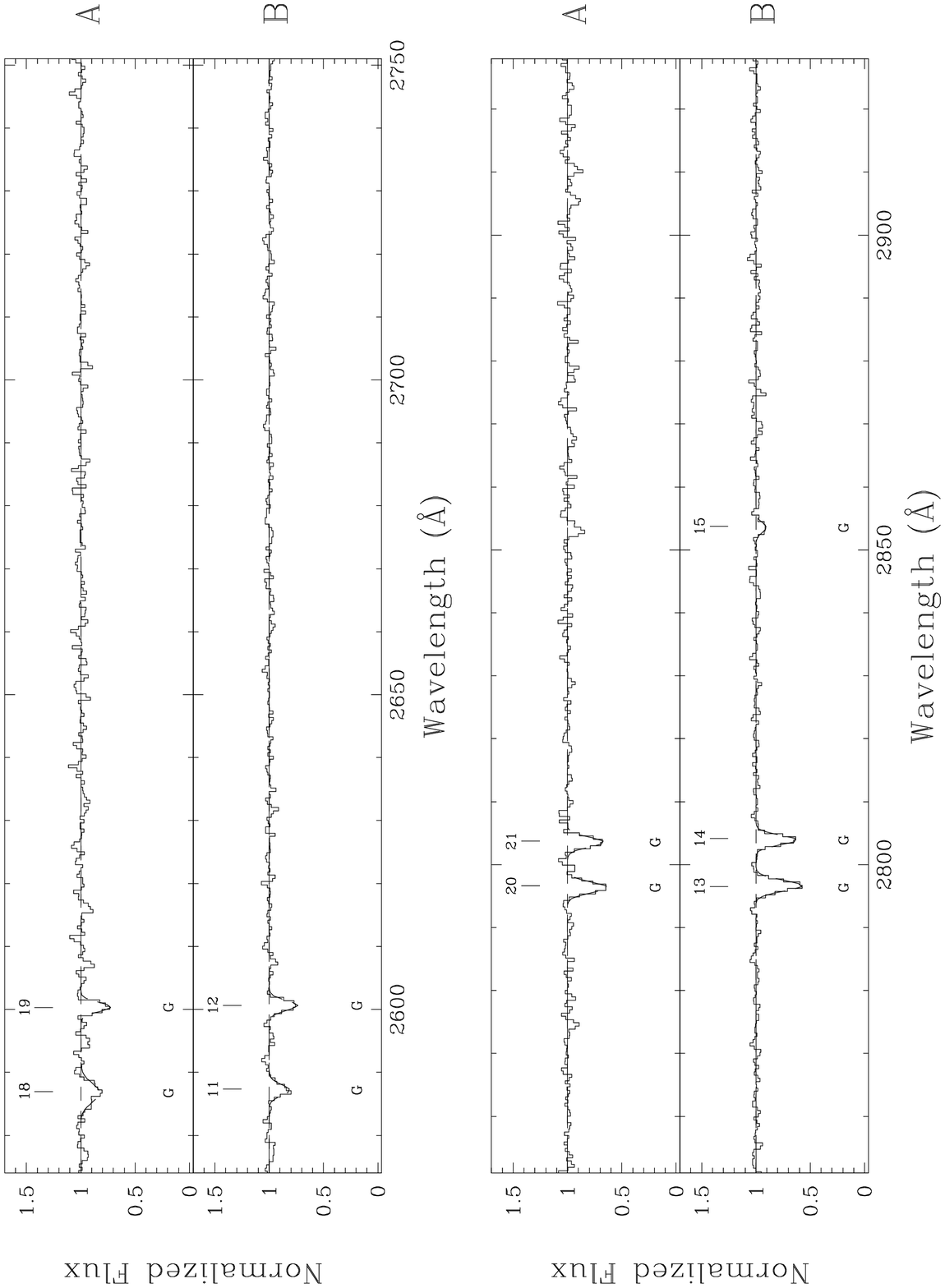}}
\end{figure}
\begin{figure}[hp]
\vskip -1.5truecm
\mbox{ \includegraphics[scale=.85, angle=0]{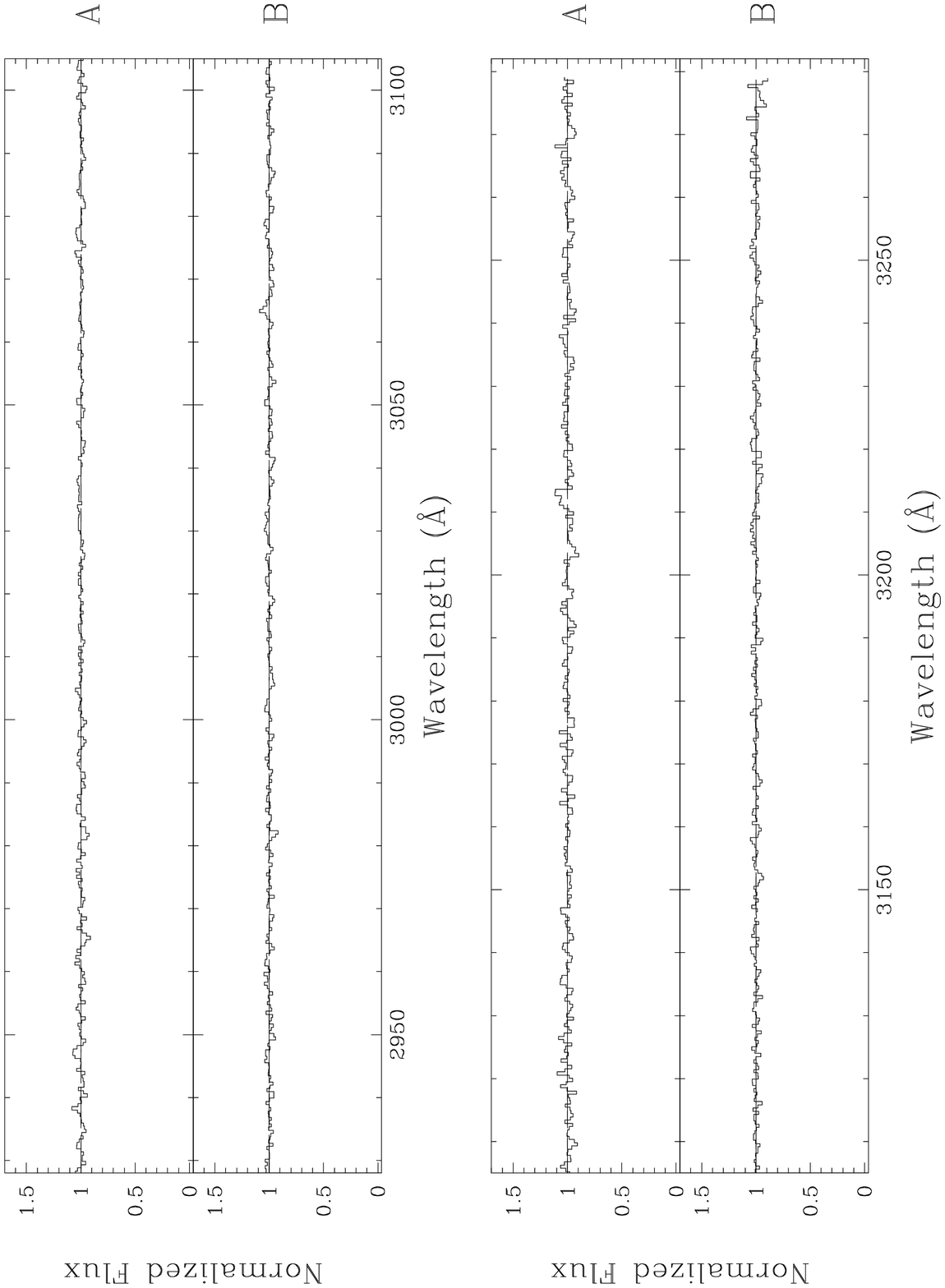} }
\caption{Expanded plot of the {\it HST} FOS G270H spectra for Q0107$-$025 A, B. The absorption features identified in Tables \ref{lineA270} and \ref{lineB270} are marked, and the fits are overplotted with a solid line. }
\end{figure}

\begin{figure}[hp]
\centering
\mbox{ \includegraphics[scale=.7, angle=0]{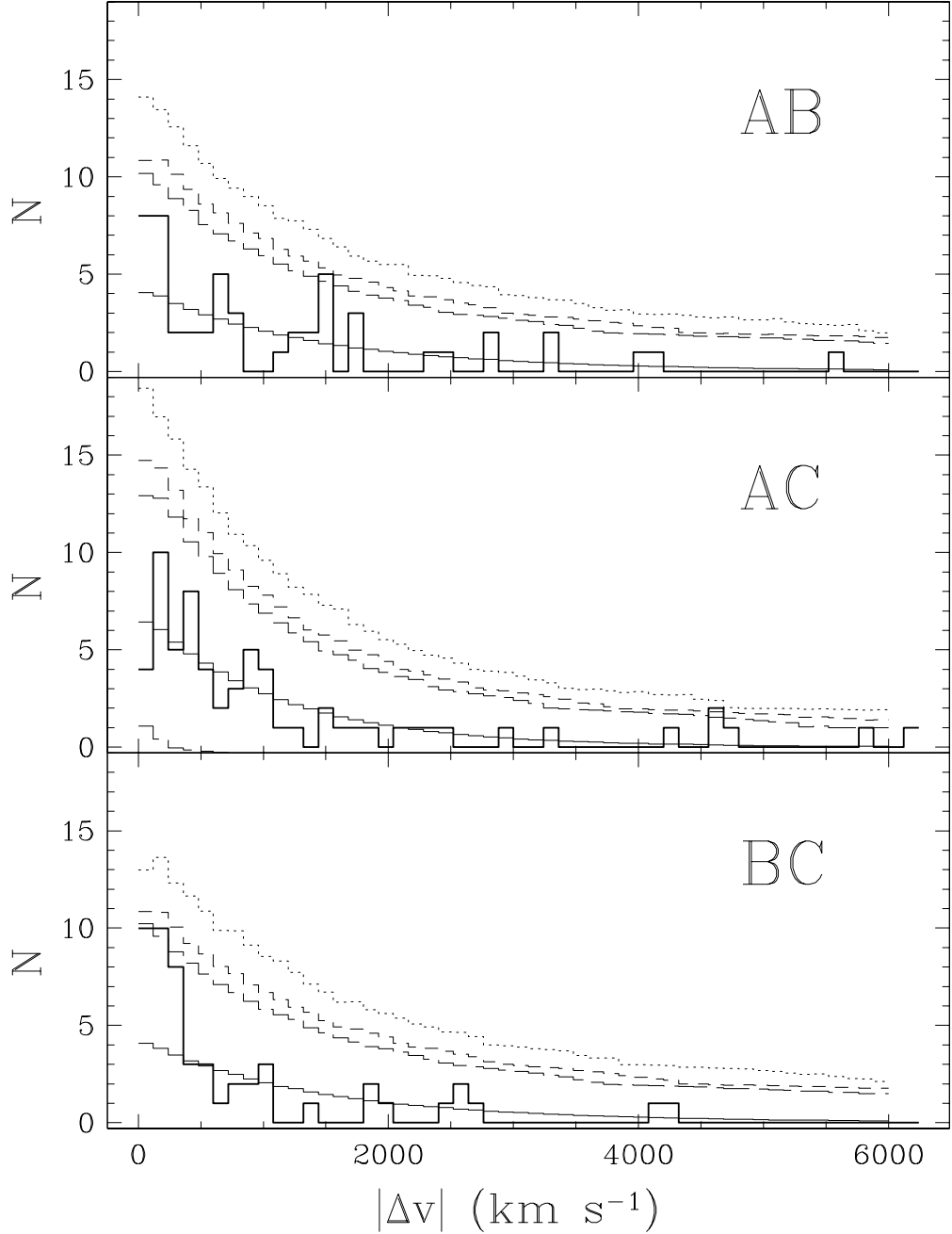} }
\caption{
Observed counts of \lya\ nearest neighbor absorber pairs for each pair of quasar sight lines ({\it thick line}). The expected mean number of absorber pairs for a random distribution in each sight line is shown by the thin solid line through the data. Variation around the random pair counts at the 90\%, 95\% and 99\% confidence intervals was determined using a Monte Carlo simulation of 10,000 realizations where the number of absorbers per sight line in each realization is drawn from a Poisson distribution with the observed number as the mean.  The bins are 120 \kms\ wide.
}
\label{nnmonte}
\end{figure}

\begin{figure}[hp]
\centering
\mbox{ \includegraphics[scale=.6, angle=0]{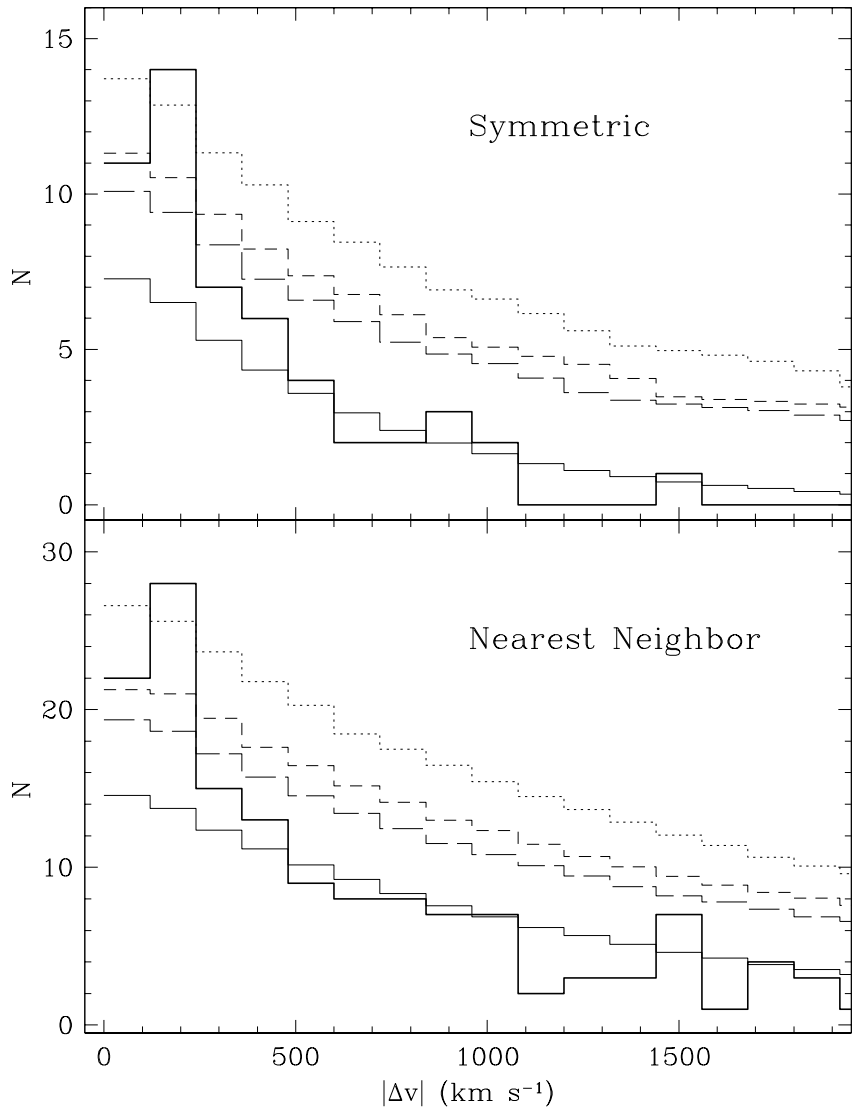} }
\caption{
The observed counts of symmetric ({\it top}) and nearest neighbor ({\it bottom}) \lya\ absorber pairs is shown by the heavy line for the combination of the three pairs of quasar sight lines.  For both panels, the expected mean number of absorber pairs for a random distribution in each sight line is the light smooth solid line through the data. The expected distribution and the 90\%, 95\%, and 99\% confidence intervals are those determined from each pair (see Figure~\ref{nnmonte}) combined in quadrature. The bins are 120 \kms\ wide.
}
\label{allpairs}
\end{figure}

\begin{figure}[hp]
\centering
\mbox{ \includegraphics[scale=0.7, angle=0]{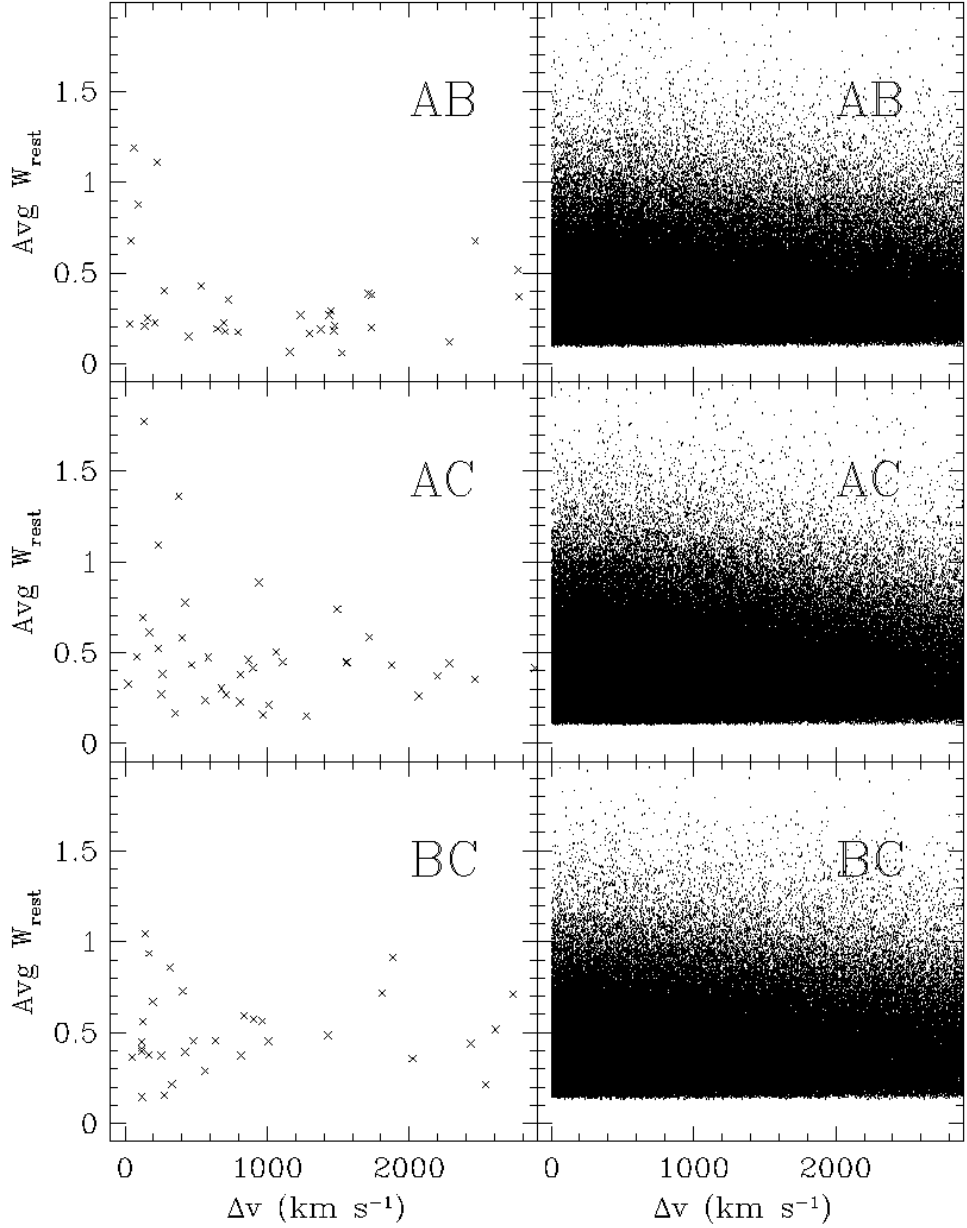} }
\caption{
For each of the three quasar paired sight lines, we plot for each 
nearest neighbor \lya\ absorber
pair the average rest equivalent width vs. its velocity separation.  The left panels are the observed absorber pairs; the right panels are the Monte Carlo simulation pairs.  For the A-B sight line pair there are five additional data points with $|\Delta v| \ge 3000$ \kms, and for the each of the A-C and B-C sight line pairs there is one additional data point with $|\Delta v| \ge 3000$ \kms.
}
\label{nnslice}
\end{figure}

\begin{figure}[hp]
\centering
\mbox{ \includegraphics[scale=.7, angle=0]{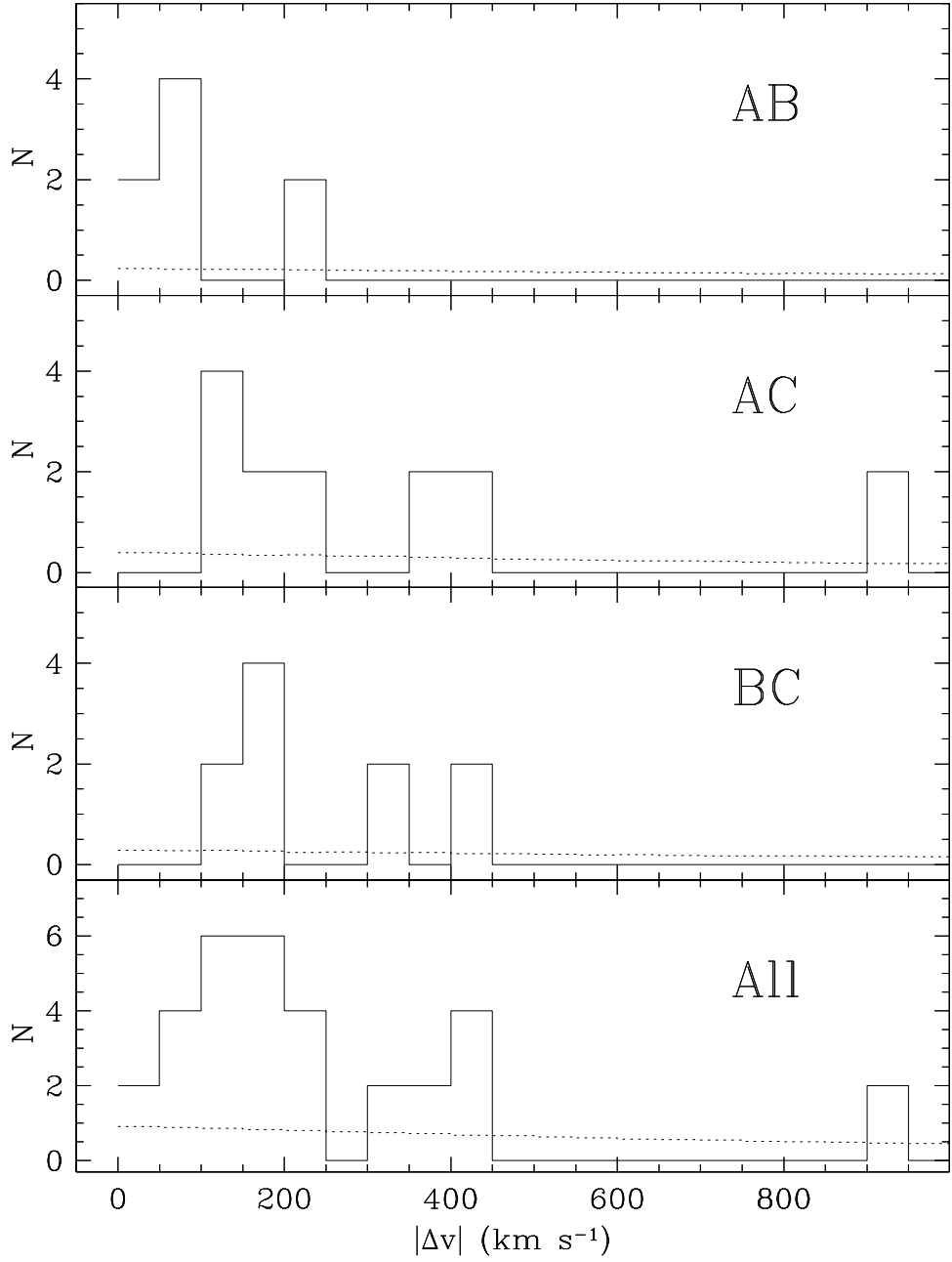} }
\caption{
Distributions of the velocity separation for the strongest nearest neighbor \lya\ absorber pairs, $W_{rest} \ge 0.6$ \AA\ ({\it solid lines}). The overplotted comparison ({\it dotted lines}) is the distribution from the Monte Carlo experiment.
}
\label{nnstrong}
\end{figure}

\begin{figure}[hp]
\centering
\mbox{ \includegraphics[scale=.7, angle=0]{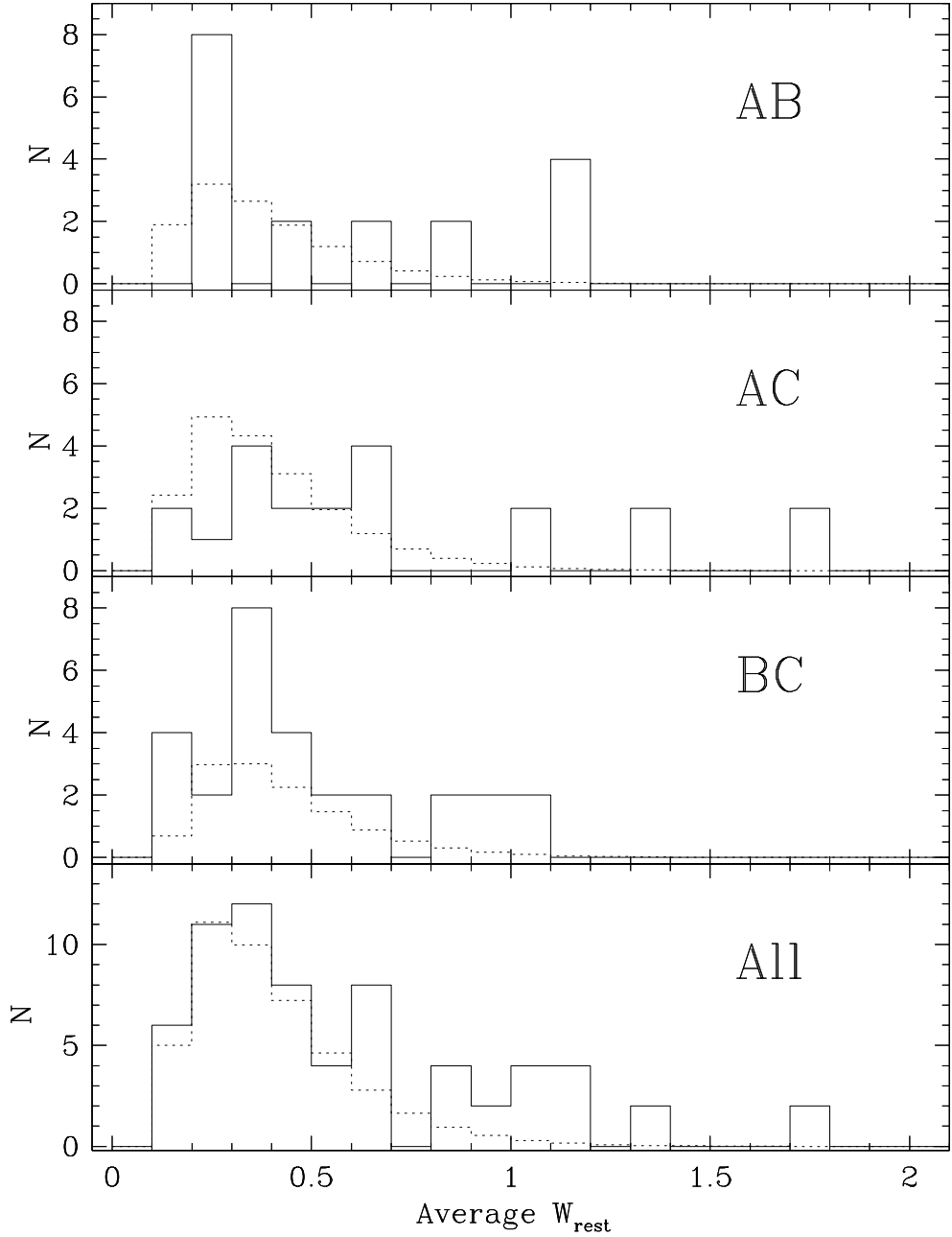} }
\caption{
Distributions of the average rest equivalent width for the closest nearest neighbor \lya\ absorbers pairs, $|\Delta v| < 400$ \kms\ ({\it solid lines}). The overplotted comparison ({\it dotted lines}) is the distribution from the Monte Carlo experiment.
}
\label{nnclose}
\end{figure}

\clearpage

\begin{figure}[hp]
\centering
\mbox{ \includegraphics[scale=.7, angle=0]{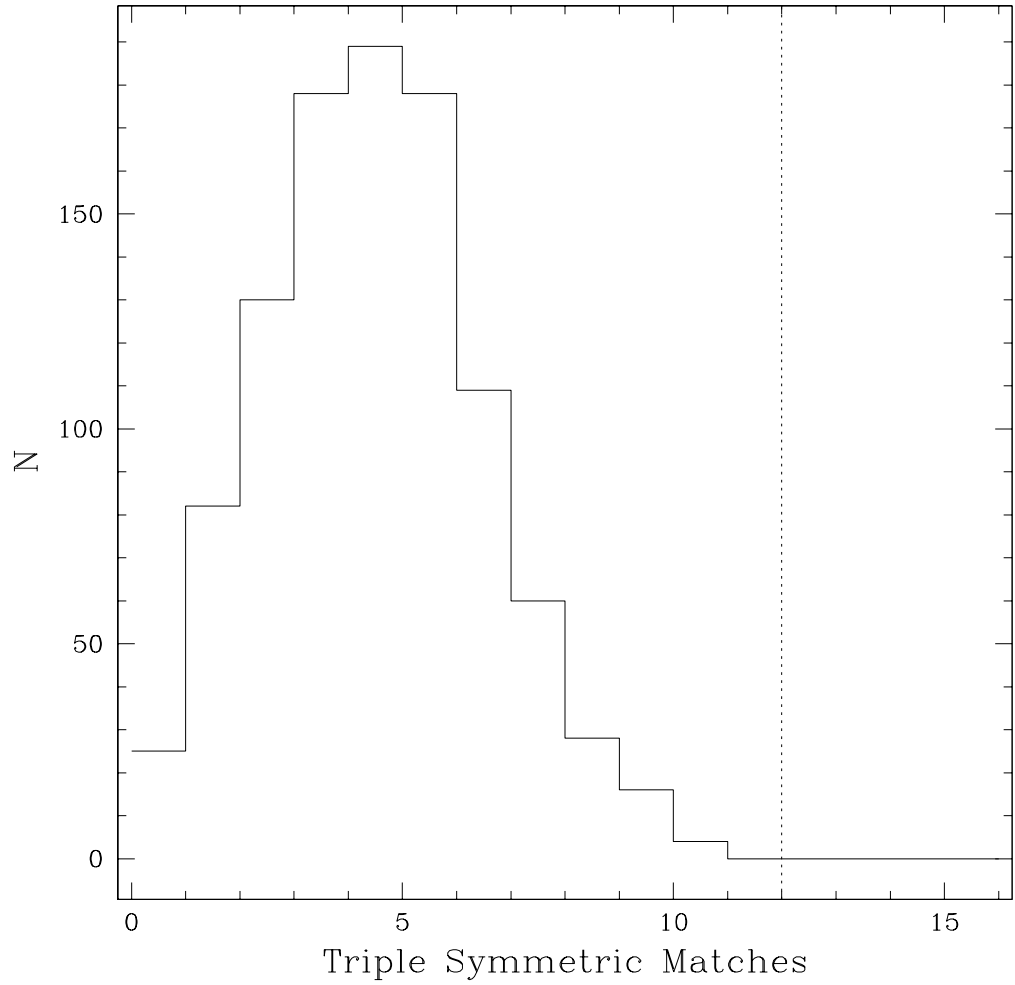} }
\caption{
Distribution of the number of symmetric triples in 1000 Monte Carlo realizations of the observed quasar lines of sight ({\it solid line}). It has a mean of $5\pm2$. The observed number of symmetric triples is indicated by the dotted line.
}
\label{xptriple12}
\end{figure}

\clearpage

\begin{figure}[hp]
\centering
\mbox{ \includegraphics[scale=.7, angle=0]{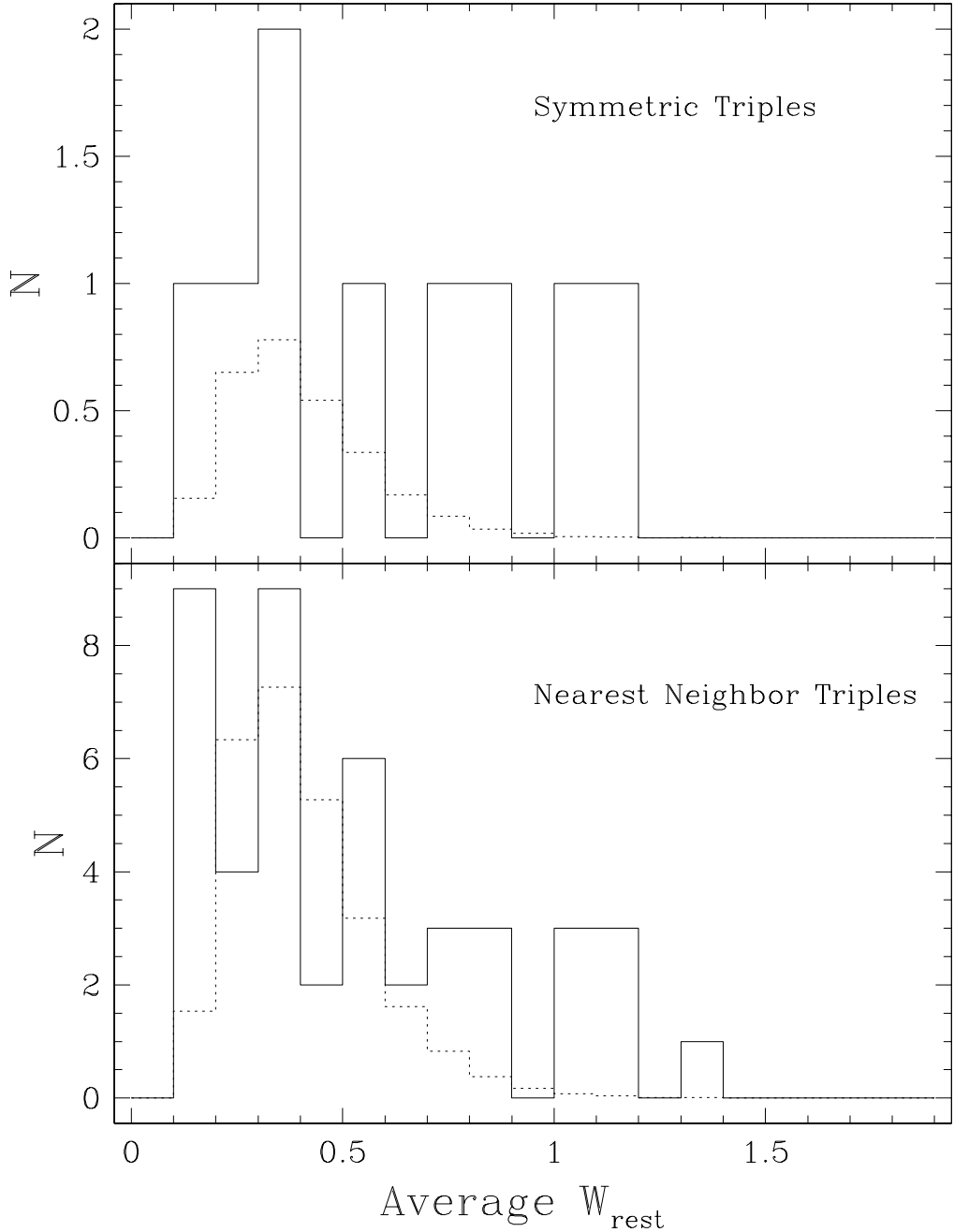} }
\caption{
Average rest equivalent width, $W_{rest}$, calculated for symmetric triples ({\it top}) and nearest neighbor triples ({\it bottom}) that have an average velocity splitting of less than 400 \kms\ (for the two closest splittings). In both panels the distribution of $W_{rest}$ is plotted for the observed triples ({\it solid lines}), and the normalized distribution for the Monte Carlo triples is overplotted.
}
\label{xptriple400}
\end{figure}

\begin{figure}[hp]
\centering
\mbox{ \includegraphics[scale=.7, angle=0]{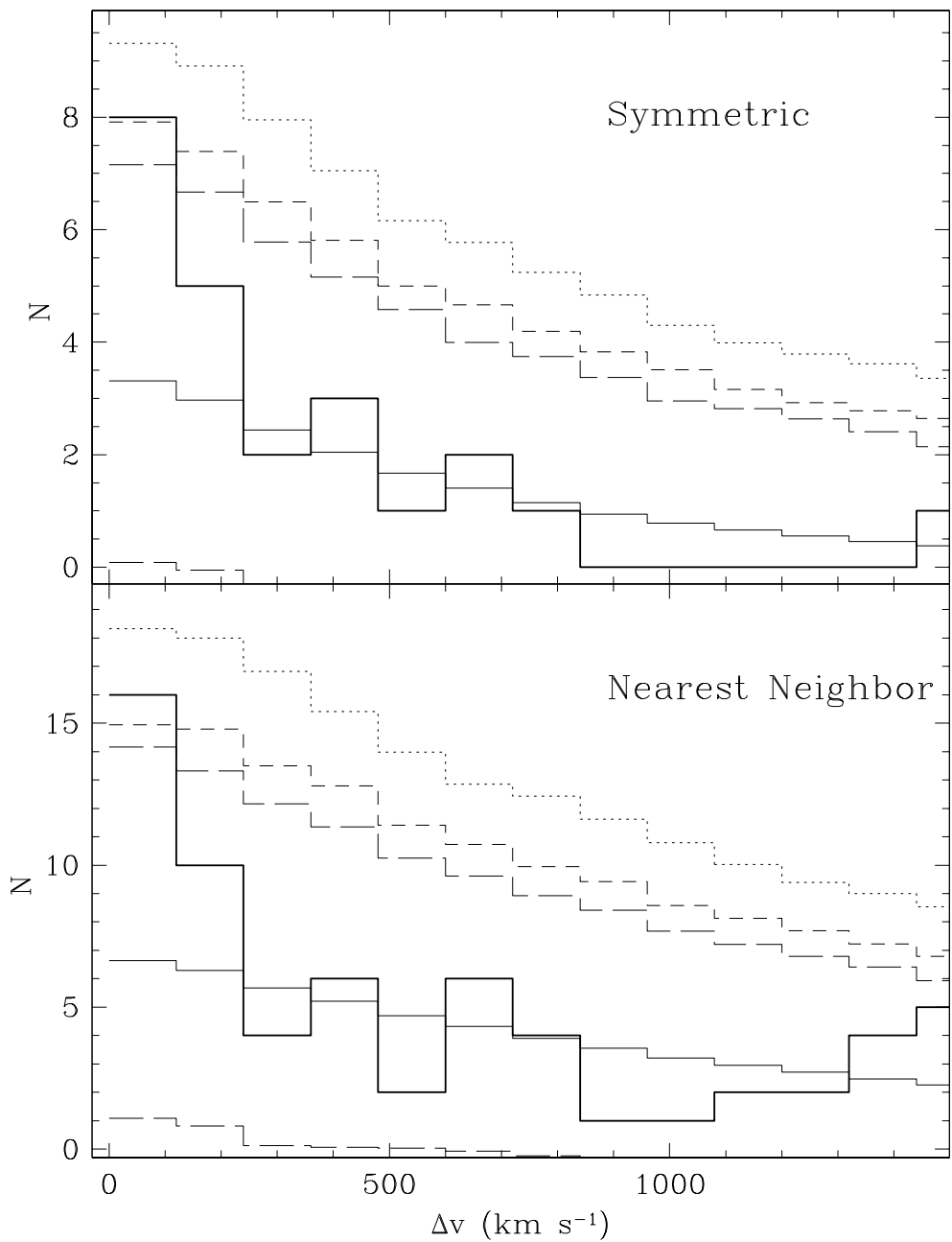} }
\caption{
Distribution of the observed symmetric ({\it top}) and nearest neighbor ({\it bottom}) \lya\ absorber pairs for the AB pair of quasar sight lines with extended wavelength coverage ({\it thick lines}). The expected mean number of absorber pairs for a random distribution of absorbers in each sight line is shown by the thin solid line through the data. Variation around the random pair counts at the 90\%, 95\%, and 99\% confidence intervals was determined using a Monte Carlo simulation of 10,000 realizations, where the number of absorbers per sight line in each realization is drawn from a Poisson distribution with the observed number as the mean.  The bins are 120 \kms\ wide.
}
\label{ABpair}
\end{figure}

\end{document}